\documentclass[useAMS,usenatbib]{mn2e}
\usepackage{threeparttable}

\usepackage{longtable}

\usepackage[T1]{fontenc}
\usepackage{aecompl}

\usepackage{amssymb}
\usepackage{mathptmx}       % selects Times Roman as basic font
\usepackage{helvet}         % selects Helvetica as sans-serif font
\usepackage{courier}        % selects Courier as typewriter font
\usepackage{type1cm}        % activate if the above 3 fonts are
\usepackage{graphicx}        % standard LaTeX graphics tool
\usepackage{multicol}        % used for the two-column index
\usepackage[bottom]{footmisc}% places footnotes at page bottom

\usepackage{natbib}
\usepackage{aas_macros}  % Journal abreviations defined in ADS bibtex
\usepackage{array}
\newcolumntype{L}{>{\centering\arraybackslash}m{2.1cm}}
\usepackage{textcomp}

\usepackage{lscape}
\usepackage[normalem]{ulem}
\usepackage[usenames,dvipsnames]{xcolor}
\makeatother
\title[Virgo Cluster and field dEs in 3D -- II. Internal dynamics]{Virgo Cluster and field dwarf ellipticals in 3D -- \\II. Internal dynamics points to tidal harassment?}
\author[A. Ry\'{s}, G. van de Ven, J. Falc\'{o}n-Barroso]{A. Ry\'{s}$^{1,2}$\thanks{E-mail:
arys@iac.es}, G. van de Ven$^{3}$ and J. Falc\'{o}n-Barroso$^{1,2}$\\
$^{1}$Instituto de Astrof\'{i}sica de Canarias, 38200 La Laguna, Tenerife, Spain\\
$^{2}$Departamento de Astrof\'{i}sica, Universidad de La Laguna, 38205 La Laguna, Tenerife, Spain \\
$^{3}$Max Planck Institute for Astronomy, K\"{o}nigstuhl 17, 69117 Heidelberg, Germany}

\begin{document}

% \date{Accepted 1988 December 15. Received 1988 December 14; in original form 1988 October 11}
\date{}

% \pagerange{\pageref{firstpage}--\pageref{lastpage}} \pubyear{2002}

\maketitle

\label{firstpage}

\begin{abstract}
We present the dynamical analysis of a sample of 12 dwarf elliptical (dE) galaxies for which we have obtained SAURON large-scale two-dimensional spectroscopic data.  We construct Jeans axisymmetric models and obtain total dynamical masses enclosed within one effective radius. We use the obtained values to show that the validity of the dynamical scaling relations of massive early-type galaxies can be extended to these low-mass systems, except that dEs seem to  contain relatively larger fraction of dark matter in their inner parts. We then demonstrate that dE galaxies have lower angular momenta than the present-day analogues of their presumed late-type progenitors and we show that dE circular velocity curves are steeper than the rotation curves of galaxies with equal and up to an order of magnitude higher luminosity. This requires a transformation mechanism that is not only able to lower the angular momentum but also one that needs to account for increased stellar concentration. Additionally, we  match the dark matter fraction of our galaxies to their location in the Virgo Cluster and see that galaxies in the cluster outskirts tend to have a higher dark-to-stellar matter ratio. Transformation due to tidal harassment is able to expain all of the above, unless the dE progenitors were already compact and had lower angular momenta at higher redshifts.
\end{abstract}


\section{Introduction}

While the total amount of (and the need for) dark matter (DM) in the Universe has been established rather accurately, its properties at galaxy scales have been more difficult to assess. Additionally, structure formation at higher galaxy masses we are now beginning to understand fairly well (see e.g. \citeauthor{cappellari:2013a} \citeyear{cappellari:2013a}a), but much less is known about the low-mass galaxies, which dominate in number. There exist detailed studies of the Local Group dwarf elliptical galaxies (e.g. \citealt{geha:2010}) because their proximity allows one to resolve individual stars and thus probe arbitrarily large galactocentric distances. This approach is, however, not feasible in the case of even the nearest galaxy clusters, and that is where dEs are found in large numbers. 

Kinematic and dynamical studies of those cluster dwarf galaxies have traditionally been hampered by the lack of high-quality data that would provide enough radial and spatial coverage (e.g. \citealt{simien:2002}, \citealt{geha:2003}, \citealt{derijcke:2005}, \citealt{chilingarian:2009}). Because of their one-dimensional nature, certain aspects and questions regarding the suspected structural complexity of these dwarfs still eluded us. A need arose to turn to integral-field spectroscopy for answers. Our survey came as a response to this challenge. It was also very timely as large-scale integral-field units (IFUs) had recently become available and were being applied to studies of giant early-type galaxies. We decided to extend those analyses to low-mass systems.

The latest evidence from photometric, stellar population and kinematic studies shows dEs as structurally complex systems and supports the scenario of them being environmentally-transformed late-types: this is based on the variety of their properties as well as clustrocentric trends arguments (e.g. \citealt{lisker:2007}, \citealt{janz:2012}, \citealt{toloba:2009}, \citealt{smith:2009}, \citealt{kormendy:2012}, \citealt{rys:2013}). These trends qualitatively agree with theoretical predictions if we employ ram-pressure stripping and harassment as the mechanisms driving the galaxies' structural evolution (e.g. \citealt{moore:1998}, \citealt{smith:2010}). 

Dwarf galaxies were for a long time omitted from galaxy classification diagrams. While dwarf late-types were included in the de Vaucoulers (1959) scheme, it was not until much later that dwarf early-types were explicitly included as well, first in \cite{sandage:1984} (as dwarf version of giant ellipticals but indicating a possible connection to dwarf late-types) and more recently in \cite{kormendy:2012} where they earned a place in the authors' updated tuning fork as a ``bulgeless extension'' of the S0 galaxies tine. \cite{cappellari:2011} for the first time presented a  galaxy classification scheme based on \textit{kinematic} properties, and later in \citeauthor{cappellari:2013b} (\citeyear{cappellari:2013b}b) dwarf early-type\,/\,spheroidal galaxies were included in this new picture, albeit based on approximate mass values. Given that these revised classification schemes go beyond simple morphology and try to incorporate the newest findings on galaxy kinematics and dynamics, it is essential to perform the same kind of detailed analysis for the low-mass systems. This will not only verify their classification status but, more importantly, help us better understand the relation to other classes and look for evidence in favor or against various proposed formation scenarios.

In this paper we analyze the dynamical properties of dE galaxies and juxtapose them with those of other galaxy types: giant early and late-types as well as late-type galaxies of luminosities comparable to those of our sample. Then, we introduce the location-in-the-cluster variable to investigate the dependence of those properties on local environment density. We also place our galaxies on the dynamical scaling relations of early-type galaxies and discuss the nature of the relations. The paper is structured as follows. Section~2 presents the sample and the photometric and kinematic analysis. In Section~3 we describe our dynamical models. We present the results in Section~4, and then discuss them in Section~5. Our summary and conclusions are in Section~6.

\section{Stellar photometry and kinematics}

\subsection{Sample selection}

A detailed presentation of our dataset can be found in \cite{rys:2013}. Briefly, we observed 12 dEs: 9 in the Virgo Cluster and 3 in the field. Virgo was chosen as it the closest and most abundant reservoir of the dE galaxy class. Our objects span a wide range of ellipticities and distances from the cluster's central galaxy M87. The sample includes a few field objects to be compared with their cluster counterparts. The name, distance to, and photometric properties of each observed object are listed in Table~\ref{observations}. One of our field objects, NGC\,3073 is only included in the angular momentum analysis since due to a high gas emission we were yet unable to create a reliable stellar mass model for the galaxy.

\subsection{Photometry}

For our analysis we used archival SDSS r-band images. The calibrated images were retrieved from the IPAC montage service website (\textit{http://hachi.ipac.caltech.edu:8080/montage}). Effective radii were obtained from a fit of a de Vaucouleurs R$^{1/n}$ growth curve to the aperture photometry profiles, as detailed in \cite{falcon:2011b}. We first masked stars and other undesired objects in the original image using the SExtractor software of \cite{bertin:1996}. We then created circular aperture profiles, which were used to build growth curves and the effective radii. The errors on the measured radii and the \textit{n} values were obtained through Monte-Carlo simulations varying noise and background level.

\begin{table*}
\caption{Properties of the observed objects: (1) name, (2) heliocentric distance based on surface brightness fluctuation (SBF) or globular cluster luminosity function (GCLF) distances  for Virgo galaxies where available (\protect\citealt{mei:2007}$\mathrm{^{1}}$, \protect\citealt{jerjen:2004}$\mathrm{^{2}}$ or \protect\citealt{jordan:2007}$\mathrm{^{3}}$) otherwise an average Virgo distance assumed$\mathrm{^{4}}$, for NGC\,3073 a SBF distance from \protect\cite{tonry:2001}, the distances to ID\,0650 and ID\,0918 were calculated using their radial velocities; all distances are corrected to H0 = 73 km/s/Mpc for consistency with \protect\cite{mei:2007}, (3) morphological type (from \protect\citealt{lisker:2007}), (4) ellipticity at one effecitve radius, (5) r-band effective radius $R_e=\sqrt(A_e / \pi)$ where $A_e$ is the area of the effective isophote containing half of the galaxy light and (6) the major axis of that isophote, both obtained by fitting our aperture photometry profiles with $R^{1/n}$ \protect\cite{sersic:1963} growth curves (see Sec. 4.1 of \protect\citealt{falcon:2011a}), (7) sersic index $n$ used in the $R_e$ calculations, (8) r-band apparent magnitude, (9) velocity dispersion measured within one effective radius $\sigma_e$, (10) specific angular momentum within one effective radius  $\lambda_{Re}$, (11) maximum circular velocity $V_{circ}^{max}$, (12) dynamical mass-to-light ratios from JAM mass-follows-light models $(M/L)_{dyn}$ integrated within 1\,R$_e$, (13) total r-band luminosity, and (14) for Virgo galaxies: r-band stellar mass-to-light ratios based on \protect\cite{lisker:2008} g-r colors and \protect\cite{bell:2003} color-to-$(M/L)_{pop}$ conversion.} 

%  \begin{threeparttable}
\centering
% \vspace{0.5cm}
\begin{tabular}{rrrrrrrrrrrrrr}
\hline
object   & distance           & type    &$\epsilon$&R$_e$ &R$_e^{maj}$&$n$& m$_r$ &$\sigma_e$&$\lambda_{Re}$&$V_{circ}^{max}$&$(M/L)_{dyn}$      & $log(L)$  &$(M/L)_{pop}$ \\ 
         &  $(Mpc)$           &         &          &$('')$&$('')$     &   &$(mag)$&$(km/s)$  &              &$(km/s)$        &$(M_\odot/L_{\odot,r})$&$(L_{\odot,r})$&$(M_\odot/L_{\odot})$ \\ 
(1)      &(2)                 &(3)      &(4)       &(5)   &(6)        &(7)&(8)    &(9)       &(10)           &(11)             &(12)               &(13)       &(14) \\
\hline
VCC\,0308&16.50$\mathrm{^{4}}$&dE(di;bc)&0.07      &18.7  &19.4  &1.2  &13.32  &    36.21 &         0.26 &            65.1&   4.59 $\pm$ 2.43 &       8.68&1.58 \\     %1.82
VCC\,0523&16.50$\mathrm{^{4}}$&dE(di)   &0.29      &27.9  &33.1  &2.4  &12.60  &    47.65 &         0.36 &            81.7&   4.85 $\pm$ 2.11 &       8.97&1.66 \\     %1.94
VCC\,0929&14.86$\mathrm{^{2}}$&dE(N)    &0.11      &22.1  &23.4  &2.5  &12.65  &    53.86 &         0.24 &            96.0&   6.02 $\pm$ 2.06 &       8.87&1.88 \\     %2.30
VCC\,1036&16.07$\mathrm{^{2}}$&dE(di)   &0.56      &17.2  &25.9  &2.5  &13.13  &    62.82 &         0.16 &            98.8&   7.55 $\pm$ 1.87 &       8.75&1.88 \\     %2.30
VCC\,1087&16.67$\mathrm{^{1}}$&dE(N)    &0.31      &28.6  &34.4  &2.0  &12.85  &    39.55 &         0.18 &            69.4&   4.80 $\pm$ 2.42 &       8.86&1.99 \\     %2.49
VCC\,1261&18.11$\mathrm{^{1}}$&dE(N)    &0.42      &19.7  &25.9  &1.9  &12.87  &    51.68 &         0.12 &            88.7&   5.49 $\pm$ 2.34 &       8.94&1.78 \\     %2.13
VCC\,1431&16.14$\mathrm{^{1}}$&dE(N)    &0.03      & 9.6  & 9.7  &1.5  &13.60  &    53.37 &         0.13 &            89.0&   5.49 $\pm$ 1.96 &       8.55&2.25 \\     %2.94
VCC\,1861&16.14$\mathrm{^{1}}$&dE(N)    &0.01      &20.1  &20.2  &1.9  &13.41  &    36.00 &         0.21 &            60.0&   4.28 $\pm$ 1.96 &       8.63&1.93 \\     %2.39
VCC\,2048&14.45$\mathrm{^{3}}$&dE(di)   &0.48      &16.5  &22.9  &3.0  &13.08  &    48.02 &         0.22 &            85.4&   4.76 $\pm$ 1.80 &       8.78&1.73 \\     %2.05
NGC\,3073&17.8                &dE/S0    &0.15      &16.1  &17.5  &2.5  &12.98  &    46.23 &         0.16 &               -&                  -&          -&- \\
ID\,0650 &25.9                &dE/S0    &0.10      &20.1  &21.2  &2.8  &13.73  &    43.75 &         0.21 &            70.1&   3.81 $\pm$ 1.84 &       8.93&- \\
ID\,0918 &16.3                &dE/E     &0.27      & 6.4  & 7.5  &2.5  &13.79  &    77.61 &         0.19 &           127.3&   6.21 $\pm$ 1.21 &       8.53&- \\
\hline
\end{tabular}
\label{observations} 
% \end{threeparttable}
\end{table*}

\begin{figure}
\centering
\includegraphics[width=0.99\columnwidth]{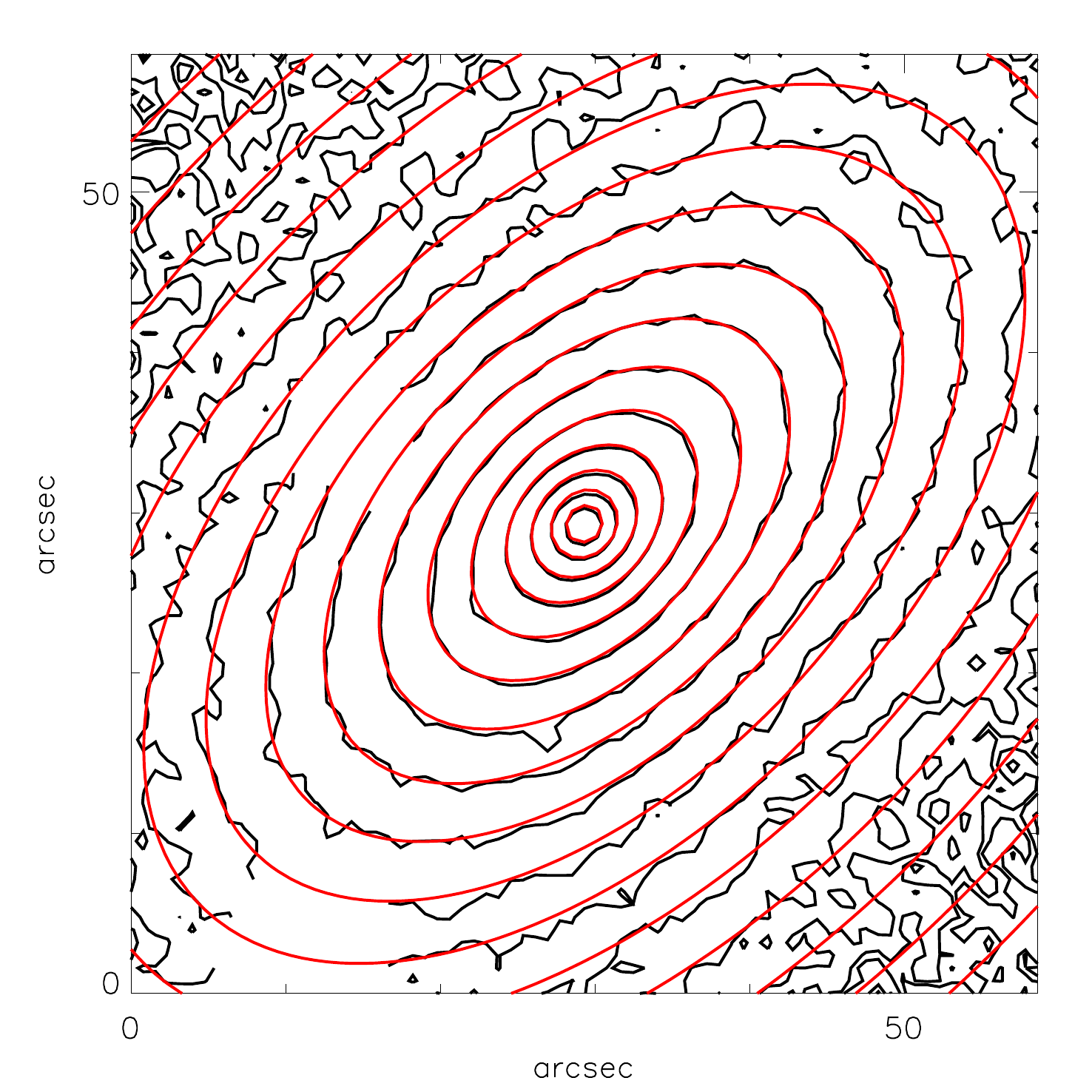}
\includegraphics[width=0.99\columnwidth]{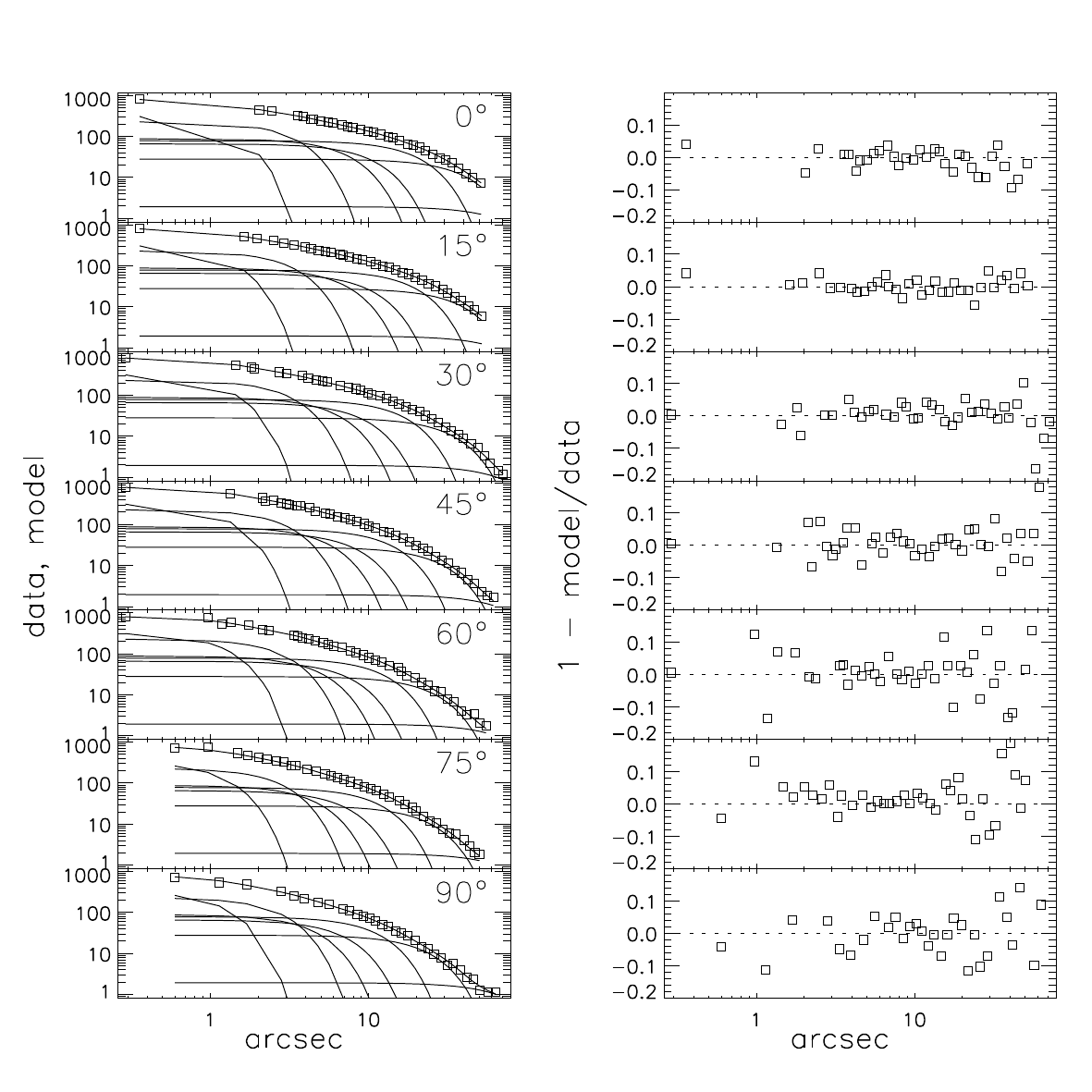}
\caption{Example MGE model of VCC1261. \underline{\textit{Top:}} a model overplotted on the input SExtractor-cleaned SDSS r-band images. \underline{\textit{Bottom left:}} comparison between the extracted 1D profiles (open squares, shown here for a few representative 5 deg-wide sectors) and their corresponding convolved model fits (solid lines). The individual gaussians used in the fit are also shown. \underline{\textit{Bottom right}}: radial variations of the relative errors for each sector. Note that galaxies with observed kinematic twist are still fitted with a symmetric MGE profile, required by the JAM routines.}
\label{MGE-example}
\end{figure}

Images of astronomical objects contain millions of pixels. Manipulating such a number of data points when, e.g. performing a deprojection, would be an arduous and time-consuming task. In order to facilitate the task one tries to come up with analytical models that would accurately describe the observed surface brightness profiles while using only a limited number of parameters.

The traditional approach was to employ ellipse fitting and describe any deviations using Fourier terms (e.g. \citealt{bender:1987}, see also the discussion in \citealt{cappellari:2002}). The method had its limitations, however, mostly because describing more complicated (i.e. deviant from ellipses) features required using a large number of Fourier terms and made the approach not well suited for modelling multicomponent objects and also significantly increased the complexity of the deprojection task.

Multi-gaussian expansion (MGE, \citealt{emsellem:1994}) is an analytical description of a 2D image based on the assumption that the intrinsic density can be described as a sum of coaxial (triaxial) gaussian distributions. Using an MGE model gives an advantage of being able to describe the potential with a single integral in lieu of a triple one (in a general case). What stems from this is that we are able to perform an analytical deprojection, as well as an analytical convolution with the (MGE) point-spread function.

A detailed description of the MGE fitting method is provided in \cite{cappellari:2002} whose fitting software was used in the present study.  Here we provide a summary which highlights its main points. We used the 2D algorithm which assumes that profiles can be accurately measured in a certain number of sectors that the galaxy is divided into. The sectors are equally spaced in angle and logarithmically in (elliptical) radius to ensure proper minimum S/N level. The profiles were measured along 36 sectors 5$^o$ wide, covering the whole galaxy image from -90$^o$ to 90$^o$. An example fit is shown in Figure~\ref{MGE-example} (contour map of the image with model contours overplotted and the comparison of the brightness profiles with their model fits overplotted, also showing the individual gaussian components). Our models were constructed using gaussian components that were as round as possible (while still able to reproduce galaxy characteristics) because the minimum allowed inclination is set by the flattest component. In Appendix~\ref{mgetables} we include tabulated models for all the galaxies.

For dynamical modelling we assume oblate axisymmetry so that the position angle (PA) of photometry and kinematics are aligned and constant, i.e. $PA_{phot}=PA_{kin}=const$. For the galaxies that show a misalignment between their photometric and kinematic major axes (VCC0523 and ID 0918) the kinematic PA was used as more representative of the galaxy properties over the whole field of view. While this produces seemingly worse MGE fits in the center, it was shown by \cite{lablanche:2012} that the approach does not affect the recovered M/L values.

The resulting models were later corrected for galactic extinction, whose values were retrieved from the NASA Extragalactic Database (NED, \textit{http://ned.ipac.caltech.edu/}) and are based on \cite{schlegel:1998}. The r-band magnitude of the Sun was assumed to be 4.67\,mag (\citealt{binney:1998}). Finally, the intensities of the MGE gaussian parameters were transformed into surface brightness ($L_{\odot}/pc^3$).

\subsection{Stellar kinematics}

A detailed presentation of our data reduction and kinematic analysis procedures can be found in  \cite{rys:2013}. Here we provide a short summary for the reader's convenience.

The observations were carried out in Jan 2010 and Apr 2011 (8 nights in total) using the WHT/SAURON instrument at the Roque de los Muchachos Observatory in La Palma, with each galaxy typically exposed for 5h. For the extraction and calibration of the data we followed the procedures described in \cite{bacon:2001} using the specifically designed XSAURON software developed at the Centre de Recherche Astrophysique de Lyon (CRAL). The data were spatially binned, using the two-dimensional Voronoi binning algorithm developed by \cite{cappellari:2003}, to achieve the required minimum signal-to-noise ratio of 30, with central bins having the S/N ratio exceeding 100.\looseness-2

Stellar absorption-line kinematics were derived for each galaxy by directly fitting the spectra in the pixel space using the penalized pixel-fitting method (pPXF) of \cite{cappellari:2004}. The method fits a stellar template spectrum convolved with a line-of-sight velocity dispersion (LOSVD) to the observed galaxy spectrum in pixel space (logarithmically binned in wavelength). The algorithm finds the best fit to the galaxy spectrum and returns the mean velocity $V$ and the velocity dispersion $\sigma$. The procedure of creating an optimal stellar template was repeated for each bin using stellar spectra from the MILES library (\citealt{sanchez-blazquez:2006}, \citealt{falcon:2011a}) containing spectra of 985 stars. Error estimates were obtained through performing Monte Carlo simulations and measuring kinematics of the different realizations of the input spectra with added noise.

% \subsection{Stellar angular momentum}
The ratio of maximum stellar velocity $V$ and velocity dispersion $\sigma$ has traditionally been used to determine the level of pressure/rotational support of a galaxy. It has, however, its limitations: very different velocity structures may still give similar V/$\sigma$ values (see section 3.1 in \citealt{emsellem:2007}). To estimate the amount of rotation in our objects we have therefore decided to use the new $\lambda_R$ parameter defined in \cite{emsellem:2007}, as it is better suited for angular momentum estimation than the traditionally employed V/$\sigma$ (ordered versus random motion). 

\section{Dynamical models}

\subsection{Jeans axisymmetric models}

Modelling methods using the solutions of the Jeans equations connect the second-order velocity moments directly to the density and the gravitational potential of a galaxy, without the need to know the phase-space distribution function, but also without the assurance of a physical distribution, e.g. the wings of the underlying distribution function can be negative but otherwise have a positive second-order velocity moment. Since in nearly all cases there are fewer velocity moments than there are equations, additional assumptions have to be made regarding the degree of anisotropy. Jeans models can range from very simple spherical models with assumed isotropic velocity distribution to general solution of the triaxial Jeans equations (e.g. \citealt{ven:2003}). 

Axisymmetric Jeans models are a simple yet realistic dynamical modelling method which can be applied to the measurements of both the M/L and the amount of rotation in galaxies (e.g. \citealt{cappellari:2008}). We fit Jeans anisotropic MGE (JAM, \citealt{cappellari:2008}) models to observed second velocity moment $\sqrt{V^2+\sigma^2}$ with the following parameters: inclination $i$, mass-to-light ratio (profile) $M/L$, meridional plane velocity anisotropy (profile) $\beta$ including $\beta_0$, $\beta_{inf}$, break radius $R_{\beta}/R_e$\footnote{We define the $\beta$ profile as: $\beta = \beta_0 + (\beta_{inf} - \beta_0) \cdot R^2/(R_{\beta}^2+R^2)$ where $\beta_0$ is the anisotropy in the center of the galaxy, $\beta_{inf}$ the anisotropy value at infinity, and $R_{\beta}$ is the break radius. The profile was varied when different models were being tested, however, in the adopted models $\beta$ was assumed constant.}, and a black hole (BH) mass. A surface brightness profile in the form of an MGE model table is also specified.

The luminous and potential gaussians are then defined, the latter by assuming a certain M/L ratio as well as its variation with radius. The JAM script then calculates a prediction for the projected second velocity moments $V_{RMS} = \sqrt(V^2 + \sigma^2)$. It implements the solution of the anisotropic Jeans equations presented in equation (28) of \cite{cappellari:2008}. Projected and intrinsic (enclosed) mass profiles are subsequently calculated, as is the mass profile - dynamical M/L and velocity anisotropy within 1 $R_e$.

\begin{figure}
\centering
\includegraphics[width=0.8\columnwidth]{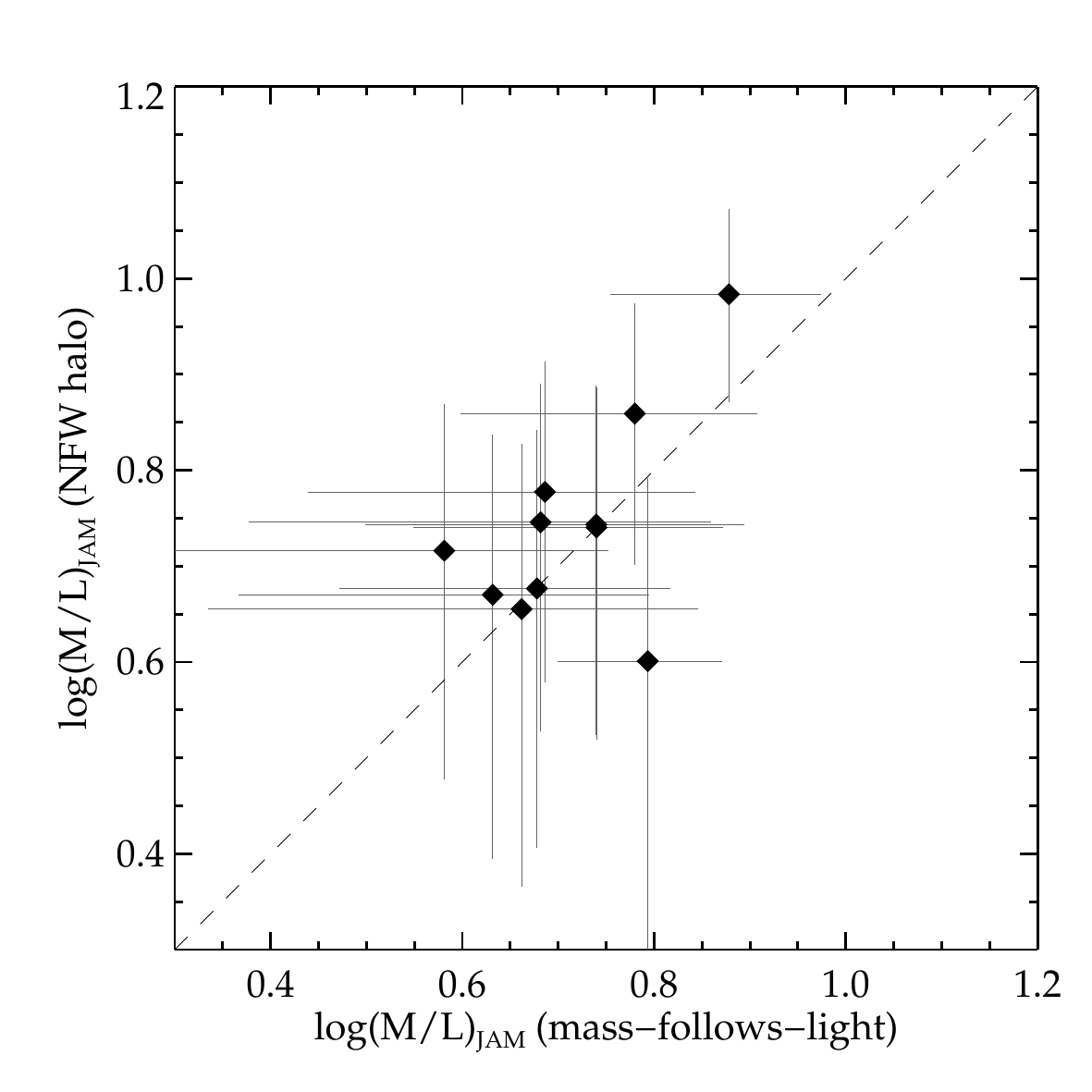}
\includegraphics[width=0.8\columnwidth]{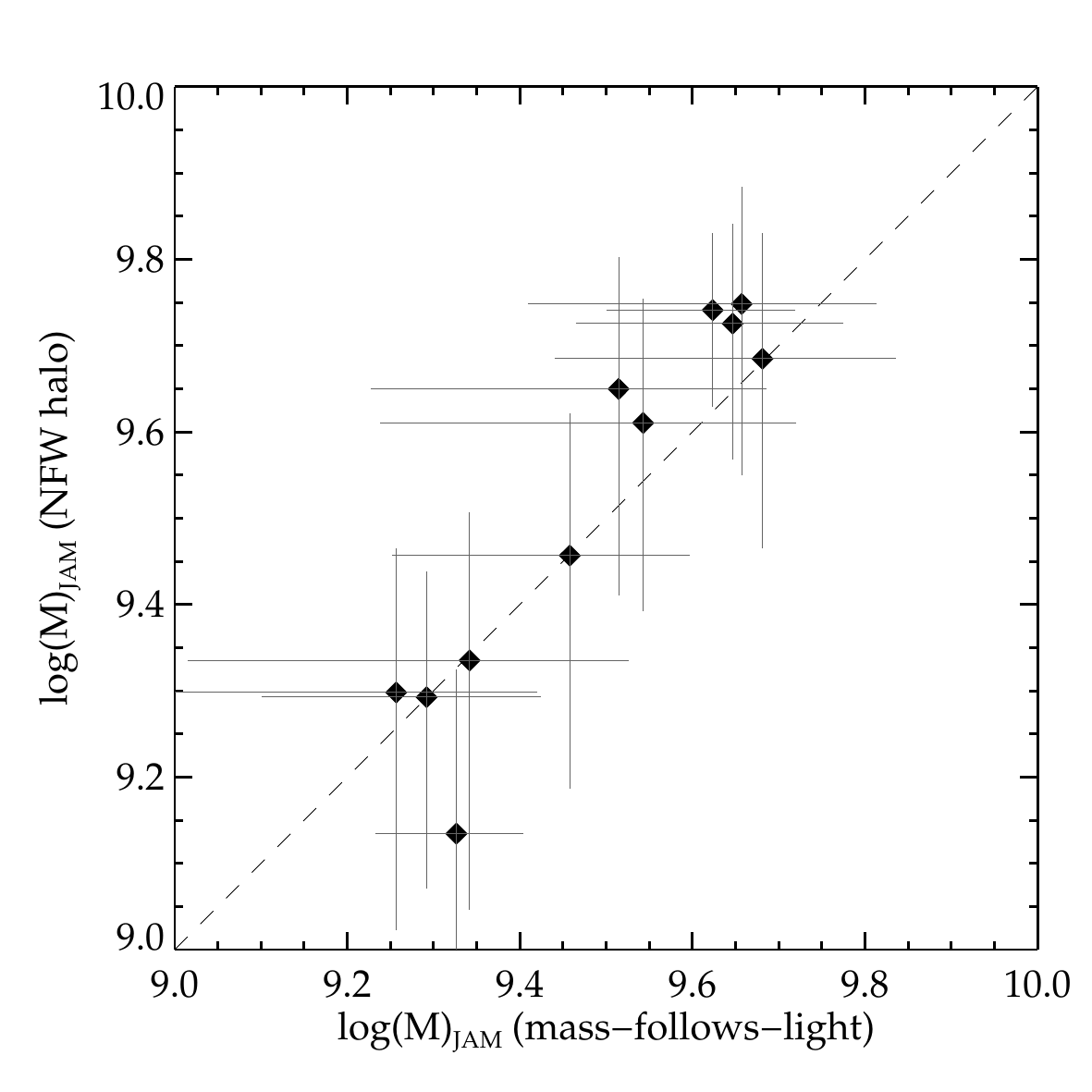}
\caption{Comparison of dynamical mass-to-light ratio (top) and mass enclosed within radius $R_e$ (bottom) estimates from mass-follows-light models with those that include an NFW dark matter halo. The two methods give consistent results to within the errors, with the scatter around the one-to-one only somewhat larger than that for more massive galaxies from the ATLAS3D sample of \protect\citeauthor{cappellari:2013a} (\protect\citeyear{cappellari:2013a}a) (see their Figure~9).}
\label{nfw-mfl}
\end{figure}

\subsection{Inclination and velocity anisotropy}

Lacking higher-order moments, we were unable to constrain inclinations of our objects, this however, does not impede our analysis since, as noted in e.g. \cite{marel:1991}, and more recently \cite{cappellari:2006}, the mass-to-light ratios do not depend on the (assumed) inclination since the effects of a change in $i$ is compensated by the change in observed velocities. Nevertheless, we tested this claim by  constructing a series of models assuming different inclination values. We found that, indeed, the results are all consistent to within the errors and do not depend on the chosen $i$. However, as noted by \cite{cappellari:2006}, this only holds true for flattened objects, those fairly round ones require an accurate $i$ estimate if large uncertainties in the measured M/L are to be avoided. Assuming a statistical\footnote{obtained through Monte-Carlo simulations assuming oblate symmetry and adopting a Gaussian distribution for the intrinsic flattening, with mean 0.59 and dispersion 0.24 following \protect\cite{lambas:1992}.} inclination for our galaxies will not guarantee that the values for individual objects are the most accurate possible. It does, however, guarantee that the average value will be close to the actual one in the case of larger samples.

The anisotropy parameter describes the velocity distribution of a given stellar system. Here it is defined as $\beta_z$=1-($\sigma_z/\sigma_R)^2$, where $\sigma_z$ and $\sigma_R$ refer to velocity dispersions in cylindrical coordinates (see section 2.4 of \citealt{cappellari:2008}). Following this definition, negative values of $\beta$ signify the prevalence of tangential orbits, and positive $\beta$ mean radial orbits dominate. While major axis kinematics puts constraints on both radial and tangential dispersion in the equatorial plane, minor axis $\sigma$ puts constraints on the behavior of the radial component. Two-dimensional kinematics can, therefore, significantly help with the anisotropy estimation.

\subsection{Stellar mass-to-light ratio}
\label{stellarMLratio}

To obtain stellar mass-to-light $(M/L)_{pop}$ estimates we used the \cite{bell:2003} color-to-$(M/L)_{pop}$ conversion formula:
\begin{equation}
log(M/L)_{pop}=a+(b*color)-0.15
\end{equation}
where $a$ and $b$ are transformation coefficients from Table~7 of the above paper, the $(g-r)$ color values come from \cite{lisker:2008}, and the $-0.15 $ is the correction to the Kroupa IMF (\citealt{kroupa:2002}). The $(M/L)_{pop}$ values used in this paper are $r$-band values. \looseness-3

\subsection{Dark-matter halo}

The inclusion of dark matter halos in dynamical models is done in various ways. A common approach when modelling the properties of dark halos is to assume that ``mass follows light'' or, in other words, that the properties of the halo are described adopting the scale lengths of the stellar component. This can be a realistic assumption under the conditions of, e.g. dwarfs being transformed by tidal stirring where most of the extended halo is lost (but see \citealt{lokas:2010} whose model galaxy remnant do retain a significant amount of DM in halos that are much more spherical than the stellar components).

In order not to put artificial constraints on our models we create two types of models: (i) mass-follows-light models that allow for a constant dark matter fraction,  and (ii) models with spherically symmetric NFW (\citealt{navarro:1996}) dark matter haloes added explicitly. In the case of NFW halos, their shapes and sizes of were determined in the following way. Given the stellar masses (\S\ref{stellarMLratio}), the expected halo mass and from that the expected virial velocity and radius were estimated using the formula of \cite{moster:2010} between stellar and dark matter halo masses of galaxies. The concentration index and from that the halo scale radius were estimated following the relations in  \cite{maccio:2008}. 

We ran a series of models with NFW haloes, varying the number of free parameters: stellar mass-to-light ratio $(M/L)_{star}$, anisotropy $\beta$, halo scale radius and velocity $r_s$ and $v_s$, as well as the presence of priors on them (see 3.5). We found that leaving all mass parameters free leads to too much degeneracy and the values are unconstrained. Using priors for $r_s$ and $v_s$ as described above leads to systematically higher values of total mass and M/L, both when compared to the mass-follows-light models as well as to the remaining spherical halo models. We concluded that the dark-to-stellar mass estimates in this mass regime are less reliable as a description of the current state of the galaxies since sizable portions of the dark halos have likely been stripped. 

We thus decided that the optimal approach in the case of NFW haloes was to fix the anisotropy to $\beta$=0.0 (isotropic), put a prior on $(M/L)_{star}$ and let the program freely fit the halo parameters $r_s$ and $v_s$. The comparison of dynamical M/L ratios and masses obtained in this way to those from mass-follows-light models is shown in Figure~\ref{nfw-mfl} and shows good agreement between the two. For the purposes of this paper (and in order to facilitate the comparison with published values for massive early-type galaxies) we adopt the values obtained from the mass-follows-light models.

\subsection{Obtaining best-fit parameters: Markov chain Monte Carlo} 

\begin{figure}
\centering
\includegraphics[width=1.00\columnwidth]{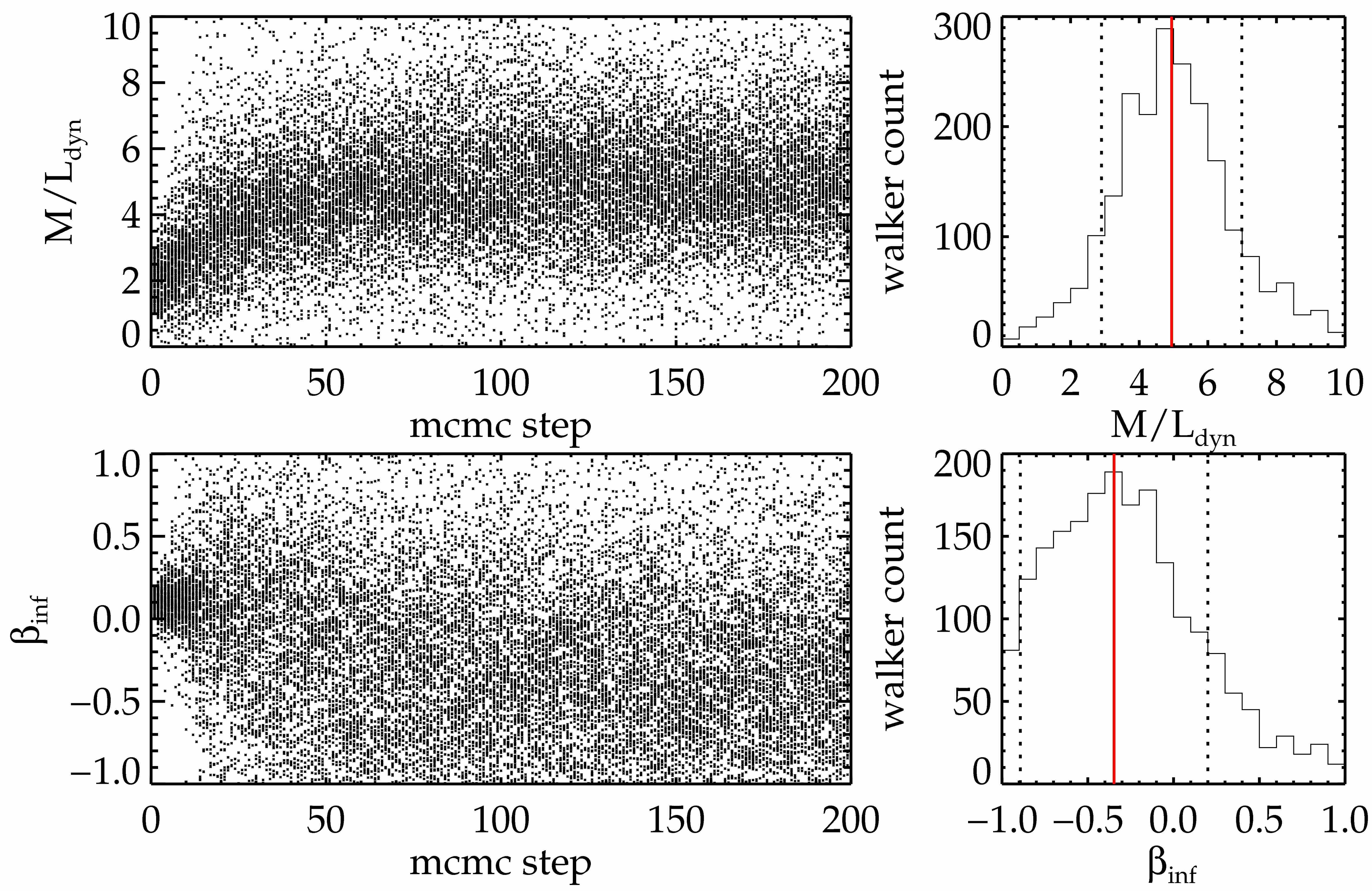}
\caption{Spread in parameters' values for \textit{walkers} as a function of step for all free parameters in the procedure. The number of \textit{walkers} per run is 200. The results converge after ca. 75 steps. In the histogram plots the recovered values and their errors are shown with red solid and black dotted lines, respectively. See also Appendix~\ref{JAMmodelstests} for details on our testing of the mcmc M/L recovery accuracy.}
\label{walkers-steps}
\end{figure}

The errors on the measured quantities were obtained through Markov chain Monte Carlo (mcmc) simulations. We used the \textit{emcee} code of \cite{foreman-mackey:2012}, a Python implementation of the ensemble sampler for Markov chain Monte Carlo method of parameter estimation, together with pIDLy\footnote{http://astronomy.sussex.ac.uk/\,\texttildelow\,anthonys/pidly/}, a code enabling the passing of the data between the emcee code and our Jeans IDL routines. 

We ran the code with 200~\textit{walkers} (i.e. members of the ensemble), whose positions/coordinates in the initial run were confined to a certain range in the parameter space. Later, the parameters were allowed to change within a very broad range and their positions in each step were determined based on the log-likelihood ($\chi^2$ value) of each fit. In the case of NFW models we multiplied the standard likelihood by a gaussian penalty term that allowed us to bias the solution towards the prior values. In the case of $M/L_{star}$ this was the color-based $M/L_{pop}$ and its error. For the halo parameters $r_s$ and $v_s$ we adopted priors based on the stellar-to-halo mass relation from \cite{moster:2010} combined with the concentration-mass relation from \cite{maccio:2008}. We let the code perform 400~runs, during which we saw the walkers' positions converge to the global minima. See Figure~\ref{walkers-steps} for an example plot showing the spread of parameters at each of the steps (only first 200 are shown in the figure since the walkers already converge after around 75 steps). The median values of the last 50 steps was taken as the best-fit value, and its standard deviation as the error. Once we had this best-fitting set of parameters, we used it to run the JAM script once again to calculate the stellar and dark mass profiles, the anisotropy profile, and the total enclosed mass within 1 R$_e$. The recovered $(M/L)_{dyn}$ values are compared among themselves for different input inclinations and anisotropy values to look for biases. We find that regardless of the initial assumptions the outcome values agree well to within the errors. \looseness-2

\begin{figure*}
\center~
\includegraphics[width=0.99\columnwidth]{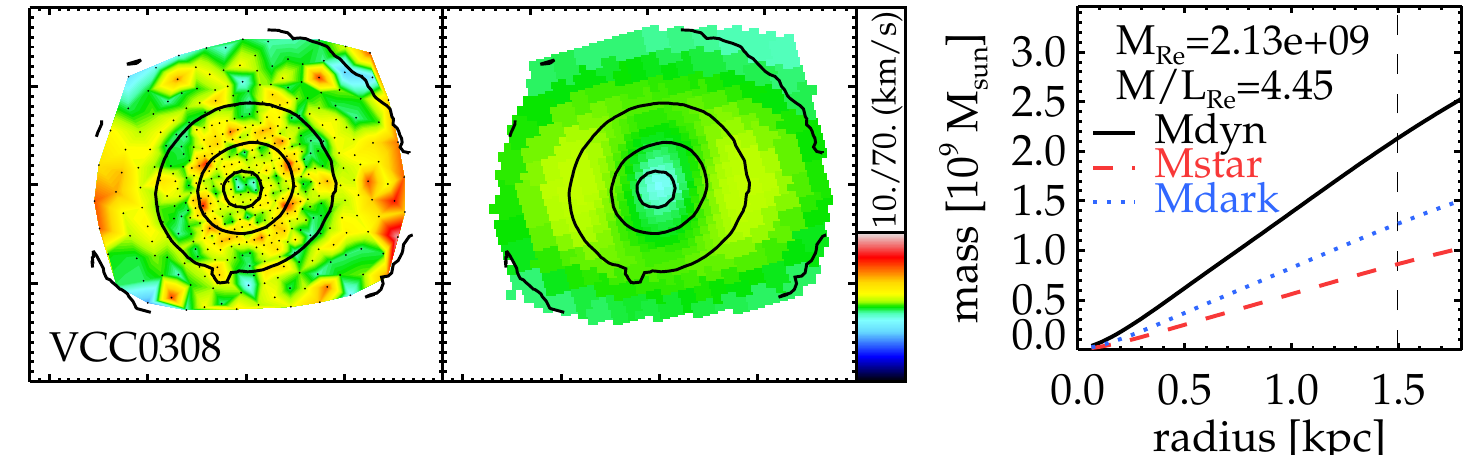}\hspace{0.6cm}
\includegraphics[width=0.99\columnwidth]{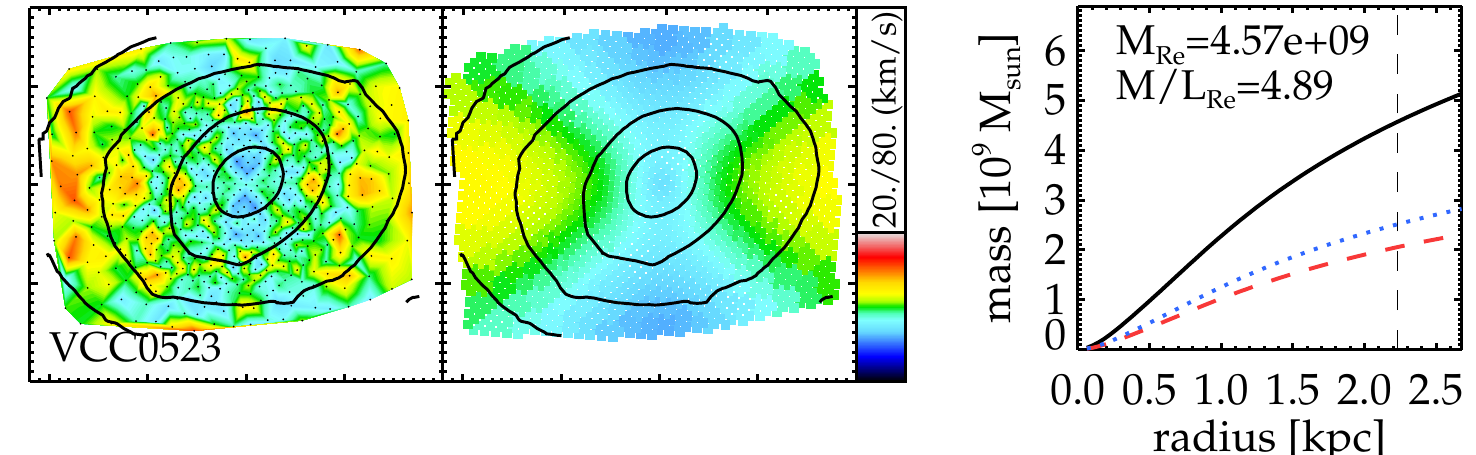}\vspace{0.2cm}
\includegraphics[width=0.99\columnwidth]{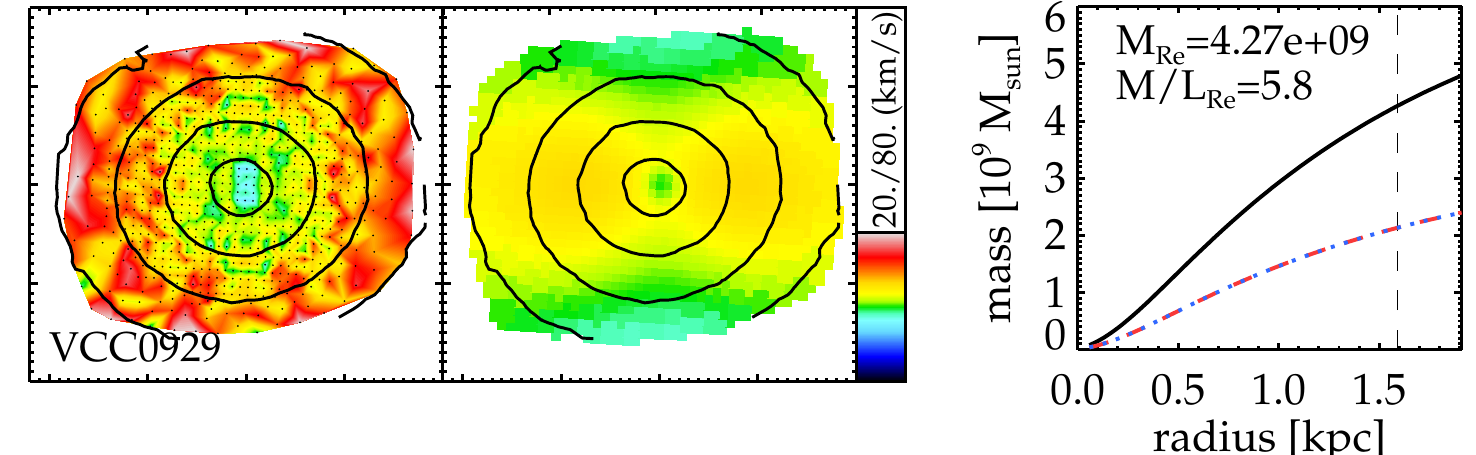}\hspace{0.6cm}
\includegraphics[width=0.99\columnwidth]{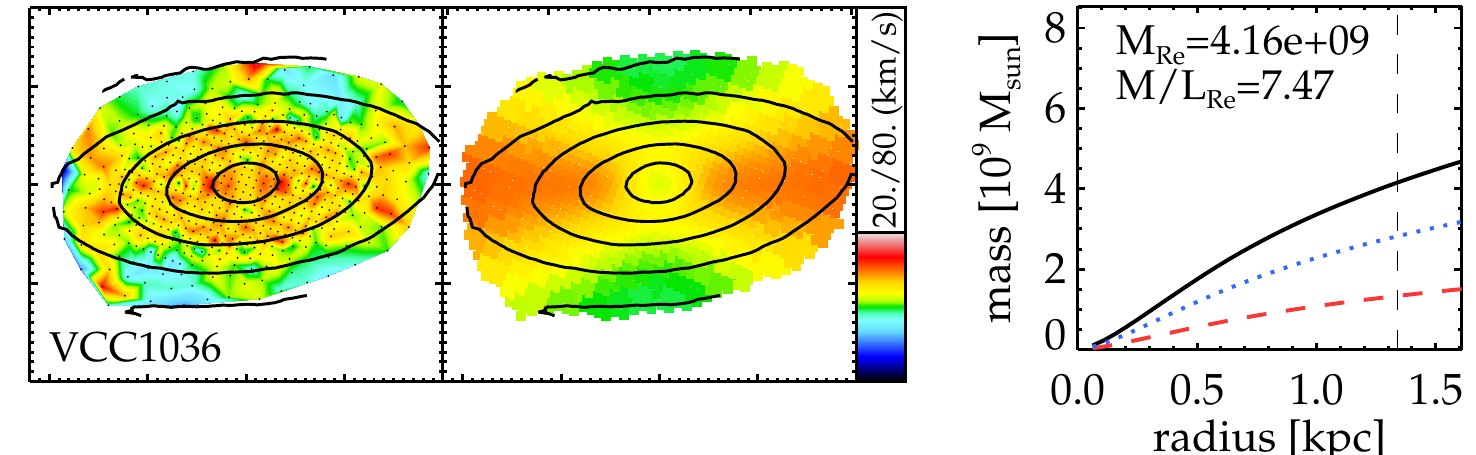}\vspace{0.2cm}
\includegraphics[width=0.99\columnwidth]{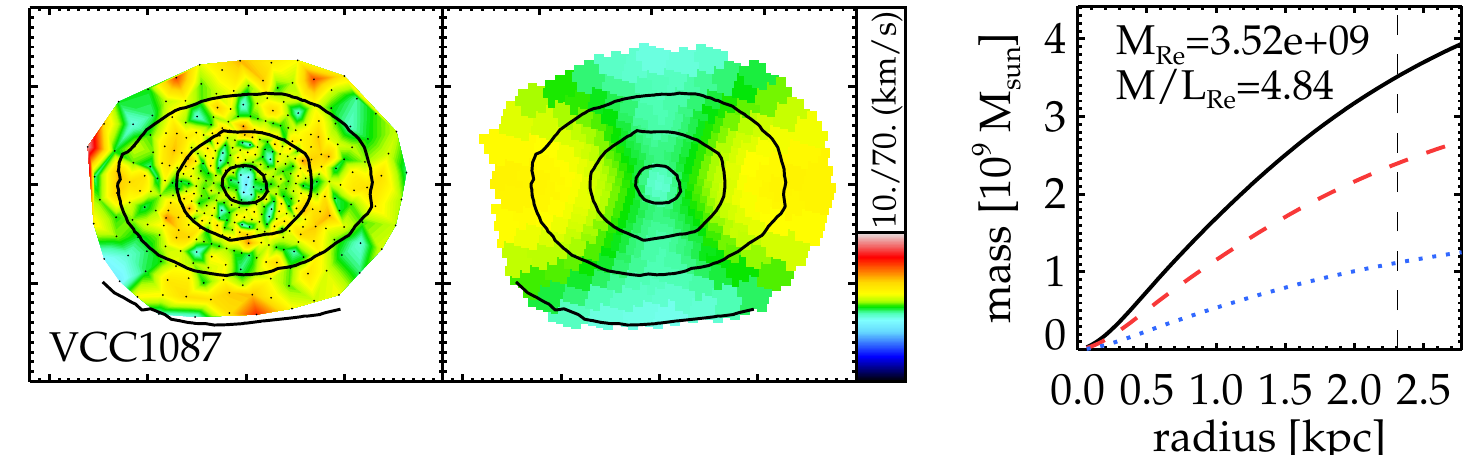}\hspace{0.6cm}
\includegraphics[width=0.99\columnwidth]{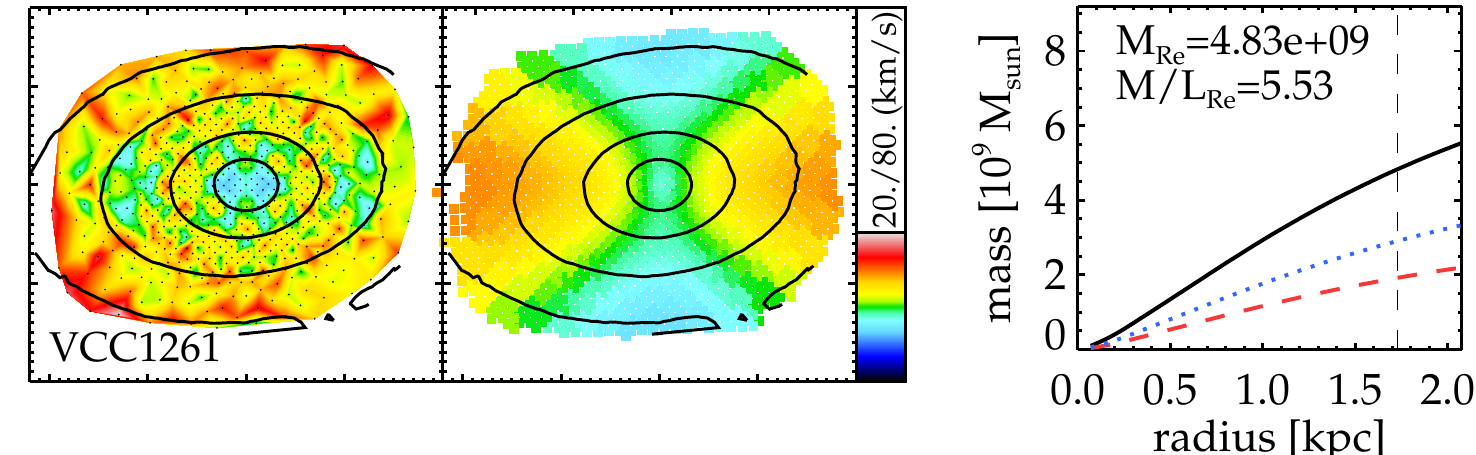}\vspace{0.2cm}
\includegraphics[width=0.99\columnwidth]{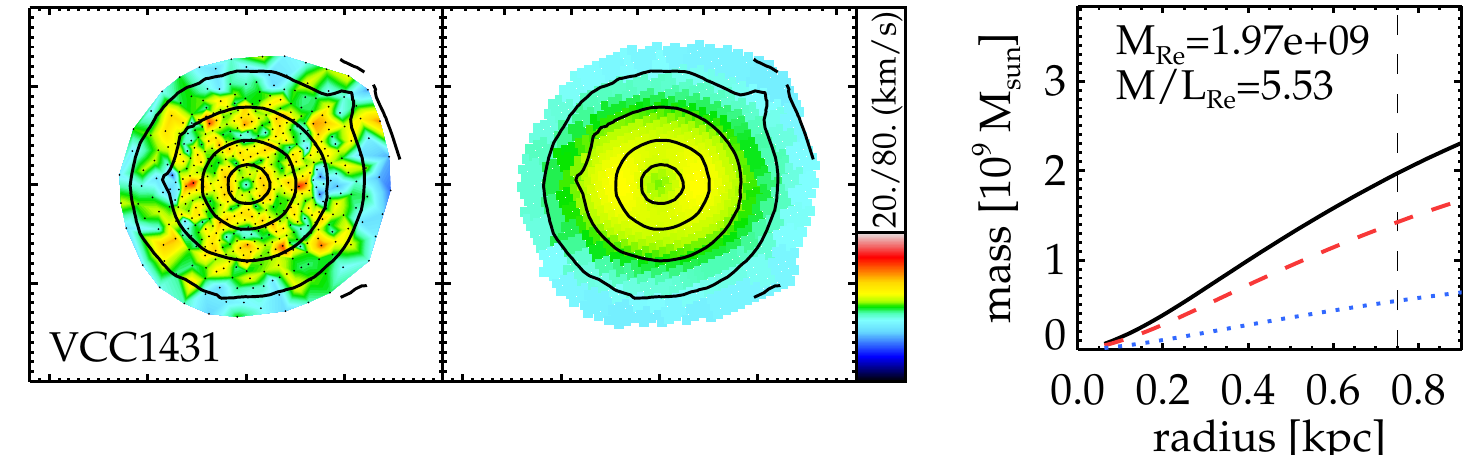}\hspace{0.6cm}
\includegraphics[width=0.99\columnwidth]{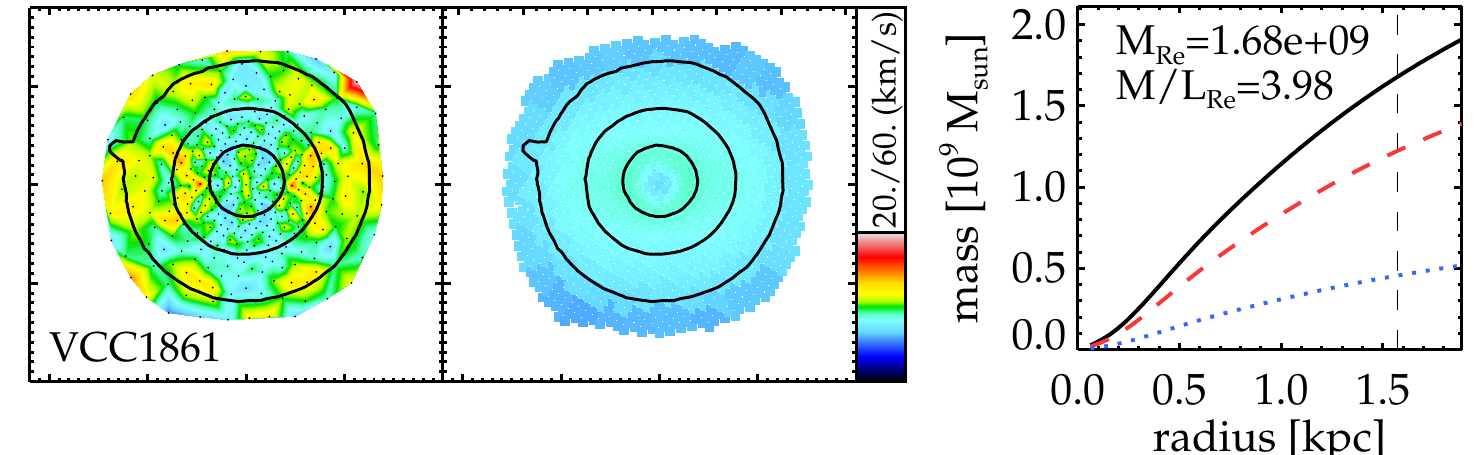}\vspace{0.2cm}
\includegraphics[width=0.99\columnwidth]{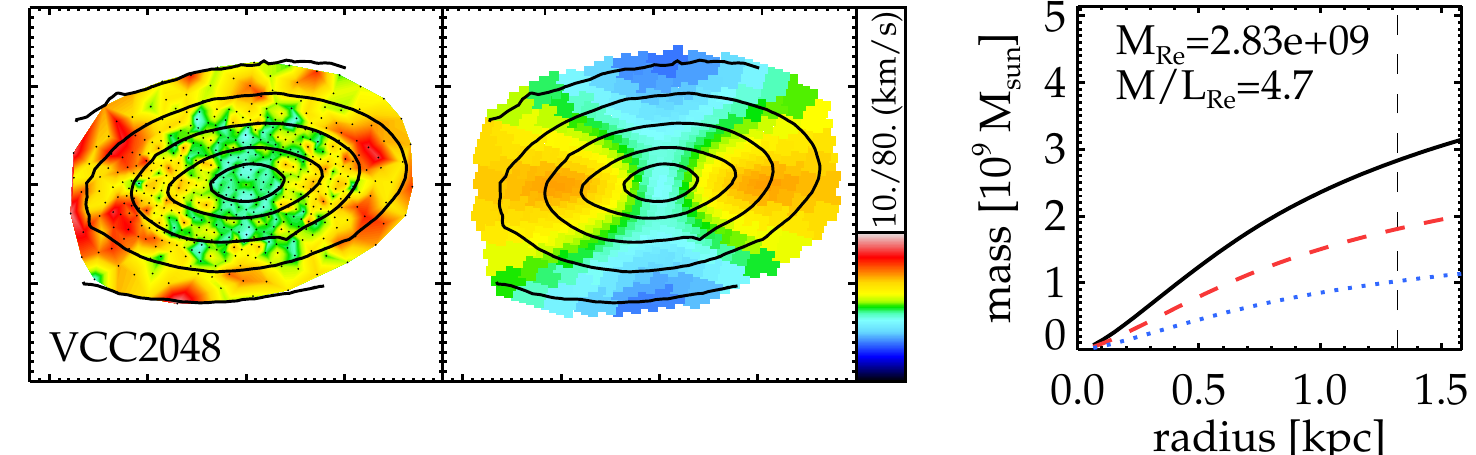}\hspace{0.6cm}
\includegraphics[width=0.99\columnwidth]{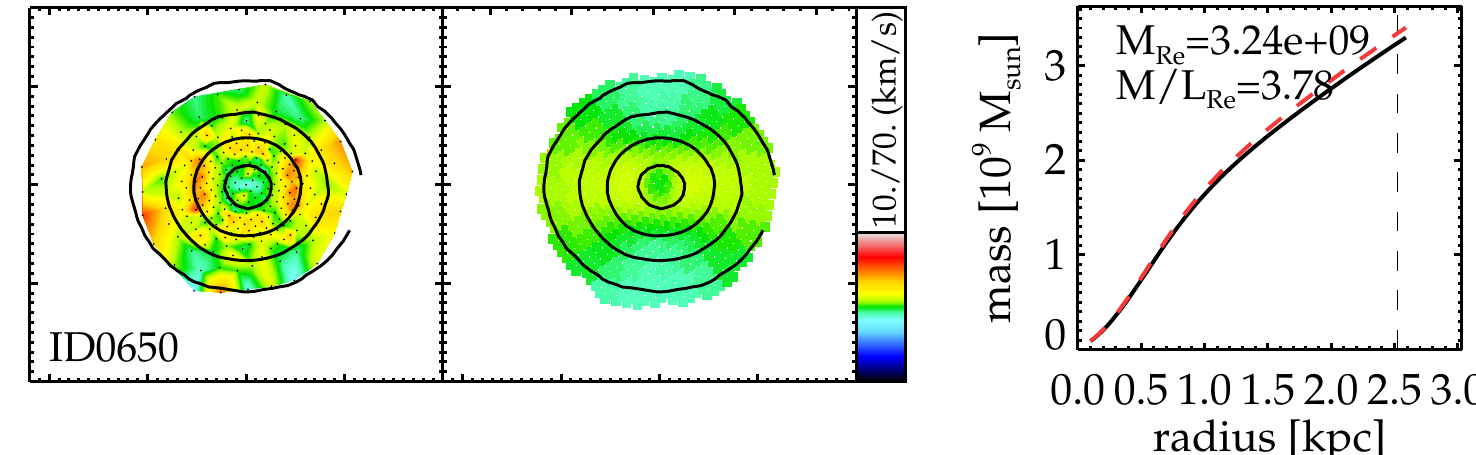}\vspace{0.2cm}
\includegraphics[width=0.99\columnwidth]{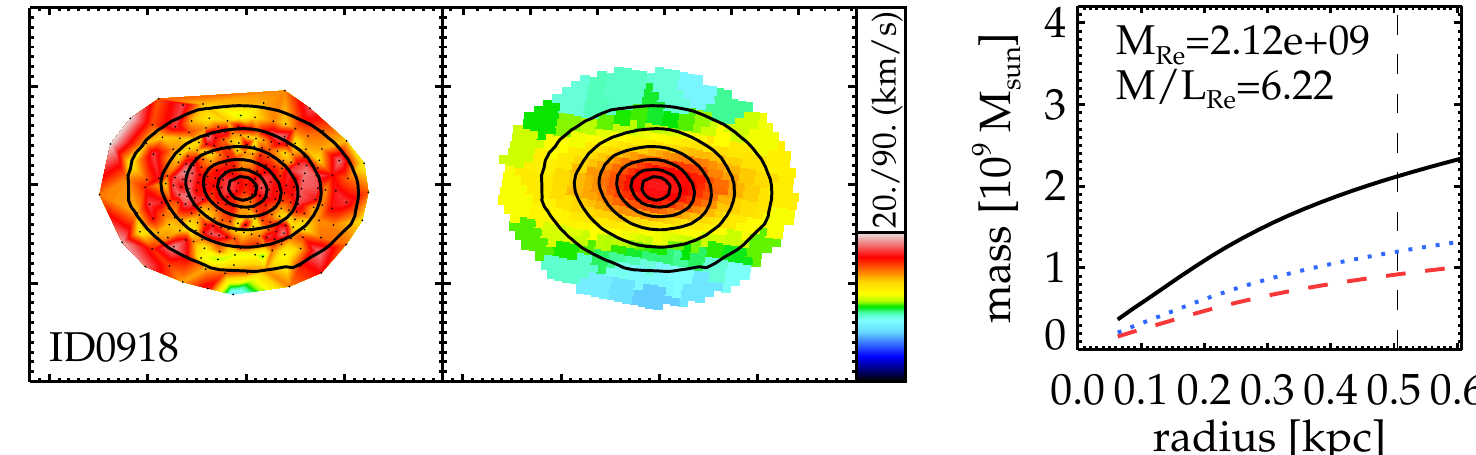}
\caption{\label{modelsplots} Best-fit mass-follows-light anisotropic axisymmetric Jeans models for our sample. For each galaxy we plot the observed and modelled $V_{rms}$ (with the range adjusted to emphasize individual characteristics) as well as the resulting enclosed mass profile (stellar, dark, and total mass). The vertical dashed lines indicate the major axis half light radii. The maps are 40x40 arcsec in size.}
\end{figure*}

Since the errors we obtained were quite high, we decided to run a series of tests to determine the accuracy of our best-fit and error recovery method, details of which can be found in Appendix~\ref{JAMmodelstests}. We find that for the models that mimic the ratios of second velocity moment to its error ($V_{rms}/\Delta V_{rms}$) of normal early-type galaxies (ETGs) we are able to recover input $(M/L)_{dyn}$ very accurately and the velocity anisotropy with somewhat larger errors. For model galaxies resembling our data the recovered $(M/L)_{dyn}$ values are accurate, albeit with large error bars (comparable to those we obtain with real data). The anisotropies in this latter case are largely unconstrained because for our dwarfs the $\sigma$ levels are so low that even with our excellently small error bars the $V_{rms}/\Delta V_{rms}$ ratio is low and causes the errors on the recovered values to span almost the full allowed range. Nevertheless, the tests confirm that the method allows us to obtain reliable $(M/L)_{dyn}$ estimates since no systematic differences between input models and output results are found.

\begin{figure*}
\center~
\includegraphics[width=0.66\columnwidth]{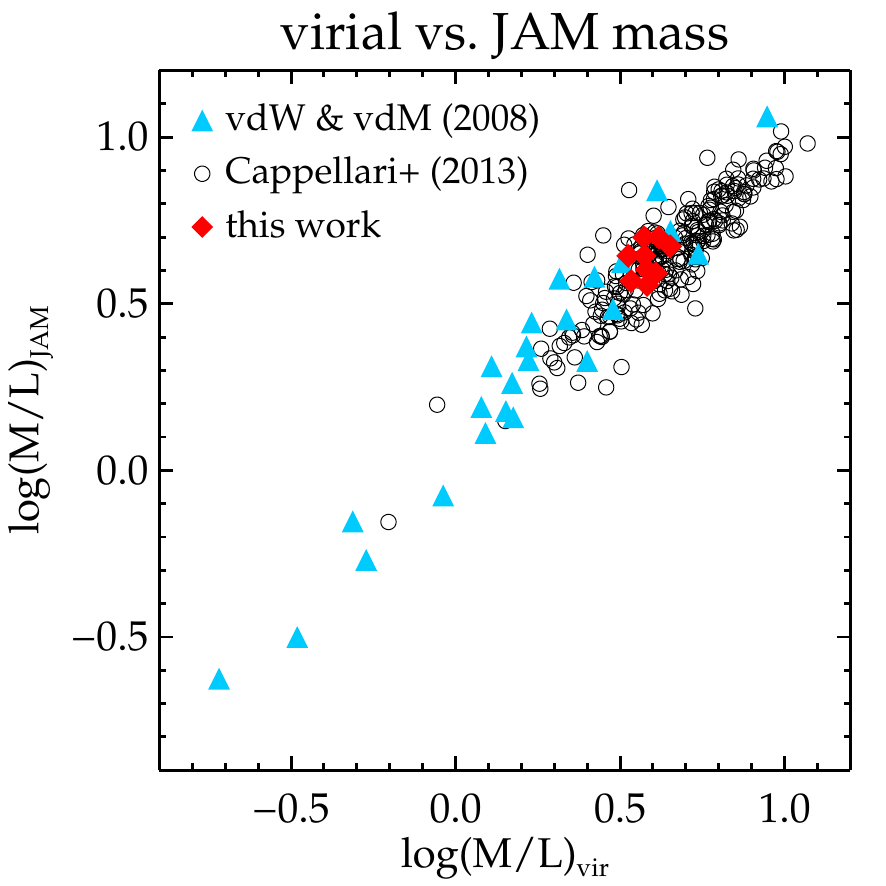} 
\includegraphics[width=0.66\columnwidth]{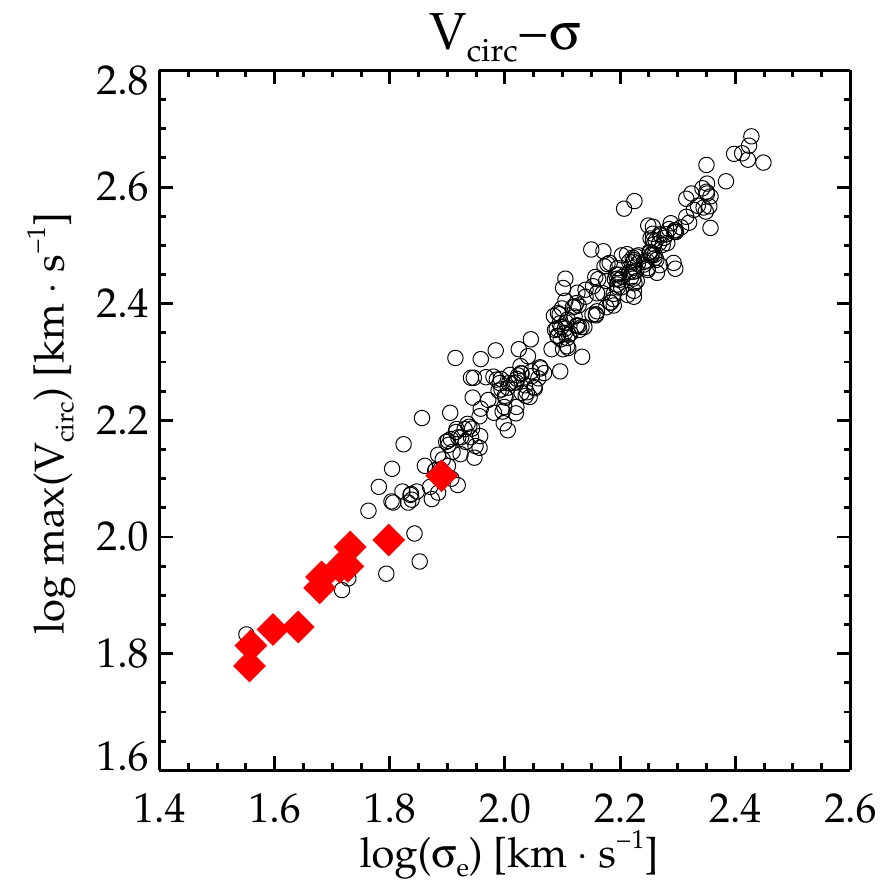}
\includegraphics[width=0.66\columnwidth]{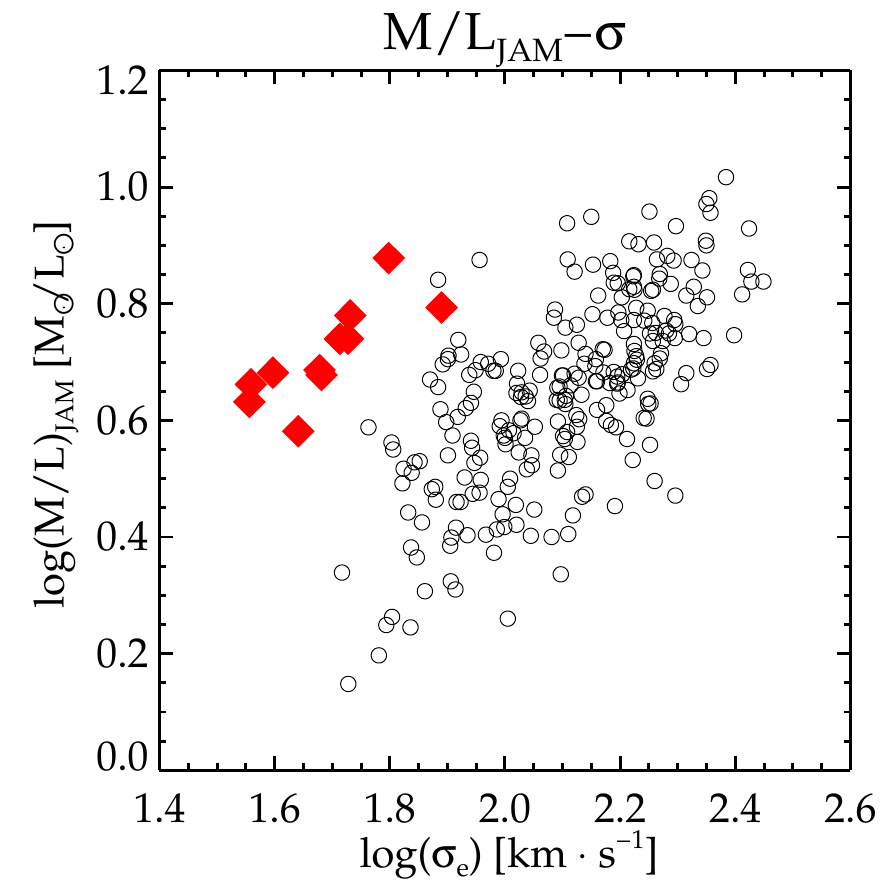} 
\includegraphics[width=0.66\columnwidth]{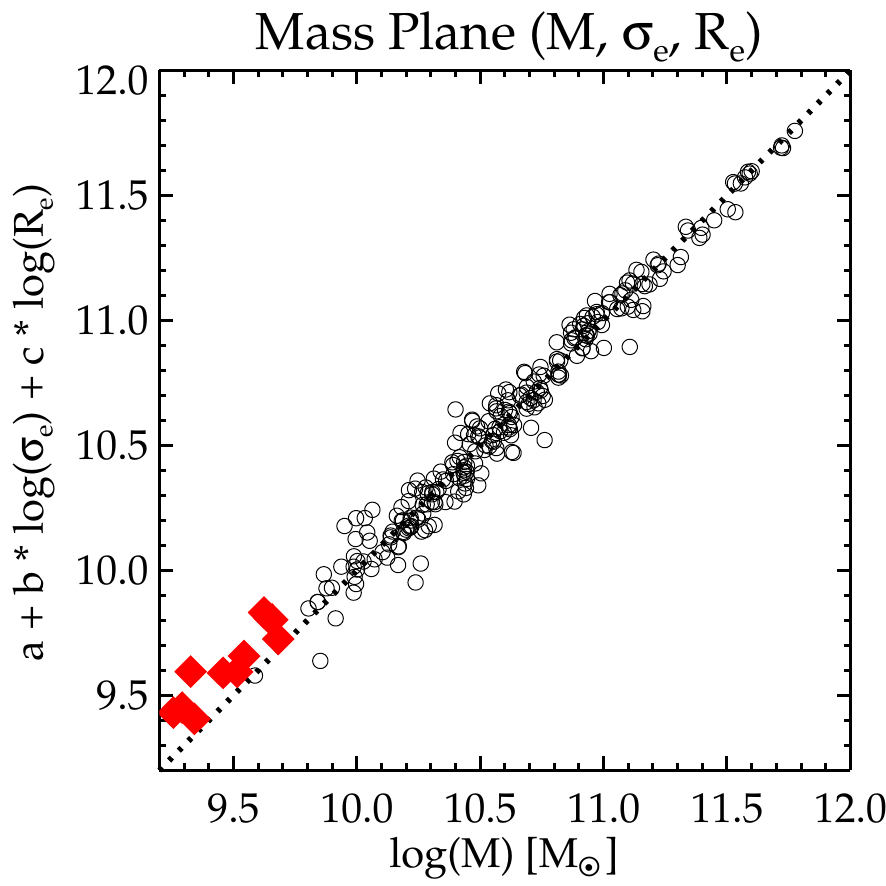} 
\includegraphics[width=0.66\columnwidth]{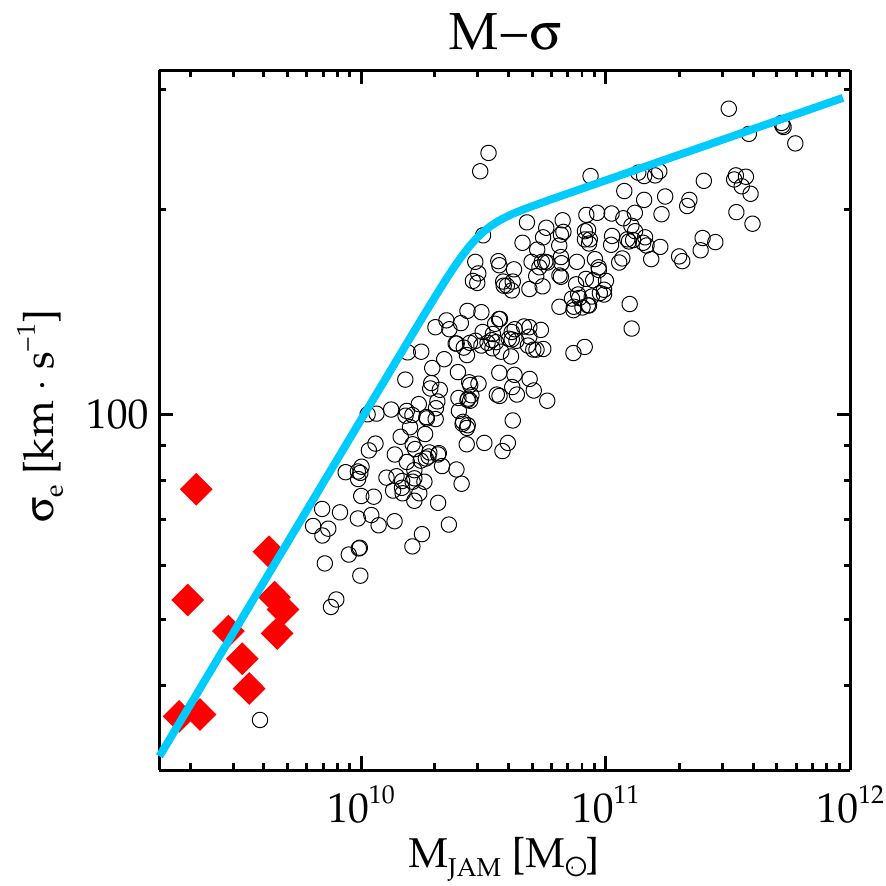} 
\includegraphics[width=0.66\columnwidth]{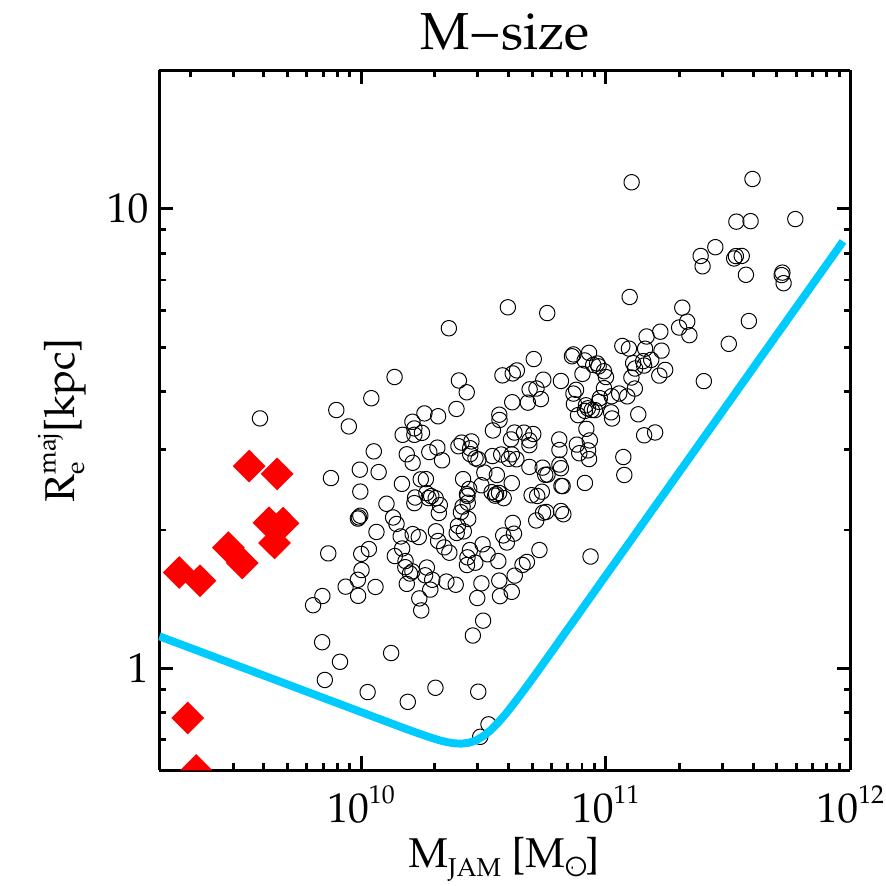}

\caption{\label{scaling_relations} \underline{\textit{Top left}.} Comparison of dynamical M/L ratios obtained from JAM models and their estimates from the virial theorem. The samples of \protect\citeauthor{cappellari:2013a} (\protect\citeyear{cappellari:2013a}a) (ATLAS3D) and \protect\cite{vanderwel:2008} are overplotted to extend the probed M/L range. The scaling factor value for dEs appears to be the same as for massive galaxies. \underline{\textit{Top middle}.} Maximum circular velocity \textit{vs} $\sigma_e$ for a combined sample of our dEs (red diamonds) and ATLAS3D galaxies (black circles). The dwarfs fall on the same tight relation as the giant galaxies. \underline{\textit{Top right}.} (M/L)$_e$--$\sigma_e$ relation: the dE sample is clearly offset from the best-fit relation for giant ETGs. \underline{\textit{Bottom left}.} Edge-on view of the Mass Plane. The same scaling coefficients $a$, $b$, and $c$ as well as normalization factors for $\sigma$ and $R_e$ have been used as in \protect\citeauthor{cappellari:2013a} (\protect\citeyear{cappellari:2013a}a). \underline{\textit{Bottom middle and right.}} Mass Plane projections of \protect\citeauthor{cappellari:2013b} (\protect\citeyear{cappellari:2013b}b). With the exception of two objects (VCC\,1431 and ID\,0918), all our galaxies are outside the zone of exclusion (ZOE, as indicated with the blue lines) at small sizes or large densities (see Sec. 3.1 of \protect\citeauthor{cappellari:2013b} \protect\citeyear{cappellari:2013b}b).}
\end{figure*}

\section{Results}

Table~\ref{observations} lists for each galaxy the integrated values within 1\,R$_e$: velocity dispersion, specific angular momentum, maximum circular velocity, dynamical and stellar mass-to-light ratio, and the total luminosity. Additionally, Figure~\ref{modelsplots} shows for each galaxy its observed and modelled second velocity moment, as well as the resulting mass (stellar, dark and total) profiles. We use these results to place our galaxies on the scaling relations of giant early-types of the ATLAS3D survey (\citeauthor{cappellari:2013a} \citeyear{cappellari:2013a}a,b). We then analyze the rotational properties of dEs in the context of of other galaxy classes, both early- and late-types and of a large mass range from the CALIFA survey (Falc\'on-Barroso et al. 2013, in prep.). Finally, we look at the changing dark matter fraction within the Virgo Cluster dEs.

ATLAS3D project's sample comprises 260 early-type galaxies within the local (42 Mpc) volume observed with WHT/SAURON (\citealt{cappellari:2011}). The goal of the ongoing CALIFA survey is to obtain IFU (PMAS/PPAK at Calar Alto's 3.5m telescope) data of $\sim$600 galaxies\footnote{out of which 158 have so far been analyzed and are presented here} in the Local Universe (0.005 \textless z \textless 0.03, \citealt{sanchez:2012}). This allows us to present the properties of dEs against those of massive systems in a homogeneous way for the largest (numerically and type-wise) sample currently possible. 

\begin{figure*}
\center~
\includegraphics[width=0.95\columnwidth]{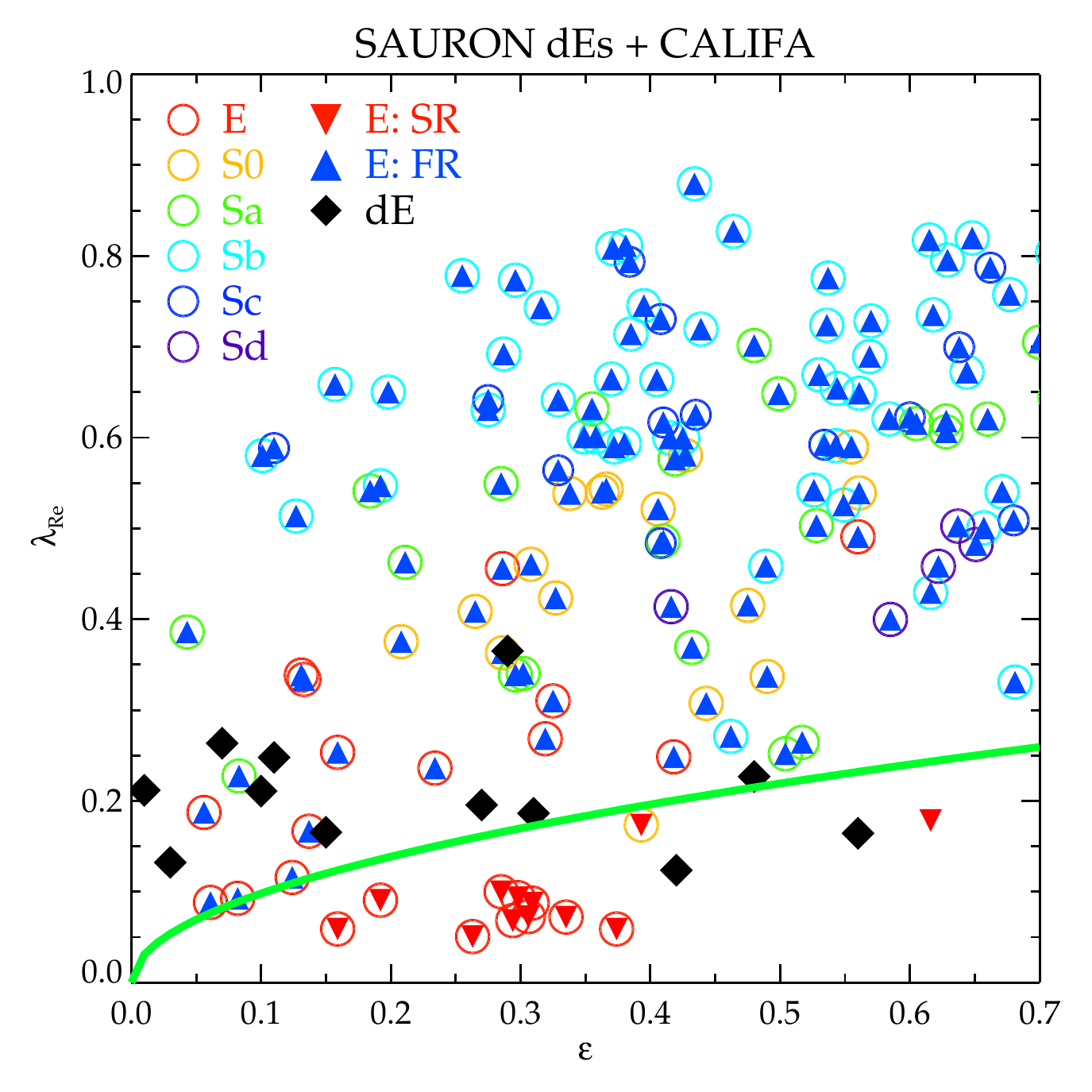}\hspace{0.6cm}
\includegraphics[width=0.95\columnwidth]{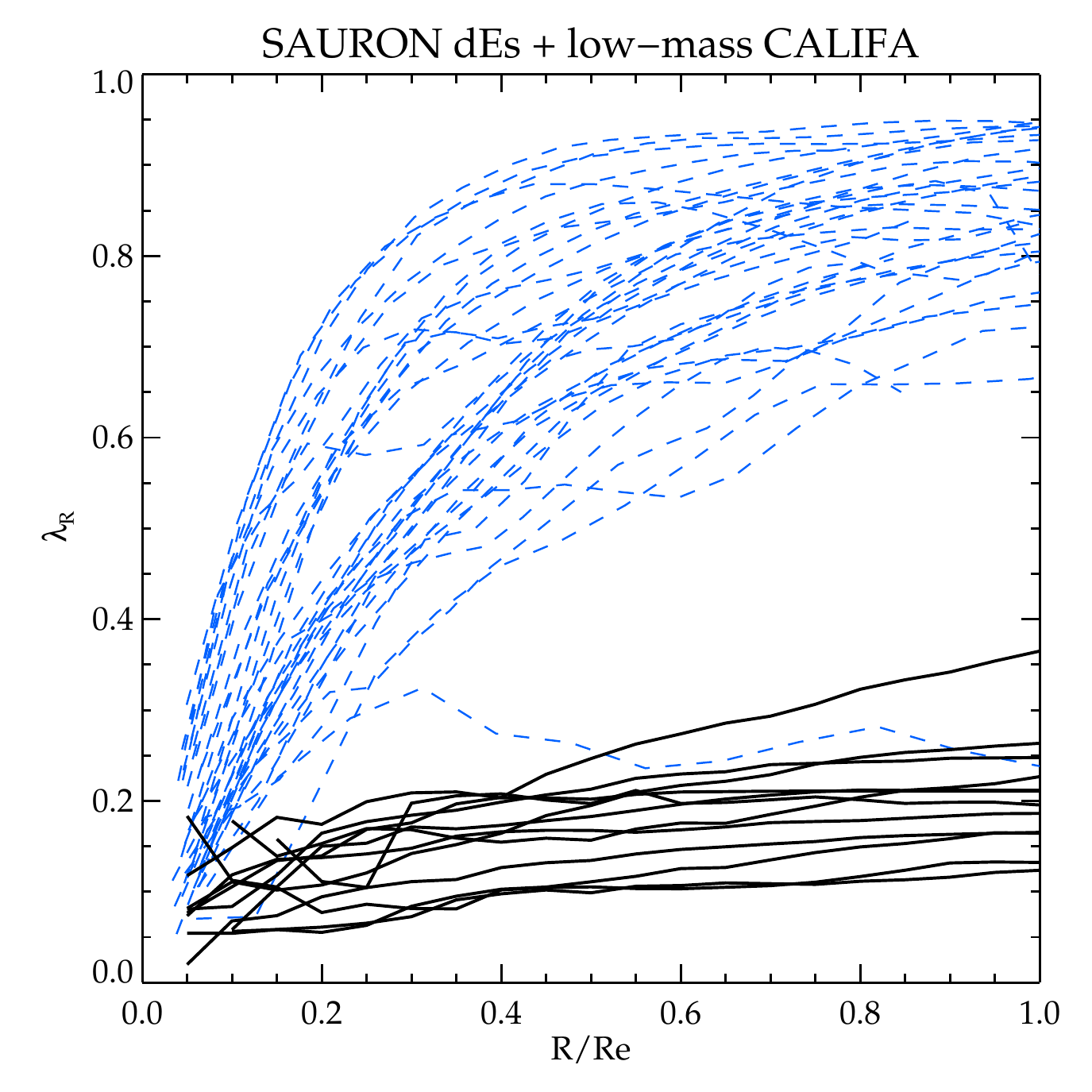}
\caption{\label{lambda-profiles} \underline{\textit{Left:}} specific angular momentum $\lambda_{R_e}$ versus ellipticity for our dEs and the currently available subset of 158 galaxies from the CALIFA survey (\protect\citealt{sanchez:2012}). The galaxies are color-coded according to their rotational properties: fast rotators are shown in blue and slow rotators in red, while the dEs are depicted in black. The green solid line corresponds to $0.31\cdot \sqrt \epsilon$, the threshold separating the fast- and slow-rotator families defined by \protect\cite{emsellem:2011}. The different galaxy types are color-coded with open circles as shown in the panel. \underline{\textit{Right:}} $\lambda_R$ profiles in units of the effective radius of dEs (solid black lines) and a subset of the CALIFA sample with stellar masses $\le 2\cdot10^{10} M_\odot$ (dashed blue lines).}
\end{figure*}

\subsection{Extending the dynamical relations of ATLAS3D}

\subsubsection{Virial mass estimator as a proxy for $\mathrm{(M/L)_{dyn}}$}

\cite{cappellari:2006} presented a simple virial estimate of the total dynamical mass through the following equation: $(M/L)_{vir}=\alpha\cdot\sigma^2_e\cdot R_e / (L\cdot G)$. The value of the scaling factor $\alpha$ was estimated to to be 5.0$\pm$0.1, which value was later confirmed through the analysis of the much larger ATLAS3D sample (\citealt{cappellari:2013a}a). Here we use this value to compute virial masses for our dE sample to see whether the calibration can be extended to those low-mass systems. In Figure~\ref{scaling_relations} (top left panel) we plot JAM vs. virial masses for both samples, also adding that of \cite{vanderwel:2008}\footnote{obtained through isotropic axisymmetric Jeans models for a sample of 25 field early-type galaxies at a median redshift z=0.97} to extend the probed M/L range. We can see that our dEs fall on the same tight relation derived for giant early-types. This may have important implications for studies where data quality or distance does not allow for direct $(M/L)_{dyn}$ calculations since they can rely on the virial estimate to a fairly high level of accuracy. 

\subsubsection{$V_{circ}-\sigma$ relation}

\citeauthor{cappellari:2013a} (\citeyear{cappellari:2013a}a) have shown that a tight relation exists between $V_{circ,max}$ and $\sigma$ (see their Figure\,17). The existence of this relation is useful for converting sigma into $V_{circ,max}$. We wanted to investigate whether the relation can be extended to lower masses. We can see in the top middle panel of Figure\,\ref{scaling_relations} that our dwarfs fall almost exactly on the extension of the relation for the ATLAS3D giant galaxies. The tight trend is not so surprising when we take into account that the maximum velocity values are reached well within the effective radius, where the visible matter still dominates.

\subsubsection{Dynamical mass-to-light ratio vs. velocity dispersion} 

\cite{cappellari:2006} and van der Marel \& van Dokkum (2007) show a tight relation between the stellar velocity dispersion and dynamical mass-to-light ratio of a given galaxy within its effective radius. Previous attempts at placing dwarf/low-mass galaxies on the relation were restricted to using Fundamental Plane $\kappa$ space proxies (e.g. \citealt{geha:2003}, \citealt{toloba:2012}). Recently \cite{geha:2010} have plotted the relation using $(M/L)_{dyn}$ from Jeans models for a compilation of samples from various sources. Here for the first time we use direct M/L estimates that come from dynamical models on a \textit{homogeneous} sample of galaxies, for which the instrument, data reduction, kinematic analysis, and dynamical modelling approach are the same. Our sample can thus be viewed as a natural extension of the ATLAS3D galaxies that in a most consistent way extends their analysis to low-mass regimes. 

Our results on the $(M/L)_{dyn}-\sigma$ relation (see the top right panel of Figure~\ref{scaling_relations}) are in agreement with previous findings that dEs are visibly offset from the main sample of giant ETGs. The properties of dEs significantly differ from those of massive early-type galaxies. This might be caused by the curvature of the relation, as shown in Figure\,5 of \cite{toloba:2012} where a compilation of samples of Es, dEs and dSphs is plotted together and it is shown how the dE/dSph offset from the main relation increases with decreasing $\sigma$. The offset has been attributed to the higher DM content of dEs (e.g. \citealt{geha:2010}, \citealt{toloba:2012}). dEs are typically younger and more metal-poor than ETGs, therefore a smaller fraction of their total M/L can be explained by stellar populations. The offset seen in the relation can thus be taken as an indication of a higher DM fraction in dEs.

\subsubsection{Mass Plane}

Using our results we were also able to place our galaxies on the Mass Plane (MP) relation and its projections (\citeauthor{cappellari:2013a} \citeyear{cappellari:2013a}a,b). While the Fundamental Plane (FP, e.g. \citealt{faber:1987}) uses $\sigma$, effective radius and effective surface brightness (L, $\sigma_e$, $R_e$) and is a good tool for distance estimates, the MP replaces effective surface brightness with total dynamical mass and as such can be used as a mass and mass-to-light ratio estimator. Using the MP was recently shown by \citeauthor{cappellari:2013a} (\citeyear{cappellari:2013a}a) to significantly reduce the scatter around the relation, said to be due to differences in M/L ratios. \looseness-2

We find that our galaxies roughly fall on the same tight MP relation, although one notices a slight systematic positive offset with respect to the best-fit relation for ATLAS3D galaxies (Figure~\ref{scaling_relations} bottom left). Given the sample size it would be hard to argue in favor of a deviation, that can still be explained with measurement errors and the relation's intrinsic scatter. Similarly, we placed our galaxies on the projections of the Mass Plane: M--$\sigma$ and M--size (Figure~\ref{scaling_relations} bottom middle and right), and see that with the exception of one field galaxy they also satisfy the same zone-of-exclusion conditions. We thus conclude that (based on our data) dEs behave in the same way as giant ETGs and that the relation applies also in this low-mass regime.

\citeauthor{cappellari:2013b} (\citeyear{cappellari:2013b}b) show in their Figure~9 the mass-size distribution of a combined sample of early and late-type galaxies across a wide range of masses. We see there that the ETG sequence naturally merges with that of late-type galaxies. For low-mass systems approximate values of the stellar mass were used in the plot. Our results, using stellar mass estimates from Jeans models, strengthen those published results. 

\subsection{Stellar angular momentum}

\begin{figure*}
\center~
\includegraphics[width=0.95\columnwidth]{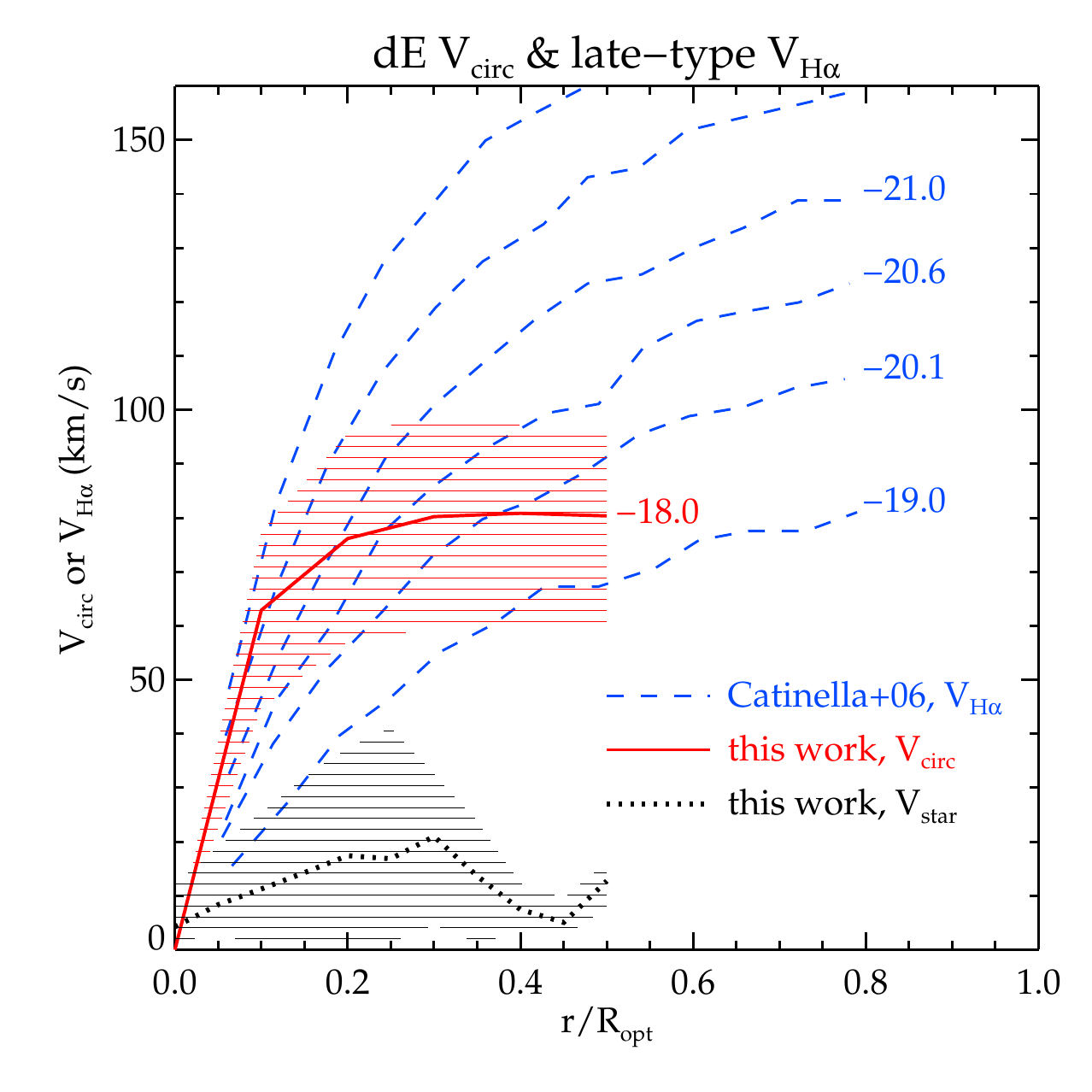}\hspace{0.6cm}
\includegraphics[width=0.95\columnwidth]{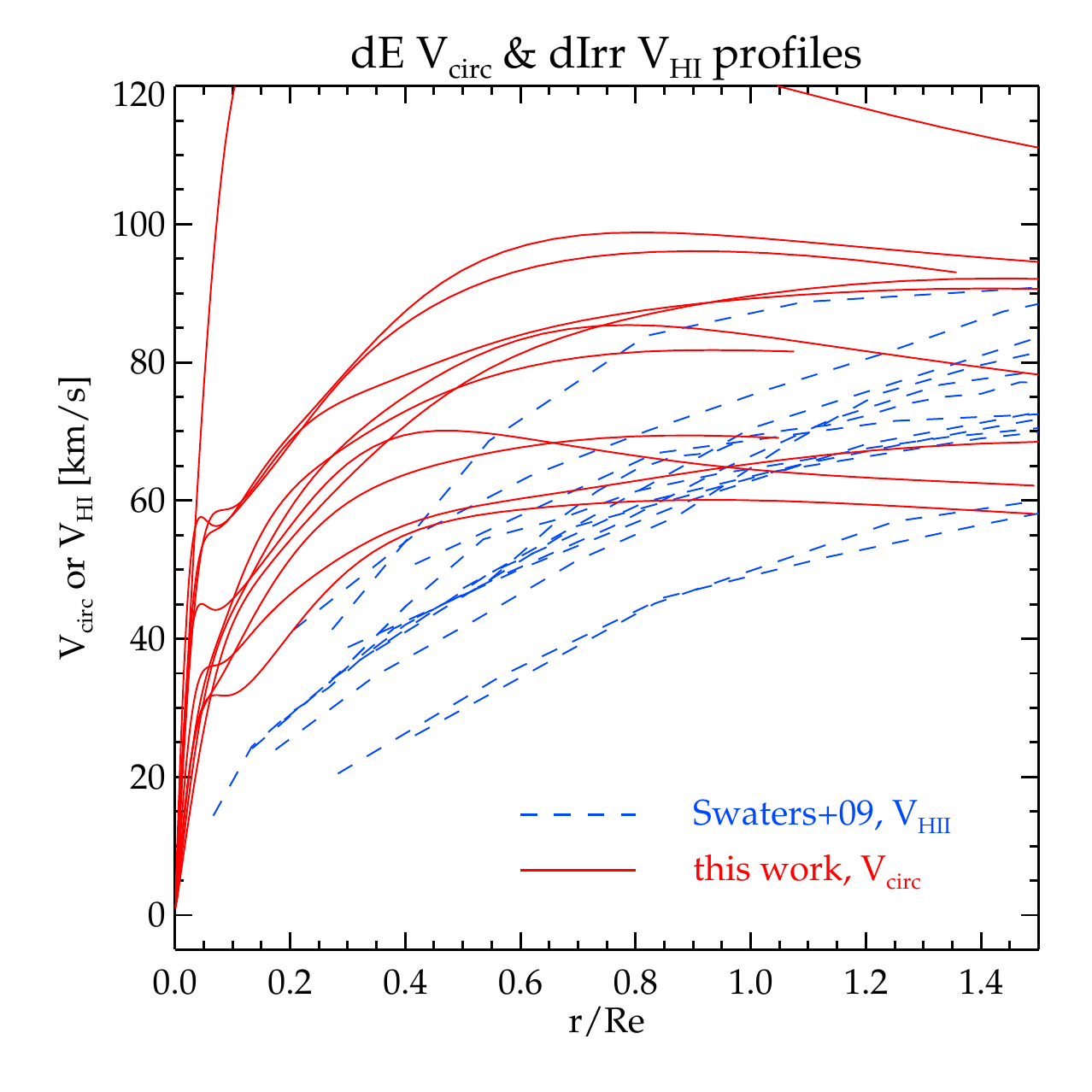}
\caption{\label{vcirc} \underline{\textit{Left.}} Average circular velocity ($V_{circ}$) curve from our 9 Virgo dE galaxies, with a mean value depicted with the red line and the red shaded region showing the extent of all values. Overplotted are \protect\cite{catinella:2006} $V_{H\alpha}$ template rotation curves for a range of absolute r-band magnitude values. For comparison we also plot (in black) an averaged inclination-corrected stellar velocity curve. \underline{\textit{Right.}} Individual dE $V_{circ}$ profiles (Virgo + field) together with the $V_{HI}$ profiles of a subsample of \protect\cite{swaters:2009} dwarf late-type galaxies. Both samples have the same mean absolute r-band magnitudes ($M_r=-17.9$).}
\end{figure*}

To estimate the amount of rotation in our objects we use the specific angular momentum $\lambda_R$. As already mentioned in Paper I and discussed in detail in e.g. \cite{emsellem:2011}, $\lambda_R$ is better suited for angular momentum estimate than the traditionally employed V/$\sigma$. While the two agree to a large degree for simple isotropic axisymmetric Jeans models (see e.g. Appendix B of \citealt{emsellem:2011}), this is no longer the case for galaxies that exhibit more complex kinematics. 

In \cite{rys:2012jenam} we compared our dE sample and SAURON ETGs of \cite{emsellem:2007} and showed that the integrated $\lambda_{R}$ values of dEs place them on or above the line that divides the ETG group into slow- and fast-rotators. Also, the shapes of dE profiles resemble those of giant fast rotators in that they seem to rise until at least the effective radius, unlike the profiles of slow rotators which have a tendency to be nearly-flat (and around zero) or start off with a positive value and later drop. We interpreted that as a possible indication of dEs bearing a structural resemblance to the fast-rotator group.

Here we compare our findings to those of Falc\'on-Barroso et al. (2013, in prep.) for the CALIFA sample, which gives the advavntage of probing a much larger range of galaxy types. In the left panel of figure~\ref{lambda-profiles} we plot the integrated $\lambda_{Re}$ values of the CALIFA and our galaxies, with the former color-coded according to both their morphological and kinematic classification. What we see is that the dEs have lower $\lambda_{Re}$ than the vast majority of the massive galaxies, with the exception of some giant ellipticals. 

To be able to interpret the plot in the context of dE formation scenarios, we restricted the mass range of the comparison sample. The first reason was that we should not be looking at the most massive objects, known to differ not only from dEs but also from medium-size ETGs in their properties. Also, among the proposed transformation scenarios only galaxy harassment is predicted to be able to remove significant (up to 90\%) portions of galaxy mass, while processes such as ram-pressure stripping are expected to only help remove the gas content, leaving the stellar mass largely unaffected. The progenitor galaxies' masses should therefore not be significantly higher than those of today's dEs. The right panel of Figure~\ref{lambda-profiles} shows a subsample of the CALIFA galaxies with stellar masses up to $\mathrm{2\cdot10^{10}}$\,$\mathrm{M_\odot}$ (i.e. up to $\sim$10 times those of our dEs). The angular momentum of the progenitor sample is substantially higher than the values we find for our dwarfs.

\subsection{Circular velocities}

Both random and ordered motions provide support against gravity in stellar systems. To learn about the underlying mass distribution we therefore need to include contributions from both of them. This is particularly important in the case of early-type galaxies where the stellar velocity dispersion levels are significant relative to the velocities themselves. To directly compare gas ans stellar rotation curves one therefore has to include the dispersion component.

Our circular velocities directly follow from our JAM model values and the resulting average profile for Virgo galaxies is presented in the left panel of Figure\,\ref{vcirc} (see Appendix~\ref{circvel} for the profiles - with errors - of individual galaxies). We then compared them to the model values of late-type H$\alpha$ rotation curves at different magnitudes from \cite{catinella:2006}. Once $V_{circ}$ is employed we can see that, in fact, the profile shape places the dwarfs above the profile predicted from their average magnitude (cf. \citealt{toloba:2011}). For comparison we also include an averaged stellar velocity curve to show the significant difference between the two. We note here that the \cite{catinella:2006} values are not corrected for dispersion, even though it should not have a strong effect on the measured rotation values of the ionised gas, for which we expect the rotation to still dominate over the dispersion.\looseness-2

On the other hand, in HI velocity profiles the effect of dispersion will be negligible. We therefore compared our results with those of \cite{swaters:2009} for late-type dwarf galaxies, obtained from HI velocity curves. For the comparison we used their high-quality subsample, i.e. galaxies with inclinations $39^o \le i \le 80^o$ and with rotation curves flagged as reliable. We exclude four faintest galaxies from the \cite{swaters:2009} sample so that both samples have the same mean r-band absolute magnitude ($M_r=-17.9$) and span a very similar magnitude range. We can see in the right panel of Figure~\ref{vcirc} that the profile shapes differ substantially. Whereas for our galaxies the rotation curves steepen quickly and reach their maximum within one effective radius, the curves of late-type galaxies keep rising beyond one $R_e$ and only flatten out at a few disk scale lengths (see Figure~4 in \citealt{swaters:2009}). 

The above result also confirms earlier findings by, e.g. Broeils (1992) that the rotation curve shape depends on both morphology and luminosity. The curves of dwarf late-types fit the picture for massive late-type galaxies, but the properties of dEs do not seem to follow the same trends: their curves are steeper in the centers and their absolute values are higher than those of late-types of similar magnitude.

\subsection{Dark matter fraction vs. Virgocentric distance}

For Virgo objects, we took the (M/L)$_{dyn}$ values we obtained and tied them to the location of the galaxies in the cluster in order to determine whether any correlation exists with local environment density. This was motivated by the claim that under the galaxy harassment scenario, in denser environments galaxies will suffer a greater mass loss, with DM lost more easily than stellar matter due to it being in the outer parts of galaxies and therefore bound less strongly (see the simulations of \citealt{moore:1998} and \citealt{smith:2010}). The results of this analysis are shown in Figure~\ref{dmtrend}.

So far most distance trend analyses have been based on projected distances, i.e. did not take cluster depth into account. This can naturally lead to biased results unless complete samples are used for which one should be able to \textit{statistically} deproject observed distances to assess their 3D distribution. One needs to be careful when interpreting a 2D distribution since the distance information we get from it only provides us with a \textit{minimum} intrinsic value, thus for example a central object could in fact be anywhere in a cluster along the line of sight. Naturally, these errors get smaller as the distance from the cluster's center increases. 

To obtain this 3D information we compiled literature data of the line-of-sight distances for our objects where available (7 out of 9). These come from surface brightness fluctuation calculations of \cite{mei:2007} and \cite{jerjen:2004} (4 and 2 objects, respectively) and globular cluster luminosity function distances of \cite{jordan:2007} (1 object). To the remaining two galaxies we assigned the average Virgo distances from \cite{mei:2007}. While the uncertainties on the distances are rather large, they are by definition smaller or equal to the uncertainties introduced through relying on projected values.

In Figure~\ref{dmtrend} we plot the dynamical-to-stellar mass ratios vs. 3D distances. We see that galaxies with higher $M_{dyn}/M_{star}$ are preferentially found in the cluster outskirts, whereas central objects show ratios that are visibly lower. All our objects have comparable stellar masses so there should be no differences between them introduced by the fact that objects of different masses are affected differently by the environmental forces. This trend agrees with the \cite{moore:1998} claim and is its first observational confirmation. However, given the limited number of studied galaxies and the uncertainties on the mass estimates, the results presented here are only indicative of a possible trend and await confirmation with larger samples.

\section{Discussion}

\begin{figure}
\center~
\includegraphics[width=0.95\columnwidth]{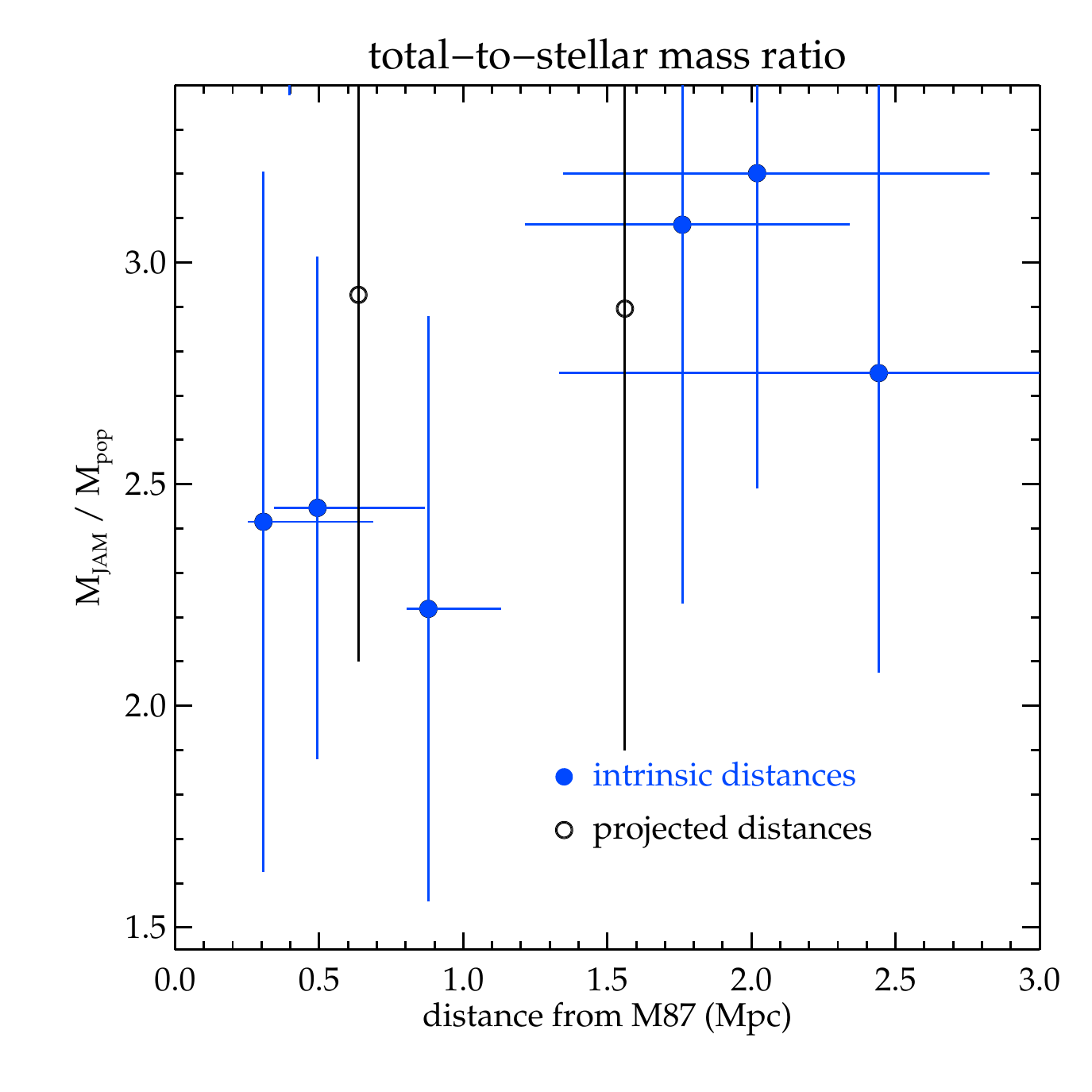}
\caption{\label{dmtrend} Dark-to-stellar matter ratio as a function of Virgocentric distance. Intrinsic distances are plotted where available, otherwise projected distances are provided. A tentative trend is seen: the ratio increases with increasing distance.}
\end{figure}

dEs have now generally come to be considered descendants of environmentally-transformed late-type galaxies (e.g. \citealt{conselice:2001}, \citeyear{conselice:2003}, \citealt{vanzee:2004a}a, \citeyear{vanzee:2004b}b, \citealt{lisker:2006a}a, \citeyear{lisker:2006b}b, \citeyear{lisker:2007}, \citeyear{lisker:2009}, \citealt{michielsen:2008}, \citealt{kormendy:2009}, \citeyear{kormendy:2012}, \citealt{toloba:2009}, \citeyear{toloba:2011}, \citeyear{toloba:2012}, \citealt{geha:2010}, \citealt{janz:2012}, \citealt{rys:2013}; but see \citealt{graham:2013} or \citealt{ferrarese:2006} for an alternative view). While it is hard to speculate about the exact properties of their progenitors, we do turn to low-redshift late-types for their possible characteristics. Naturally, the assumption that the progenitors of dEs resemble \textit{today's} late-type galaxies is quite likely not true. Today's late-types have also been evolving over the last few Gyrs and it is more correct to say that the two classes share a common progenitor. The differences in their current appearance and structure would then be caused precisely by the different environmental conditions in which they have found themselves (see also \citealt{lisker:2013} for a discussion on this topic). However, since an equally detailed analysis of high-redshift objects is currently not feasible, we need to rely on low-redshift clues in our search. The dataset used for the analysis presented here is the best currently available in the literature in terms of data quality, mass coverage and homogeneity.

The possible low-redshift progenitor galaxies include irregular and/or low-mass spiral galaxies. As can be seen in the profiles in Figure~\ref{lambda-profiles} the potential candidates all have angular momentum values higher than those of our dEs. What this implies is that we can exclude the scenarios in which only ram-pressure stripping is responsible for the transformation since it preserves angular momentum. A combination of mechanisms (the way we understand them today) must have come into play to create dEs.

The shapes of the velocity curves have important implications for our understanding of the progenitors-descendants issue. We can see that not only do we need to account for the loss of angular momentum but also for the fact that dE circular velocity curves are steeper than those of their presumed late-type progenitors. The implication is that either the true progenitors were already more compact at higher redshifts or that whatever transformation mechanism has acted on those galaxies managed to produce those effects. 

As for the former, it is interesting to note that blue compact dwarf galaxies (BCDs) have also been noted to have rotation curves steeper than dwarf irregulars of similar mass \citep{vanzee:2001}. The authors of the study also noted the difference of the angular momenta of BCDs and dEs, albeit it was then based on a small and radially limited sample of dEs which showed little to no rotation. Today we have plenty of evidence for the presence of stellar rotation in early-type dwarfs but it is still significantly lower than the values measured by \cite{vanzee:2001} from HI gas. Angular momentum still needs to be removed somehow.

As to the latter possibility, we can think of a scenario in which a more massive galaxy loses some portion of its mass due to tidal harassment. The tidal effects would then likely drive the remaining gas inwards, thus inducing a burst of star formation which could produce a ``bulge-like'' inner stellar distribution. This would take care both of the angular momentum loss as well as the change in the circular velocity curves steepness since the latter reflects the concentration of a galaxy: compare the individual curves of our galaxies where the steepest one belongs to the galaxy with the smallest effective radius. In fact, this could also take place efficiently even due to a  redistribution of stellar mass if e.g. a bar is formed as a result of harassment, as seen in the simulations of \cite{mastropietro:2005}.

If the paradigm of environmentally-induced secular evolution of dwarf galaxies is correct, then, as mentioned earlier, some dependence of certain galaxy properties (e.g. rotational support, DM fraction) on environmental local density is expected. The DM fraction -- intrinsic Virgocentric distance trend presented here (Figure~\ref{dmtrend}) is a tentative confirmation of those theoretical predictions.

Trend analyses that rely on projected (angular) distances from an assumed cluster center - and that is the majority of available results - ignore the depth argument and necessarily introduce noise to their results. For example, for galaxies located very centrally in projection we cannot be sure of their actual location, which could happen to be anywhere along the line of sight (albeit the probability for a central location is of course the highest). Only with large/complete samples and statistical deprojection are we able to get  more accurate trend results. Since the intrinsic distances are by definition at least as accurate as the projected ones, the intrinsic values should thus be used whenever possible.

However, such an analysis is thwarted by a number of issues. What can potentially dilute trends is the fact the the cluster obviously lacks spherical symmetry. Still, to the first-order approximation clustrocentric distance can be used as a measure of local density. However, how much a given galaxy has been affected by the environment does not depend on its current location within the cluster but rather on its trajectory within the cluster and how much time it has already spent in the cluster's high-density regions. Unfortunately, the orbital configuration of Virgo cluster galaxies is yet unknown and additionally we might also be dealing with ``splashback'' galaxies. Second, we cannot unequivocally determine what type of preprocessing (if any) took place before galaxies entered the cluster environment. 

Performing a wide-ranging analysis, one that combines morphology, stellar populations, kinematics and dynamics, could potentially help us shed more light on the issue. Still, it would be of vital importance to obtain orbital parameters of the galaxies. While it is not feasible directly due to the lack of proper motion information, what could potentially be of interest would be performing a statistical analysis using a combination of the timing argument and phase-space distribution functions such as that performed for the M31 group by \cite{watkins:2013}. Nevertheless, large samples are needed to make such an analysis significant and robust.

\section{Summary}
We use our published stellar kinematic results of \cite{rys:2013} to analyze the dynamical properties of 12 dwarf elliptical galaxies located in the Virgo cluster and in the field. We assess their rotational support using the specific angular momentum $\lambda_R$ and their dynamical masses and mass-to-light ratios within 1 effective radius by building Jeans axisymmetric MGE models.

We place our galaxies on the scaling relations for giant early types of \citeauthor{cappellari:2013a} (\citeyear{cappellari:2013a}a,b). We find that also for dEs $(M/L)_{dyn}=5.0\cdot\sigma^2_e\cdot R_e / (L\cdot G)$ provides an adequate estimate of the total mass, and also that the maximum circular velocity is tightly correlated with the velocity dispersion $\sigma_e$ within the efffective radius $R_e$. We analyze the $(M/L)_{dyn}$--$\sigma_e$ relation and confirm that dEs are offset from the sequence of giant galaxies, indicative of a higher dark matter content in the inner parts of dEs as compared with giant early-types. When analyzing the Mass Plane and its projections, we find that our dEs fall on the extension of the trend for massive galaxies and satisfy the same zone-of exclusion criteria. We therefore conclude that the validity of the relations can be extended to these low-mass systems.

We find that the specific angular momenta of dEs are significantly lower than those of the late-type galaxies of the CALIFA survey. This leads to the conclusion that when designing possible transformation paths we need to include processes that lower stellar angular momentum. We also compare our dEs circular velocity profiles to those of massive \citep{catinella:2006} and dwarf \citep{swaters:2009} late-type galaxies and find that dE curves are steeper and their absolute values higher than those of late-type galaxies of comparable luminosity. Therefore, we also need to account for this increase in concentration in our search for late-to-early type transformation mechanisms. Furthermore, we investigate whether any correlation exists between the dark matter fraction  and the 3D (i.e. intrinsic) clustrocentric distance and we find that galaxies in the cluster outskirts tend to have higher dark-to-stellar matter ratios. 

What this implies is that processes like ram-pressure stripping alone are not able to explain the observed characteristics of the dE class. Tidal harassment is currently the only available scenario with which we can explain all the above findings, unless the dE progenitors were already compact and had lower angular momenta at high redshift.

\section*{Acknowledgments}
\small
We thank the anonymous referee for their useful feedback and constructive comments that have helped optimize the presentation of this work. We thank Thorsten Lisker for fruitful discussions and comments on the draft version of this paper. AR acknowledges the repeated hospitality of the Max Planck Institute for Astronomy in Heidelberg, to which collaborative visits have contributed to the quality of this work. JFB acknowledges support from the Ram\'{o}n y Cajal Program financed by the Spanish Ministry of Economy and Competitiveness (MINECO). This research has been supported by the Spanish Ministry of Economy and Competitiveness (MINECO) under grants AYA2010-21322-C03-02 and AIB-2010-DE-00227. GvdV and JFB acknowledge the DAGAL network from the People Programme (Marie Curie Actions) of the European Union’s Seventh Framework Programme FP7/2007-2013/ under REA grant agreement number PITN-GA-2011-289313. The paper is based on observations obtained at the William Herschel Telescope, operated by the Isaac Newton Group in the Spanish Observatorio del Roque de los Muchachos of the Instituto de Astrof\'isica de Canarias.
\normalsize

\bibliography{biblio}

\appendix

\section{MGE models} 
\label{mgetables}
Here we provide a table with the parameters of our MGE models for each galaxy.

\begin{table*}
\begin{minipage}[b]{0.45\linewidth}\centering
\begin{tabular}{rrrr}

\hline
\hline
i & $I_i (L\odot/pc^2/'')$ & $\sigma_i ('')$ & $q_i$ \\
\hline
\hline
\multicolumn{4}{c}{ID\,0650}\\
\hline
       1 &       1069.49 &      0.378468 &      0.850000\\
       2 &       114.806 &       1.18991 &      0.862504\\
       3 &       182.034 &       3.65471 &      0.920090\\
       4 &       57.5667 &       6.13842 &      0.924325\\
       5 &       25.9653 &       17.5111 &      0.906618\\
       6 &       2.17554 &       60.3000 &      0.997371\\
\hline
\multicolumn{4}{c}{ID\,0918}\\
\hline
       1 &       7260.45 &      0.391084 &       1.00000\\
       2 &       571.102 &       1.09222 &       1.00000\\
       3 &       495.835 &       1.86097 &      0.683477\\
       4 &       519.354 &       2.65288 &      0.760678\\
       5 &       64.9346 &       4.21111 &      0.888877\\
       6 &       88.0896 &       6.93517 &      0.631031\\
       7 &       31.3808 &       8.61897 &      0.880134\\
       8 &       23.5969 &       18.2167 &      0.600000\\
\hline
\multicolumn{4}{c}{VCC\,0308}\\
\hline
       1 &       715.373 &      0.428931 &      0.850000\\
       2 &       134.687 &       1.59259 &      0.866843\\
       3 &       136.632 &       3.88933 &      0.877579\\
       4 &       27.8445 &       8.32938 &       1.00000\\
       5 &       14.9958 &       11.9222 &      0.850000\\
       6 &       45.9837 &       17.3479 &       1.00000\\
       7 &       2.51874 &       56.4000 &      0.889721\\
\hline
\multicolumn{4}{c}{VCC\,0523}\\
\hline
       1 &       271.403 &      0.772686 &      0.600000\\
       2 &       195.478 &       2.48074 &       1.00000\\
       3 &       65.6354 &       5.04799 &       1.00000\\
       4 &       89.4374 &       9.23906 &      0.772414\\
       5 &       40.8314 &       16.3182 &      0.642152\\
       6 &       25.9185 &       27.2888 &      0.755431\\
       7 &       2.98013 &       73.8600 &      0.873126\\
\hline
\multicolumn{4}{c}{VCC\,0929}\\
\hline
       1 &       1759.10 &      0.409422 &      0.844498\\
       2 &       199.352 &       1.28584 &       1.00000\\
       3 &       81.5290 &       3.03392 &       1.00000\\
       4 &       67.2596 &       3.82072 &      0.800000\\
       5 &       144.079 &       6.22959 &      0.868817\\
       6 &       79.7003 &       11.6443 &      0.866277\\
       7 &       23.1209 &       14.2399 &      0.993129\\
       8 &       13.4919 &       28.8132 &      0.800000\\
       9 &       5.95676 &       63.6000 &      0.949449\\
\hline
\multicolumn{4}{c}{VCC\,1036}\\
\hline
       1 &       981.259 &      0.401302 &       1.00000\\
       2 &       217.888 &       1.41007 &      0.978295\\
       3 &       194.158 &       4.12913 &      0.358271\\
       4 &       160.456 &       5.90150 &      0.489067\\

\end{tabular}
\end{minipage}
\hspace{0.5cm}
\begin{minipage}[b]{0.45\linewidth}
\centering
\begin{tabular}{rrrr}
\hline
\hline
i & $I_i (L\odot/pc^2/'')$ & $\sigma_i ('')$ & $q_i$ \\
\hline
\hline
       5 &       87.1061 &       9.07012 &      0.502980\\
       6 &       38.0631 &       17.7198 &      0.343545\\
       7 &       16.4075 &       23.6829 &      0.415251\\
       8 &       17.3937 &       30.1697 &      0.449681\\
       9 &       3.19672 &       67.2000 &      0.718662\\
\hline
\multicolumn{4}{c}{VCC\,1087}\\
\hline
       1 &       233.067 &      0.653224 &       1.00000\\
       2 &       40.3476 &       1.68518 &      0.909911\\
       3 &       57.2105 &       3.18183 &      0.998769\\
       4 &       99.4211 &       4.92218 &      0.730553\\
       5 &       48.2223 &       9.78084 &      0.600000\\
       6 &       34.5194 &       12.9203 &      0.822136\\
       7 &       26.9525 &       27.5884 &      0.623184\\
       8 &       1.97365 &       81.9600 &      0.600000\\
\hline
\multicolumn{4}{c}{VCC\,1261}\\
\hline
       1 &       1152.55 &      0.580318 &      0.753789\\
       2 &       271.213 &       2.30326 &      0.826910\\
       3 &       95.0538 &       5.24197 &      0.611075\\
       4 &       69.5415 &       7.64374 &      0.654186\\
       5 &       81.9032 &       14.7032 &      0.553211\\
       6 &       28.9570 &       29.1676 &      0.576417\\
       7 &       2.01664 &       55.6677 &       1.00000\\
\hline
\multicolumn{4}{c}{VCC\,1431}\\
\hline
       1 &       458.400 &      0.504625 &       1.00000\\
       2 &       195.277 &       1.05516 &       1.00000\\
       3 &       230.792 &       2.59098 &      0.921091\\
       4 &       78.6852 &       3.50216 &       1.00000\\
       5 &       164.557 &       6.59457 &      0.956387\\
       6 &       47.3332 &       12.1505 &       1.00000\\
       7 &       2.40164 &       26.4000 &       1.00000\\
\hline
\multicolumn{4}{c}{VCC\,1861}\\
\hline
       1 &       509.637 &      0.548779 &      0.910000\\
       2 &       152.334 &       3.86571 &       1.00000\\
       3 &       44.7615 &       8.39131 &       1.00000\\
       4 &       19.6991 &       9.91900 &       1.00000\\
       5 &       14.9746 &       17.9168 &       1.00000\\
       6 &       8.88512 &       29.2838 &      0.910000\\
       7 &      0.298521 &       64.5600 &      0.910000\\
\hline
\multicolumn{4}{c}{VCC\,2048}\\
\hline
       1 &       363.571 &      0.637387 &       1.00000\\
       2 &       156.049 &       1.83101 &      0.813336\\
       3 &       223.679 &       2.49658 &      0.300000\\
       4 &       203.764 &       5.22195 &      0.317908\\
       5 &       160.523 &       5.99154 &      0.572504\\
       6 &       87.9913 &       9.78822 &      0.573087\\
       7 &       42.7346 &       24.8420 &      0.370404\\
       8 &       6.26979 &       48.7935 &      0.439262\\

\end{tabular}
\end{minipage}

\caption[mge]{MGE gaussian fit parameters: surface densities, sigmas, and axial ratios of the gaussian components.}
\label{mge}
\end{table*}

\section{Circular velocities}
\label{circvel}

Here we plot the individual circular velocity profiles together with their uncertainties based on mcmc simulations. The observed stellar velocity $V$ and velocity dispersion $\sigma$ profiles are also plotted for comparison. 

\begin{figure*}
\center~
\includegraphics[width=0.66\columnwidth]{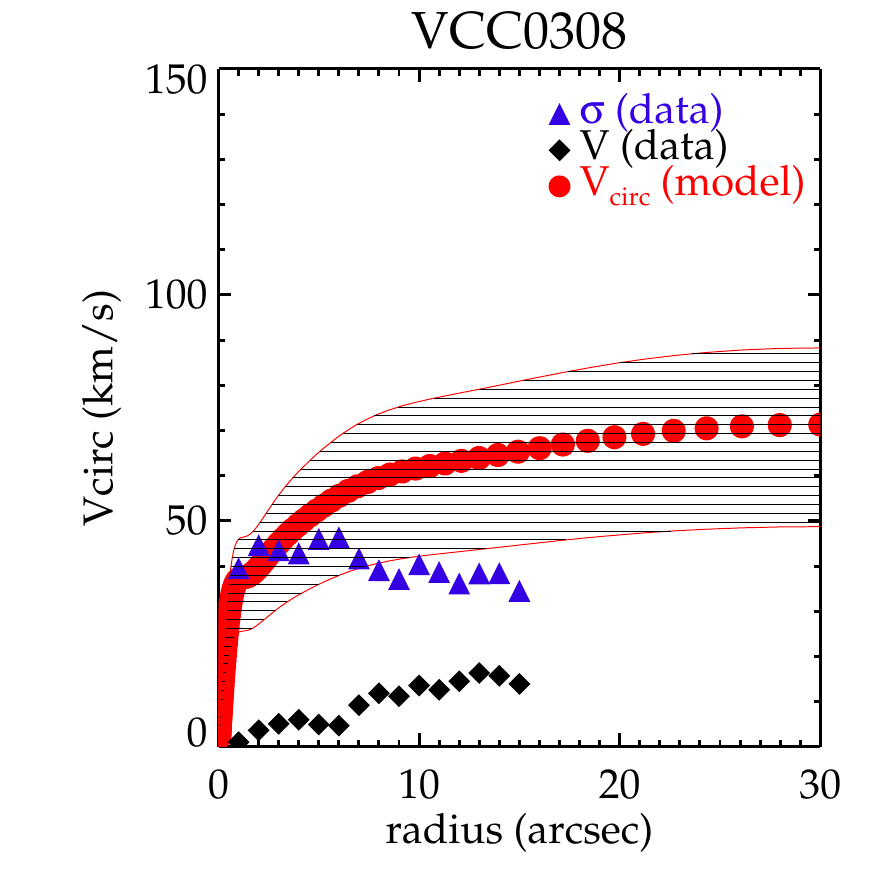}
\includegraphics[width=0.66\columnwidth]{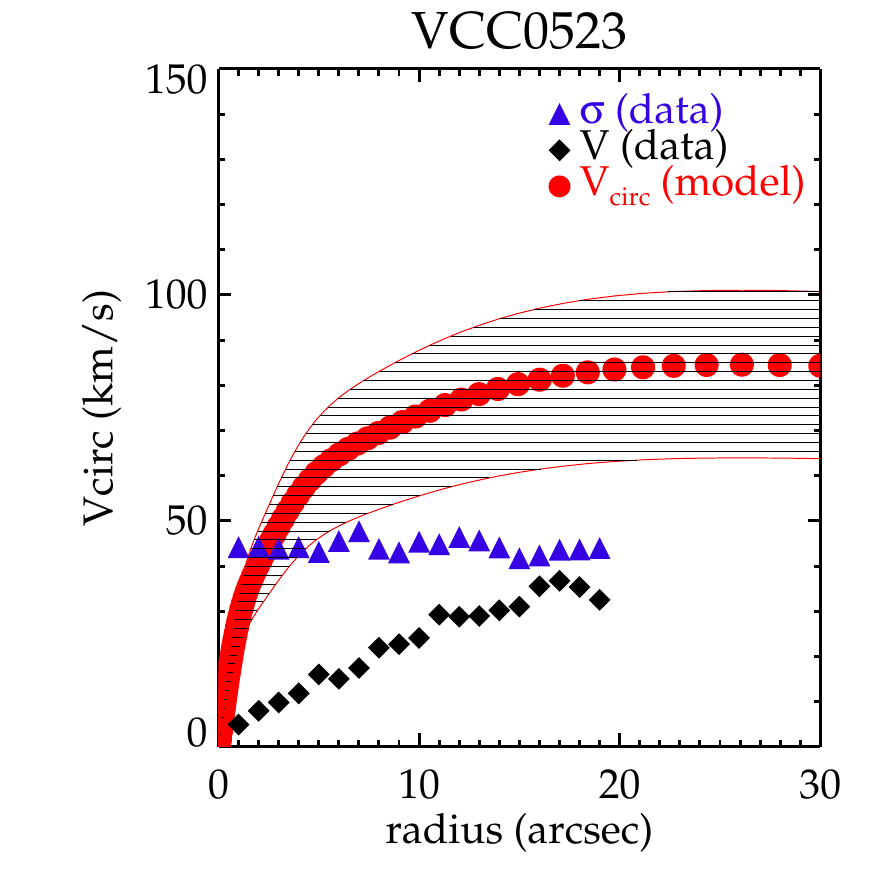}
\includegraphics[width=0.66\columnwidth]{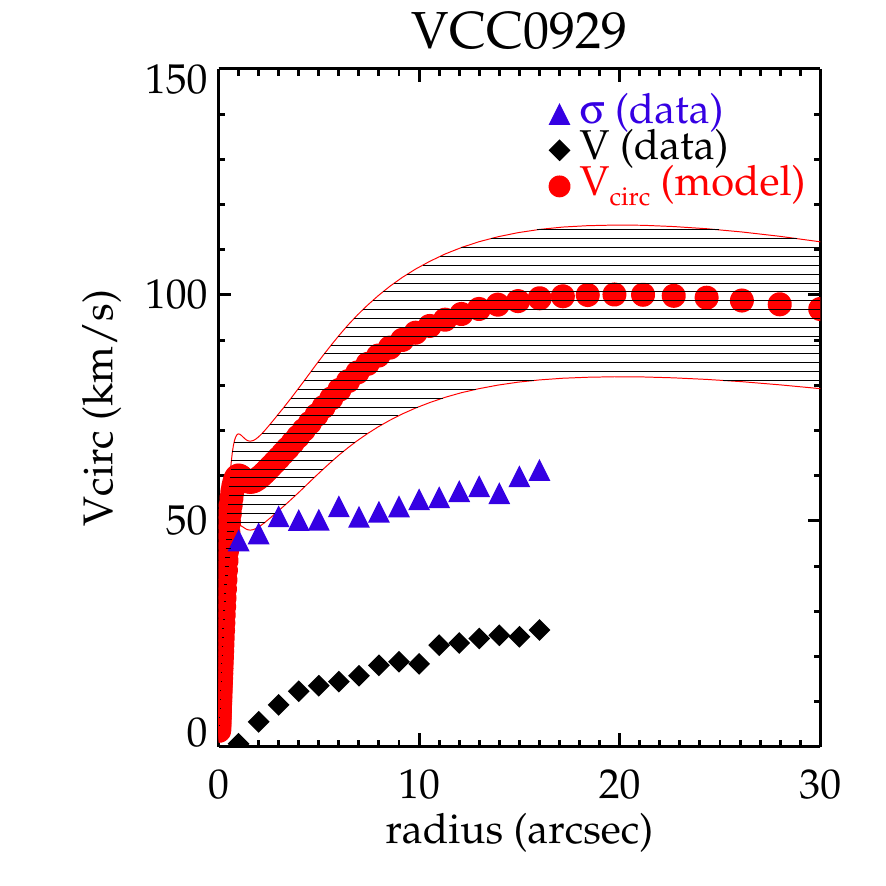}
\includegraphics[width=0.66\columnwidth]{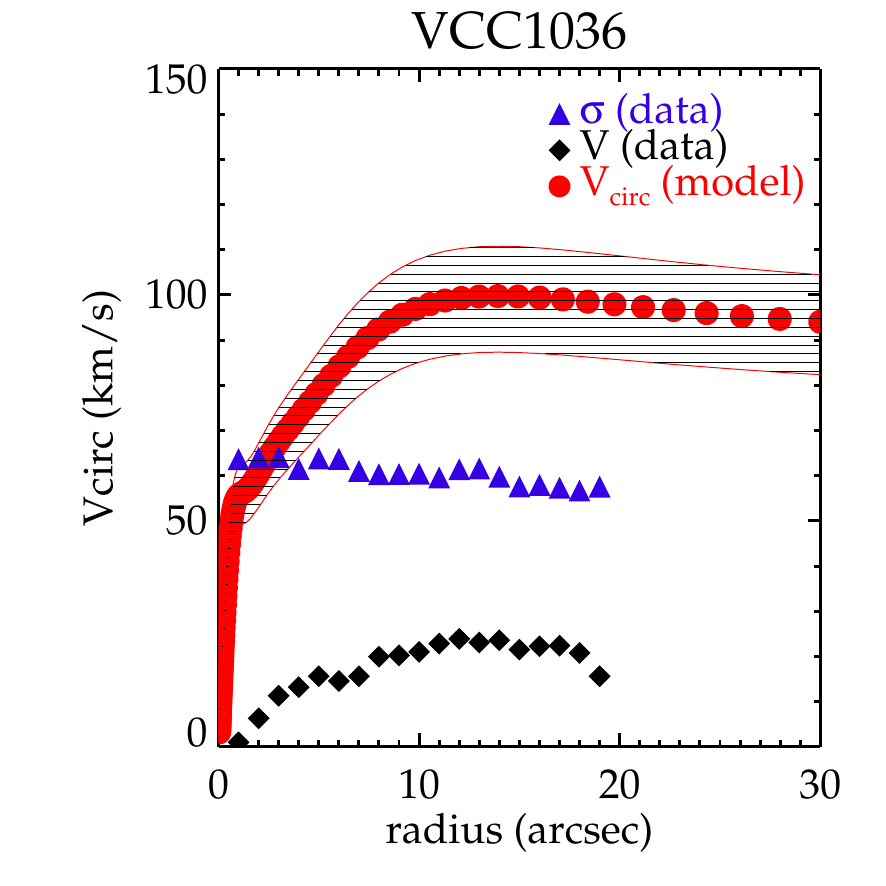}
\includegraphics[width=0.66\columnwidth]{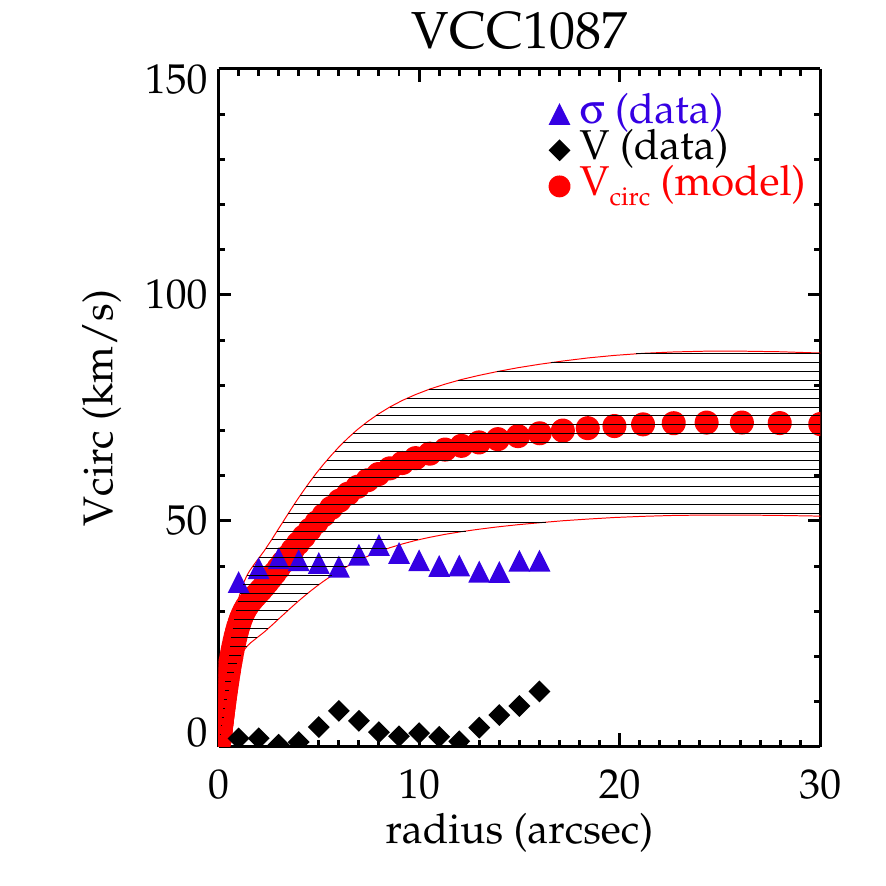}
\includegraphics[width=0.66\columnwidth]{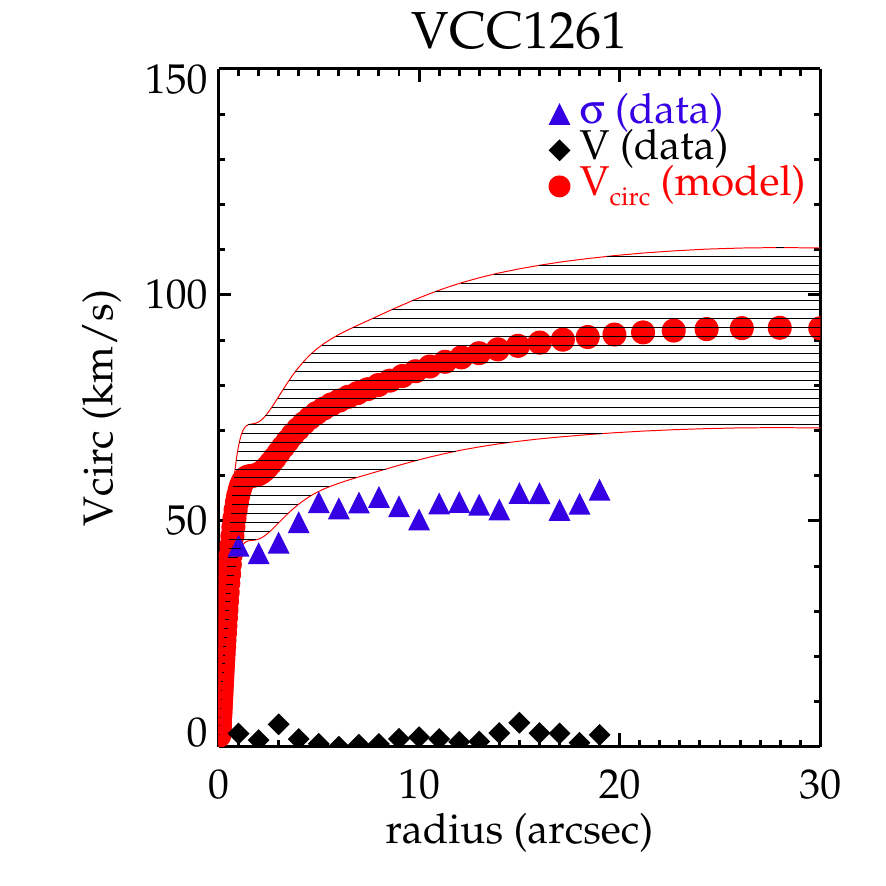}
\includegraphics[width=0.66\columnwidth]{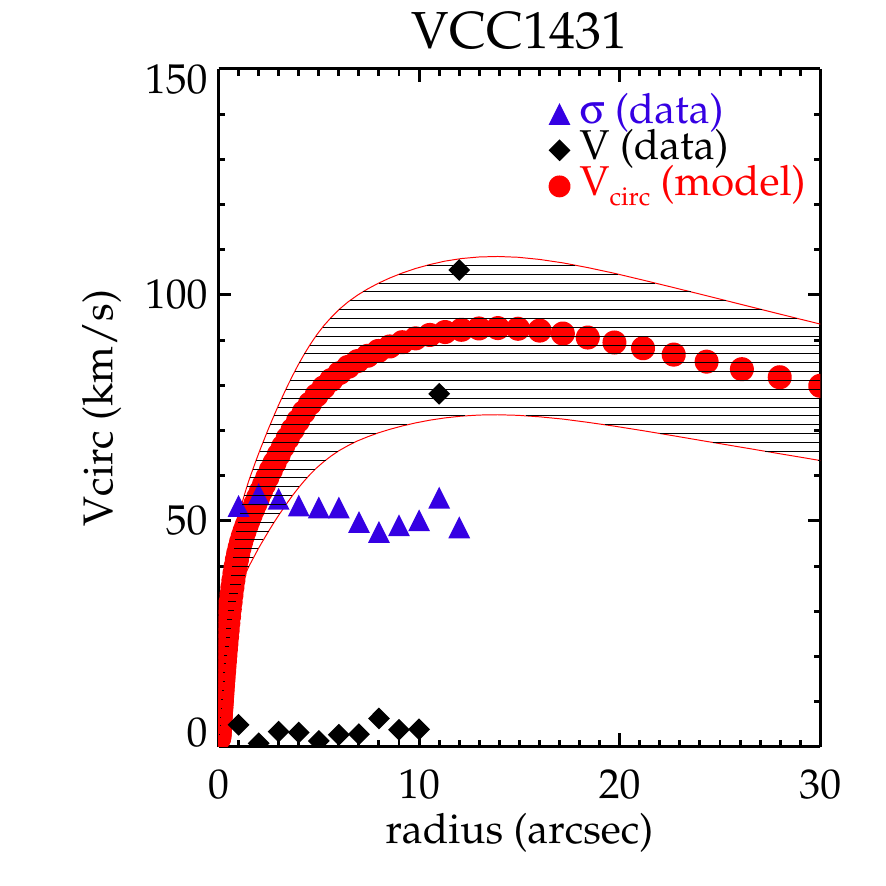}
\includegraphics[width=0.66\columnwidth]{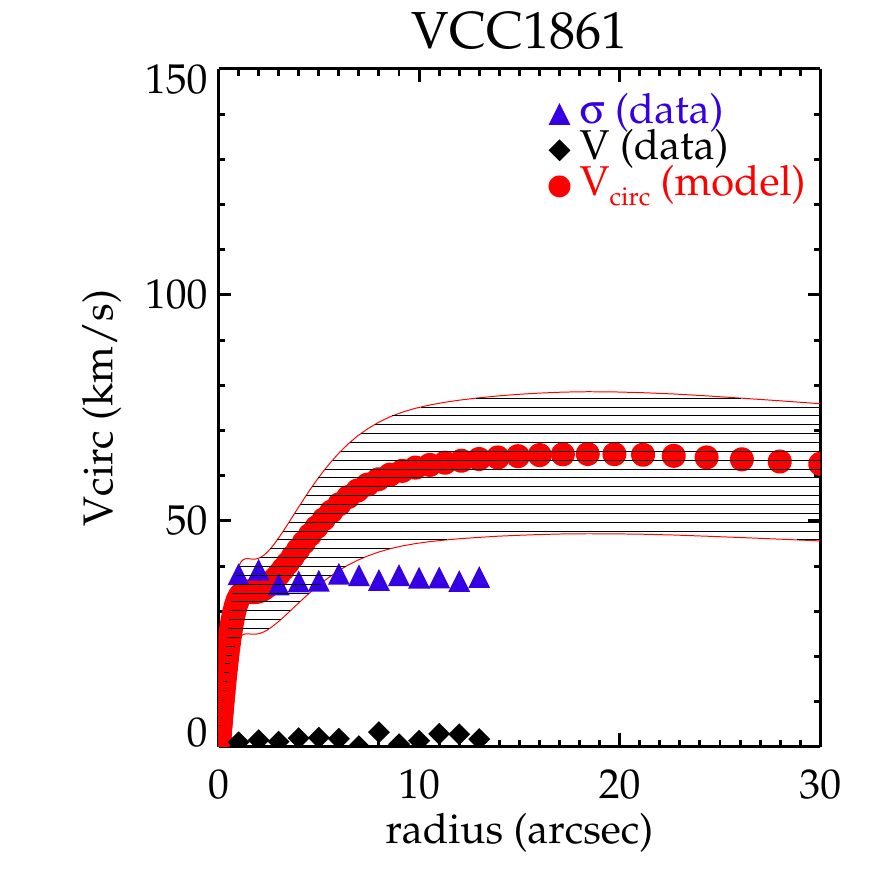}
\includegraphics[width=0.66\columnwidth]{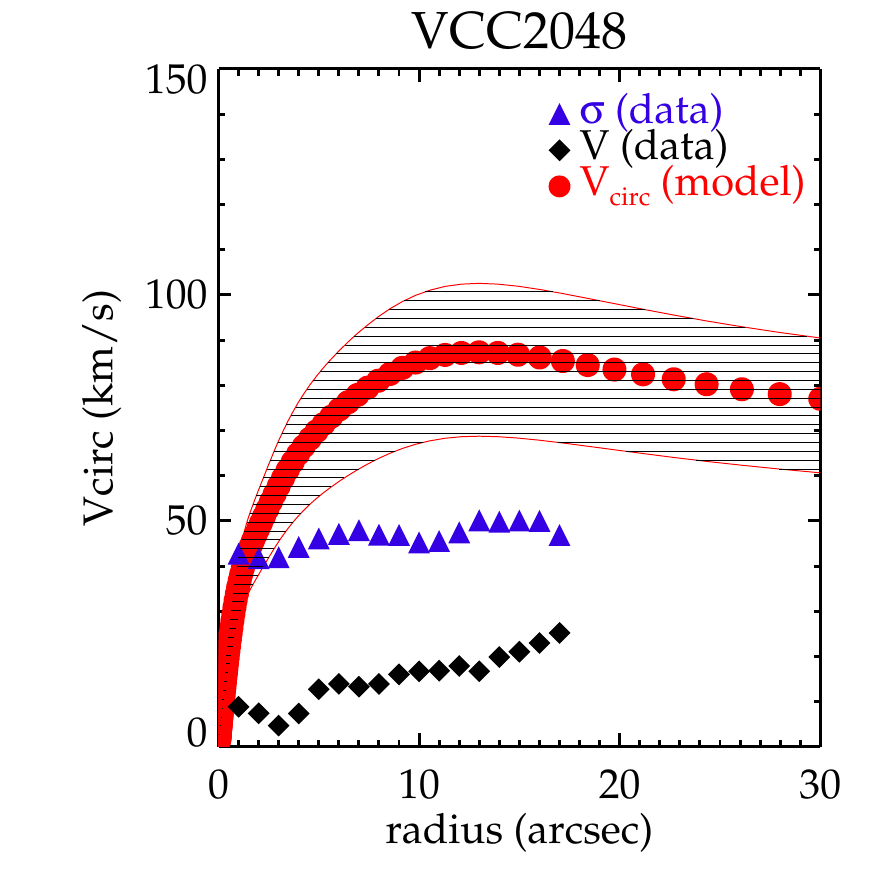}
\includegraphics[width=0.66\columnwidth]{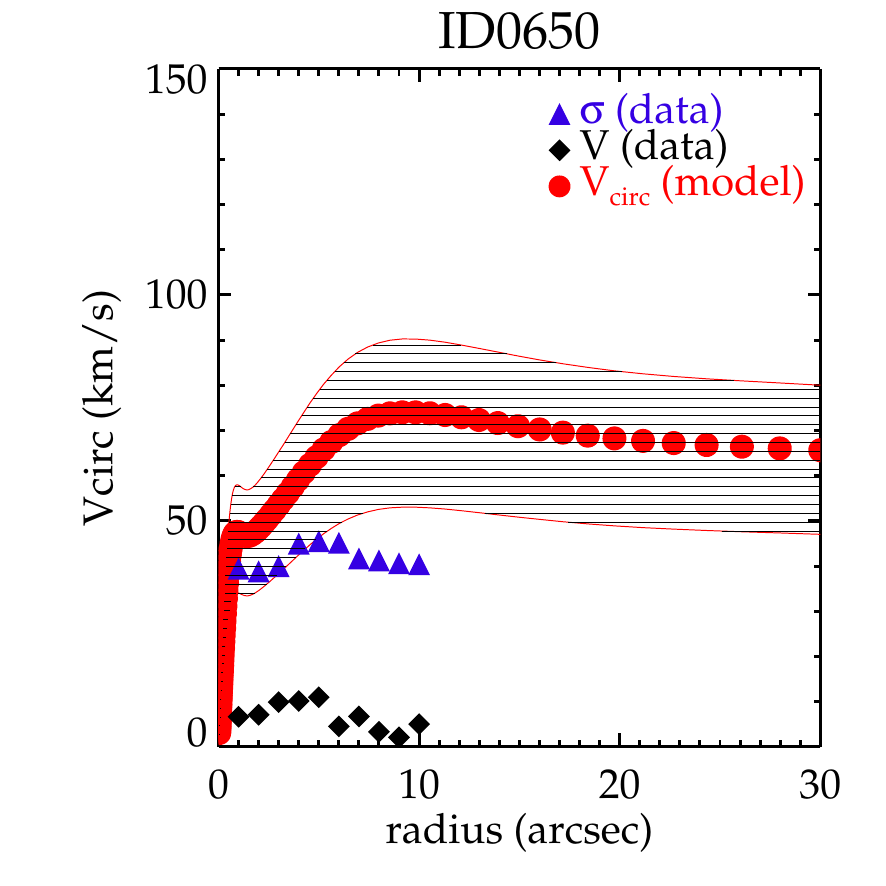}
\includegraphics[width=0.66\columnwidth]{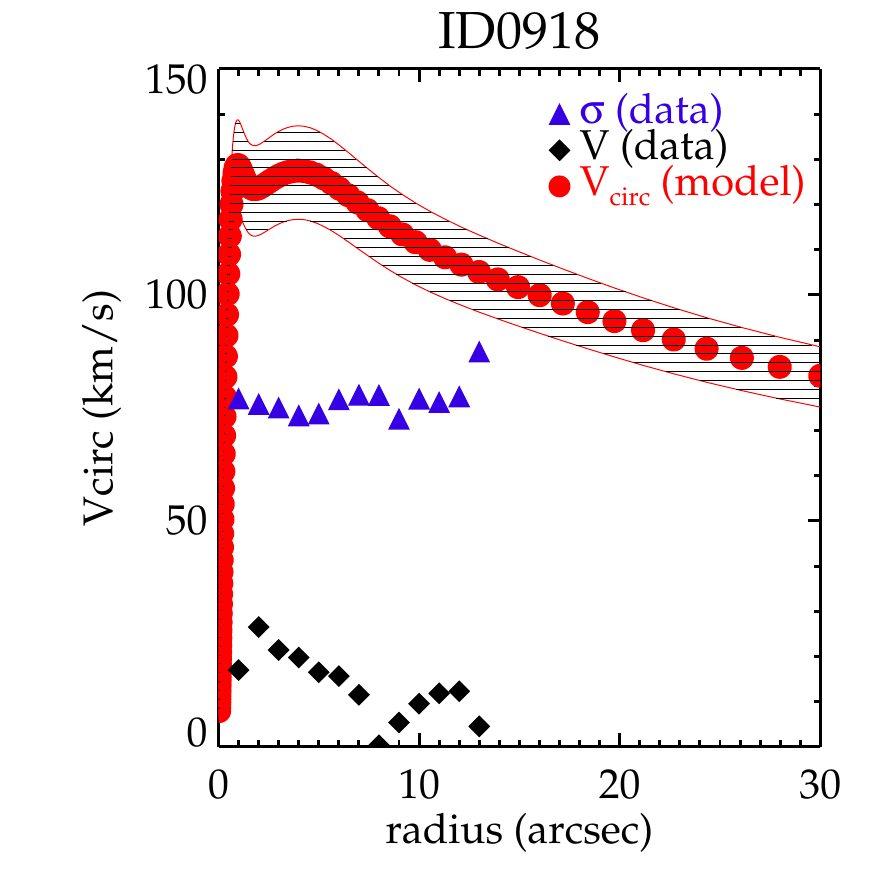}
\caption{Circular velocity ($V_{circ}$) profiles for individual galaxies. $V_{circ}$ is shown as red dots and the shaded regions indicate the uncertainties. The measured $V$ and $\sigma$ values are shown with black diamonds and blue triangles, respectively.}
\end{figure*}

\section{Jeans models tests}
\label{JAMmodelstests}
To estimate the accuracy of our dynamical model fits, we created mock galaxy data, starting with varying the relative error on the measured second velocity moment (i.e. $V_{RMS}/\Delta\,V_{RMS}$). One of our models had the same errors as our data, and for the other we assumed 10 times lower relative errors, comparable with the values of the massive elliptical galaxies from the SAURON or ATLAS3D surveys. We varied the input parameters and the recovery methods in the following ways:
\begin{itemize}
 \item isotropic model ($\beta=0.0$), recovered assuming correct inclination
 \item isotropic model ($\beta=0.0$), recovered assuming incorrect inclination (incl=90$^o$)
 \item anisotropic model ($\beta=-0.5$), recovered assuming correct inclination
 \item anisotropic model ($\beta=-0.5$), recovered assuming incorrect inclination (incl=90$^o$)
\end{itemize}
All the recovered models are self-consistent (i.e. mass-follows-light). We were interested in knowing how would the recovery depend on not only the data quality (i.e. $V_{RMS}/\Delta\,V_{RMS}$), but also the underlying anisotropy and the assumed inclination.

The \textit{walker} plots and final steps' histograms for each of the scenarios, together with the input recovered values of the $(M/L)_{dyn}$ and $\beta$ are plotted in Figure~\ref{mcmc-appendix_1}. All the values are also given in Table~\ref{mctable}.

The first important conclusion from the tests is that what prohibits us from recovering $\beta$ is not our data quality -- since the absolute errors are low -- but the fact that we are deadling with low dispersion objects. This is evidenced in the very accurate value recovery for all our high $V_{RMS}/\Delta\,V_{RMS}$ models (comparable to the highest quality data for massive early-types of the SAURON survey). For the same models but with low $V_{RMS}/\Delta\,V_{RMS}$, characteristic of our data, we are still able to accurately recover the M/L (albeit with much larger error bars) but the $ \beta$ recovery largely fails due to the errors spanning most of the allowed parameter range.

The second result deals with the fact that for all the used input/output combinations of the M/L and $\beta$, the recovered values agree well to within the errors. This gives us confidence that the values obtained from the simulations run on real data can be trusted.

\begin{figure*}
\centering
\includegraphics[width=0.33\textwidth]{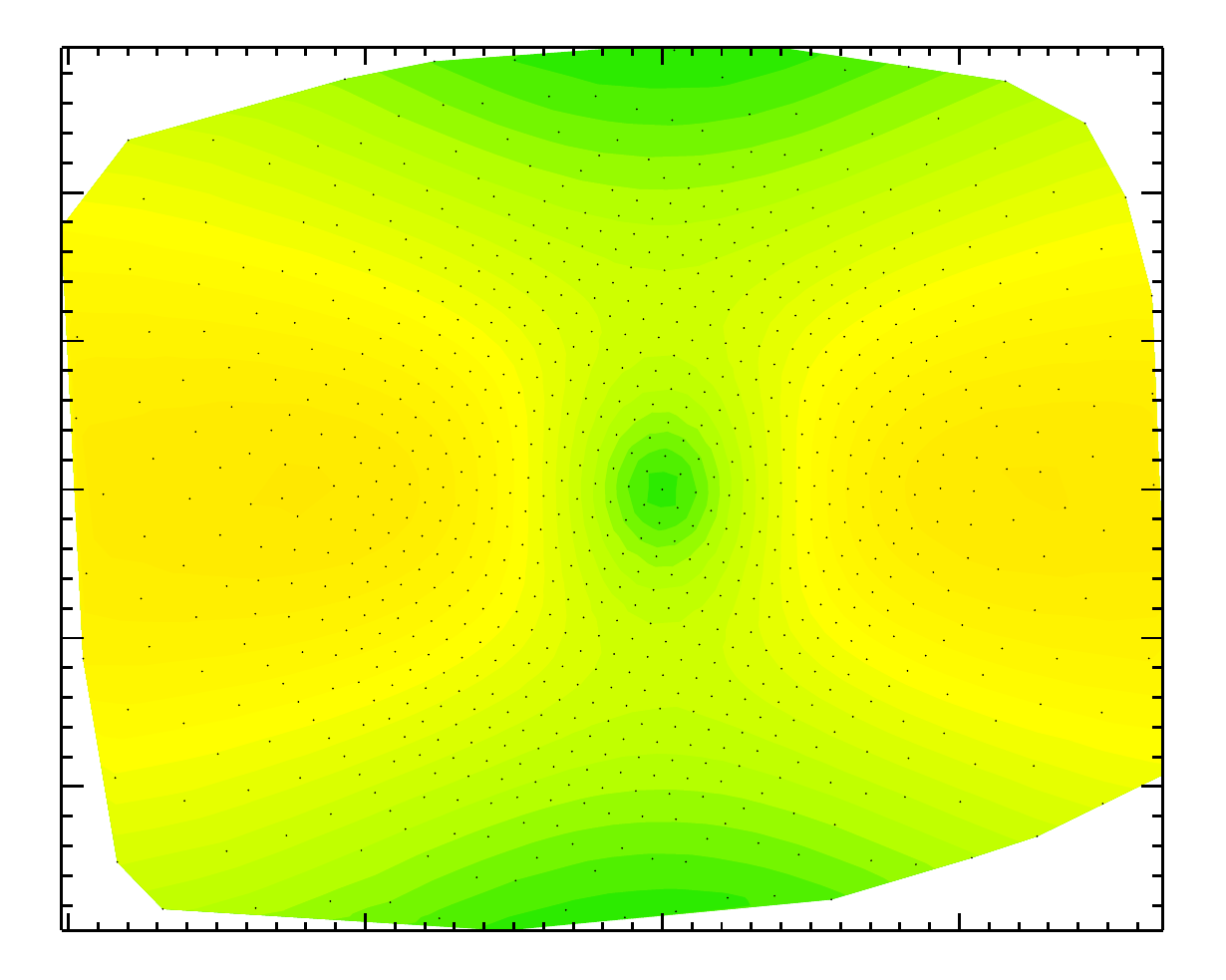}
\includegraphics[width=0.33\textwidth]{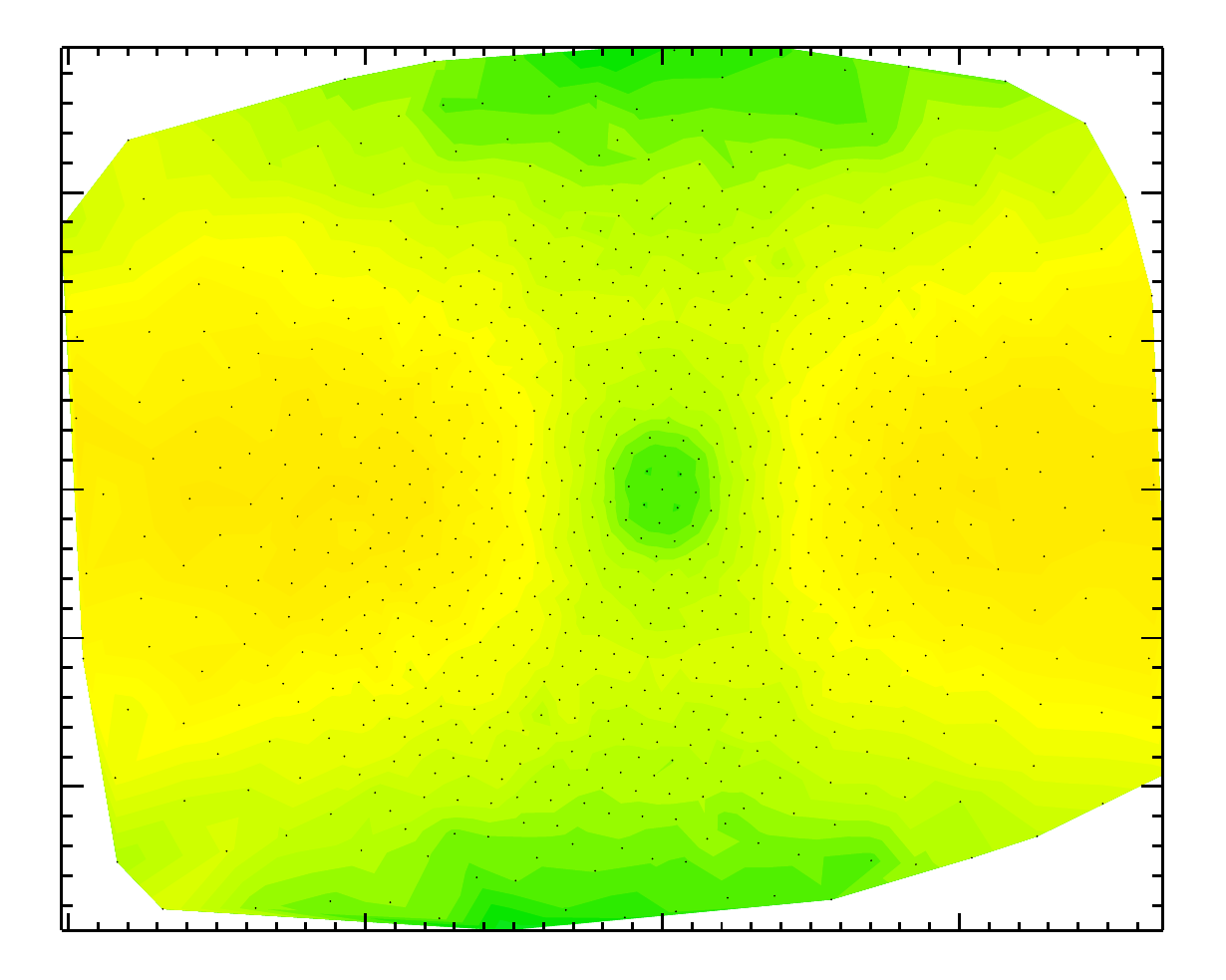}
\includegraphics[width=0.33\textwidth]{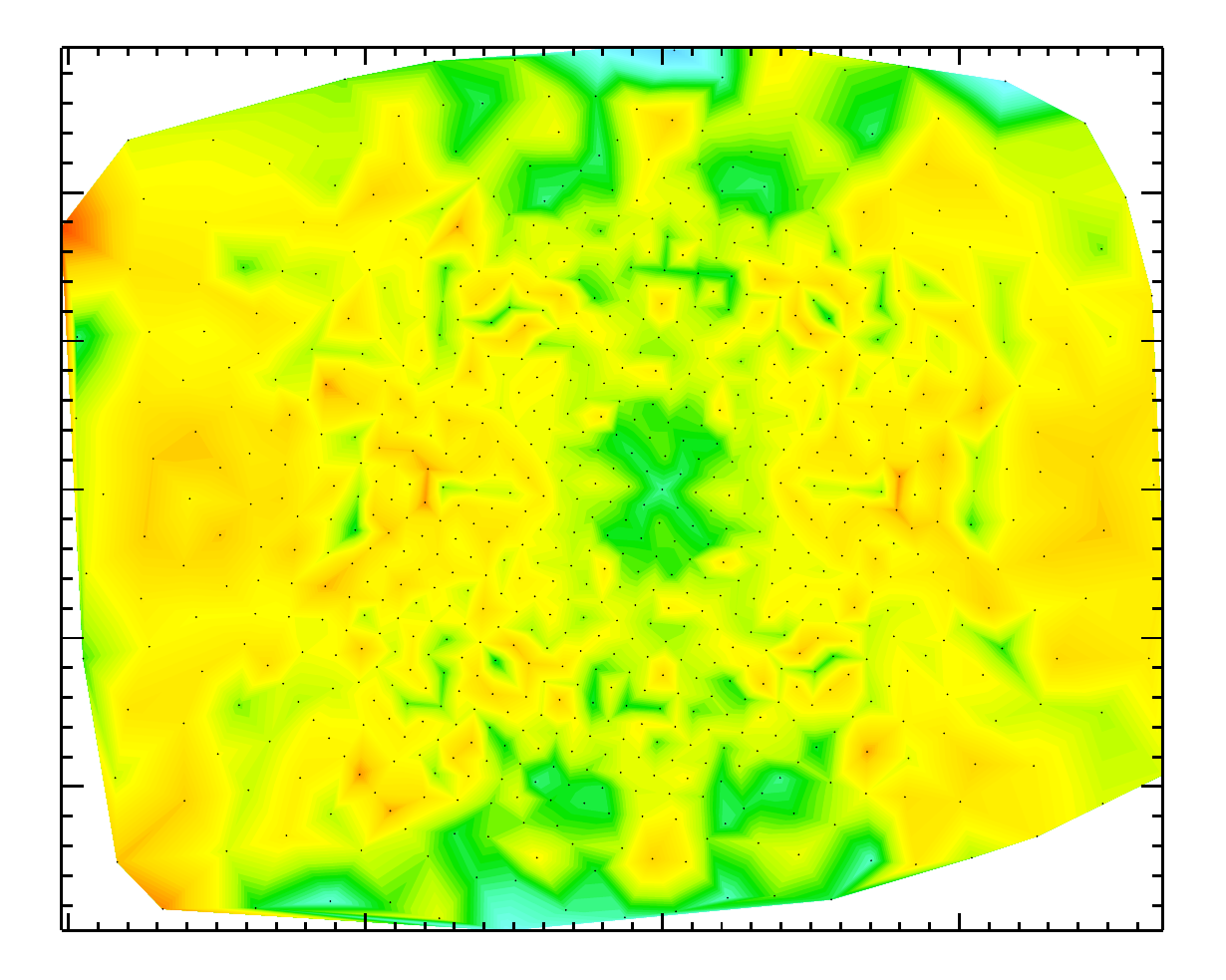}
\caption{Symmetrized second velocity moment for the model data used in the mcmc tests, here with the input value of $\beta$=0.0. From left to right we show: an unperturbed model, a high $V_{rms}/\Delta V_{rms}$ model (10\% of the errors for our galaxies), and a low $V_{rms}/\Delta V_{rms}$ model, resembling our real data. The fields are 30''x37'' in size and the plotted $V_rms$ range is 0-80 km/s.}
\label{mcmc-appendix_01}
\end{figure*}

\begin{figure*}
\centering
\includegraphics[width=0.33\textwidth]{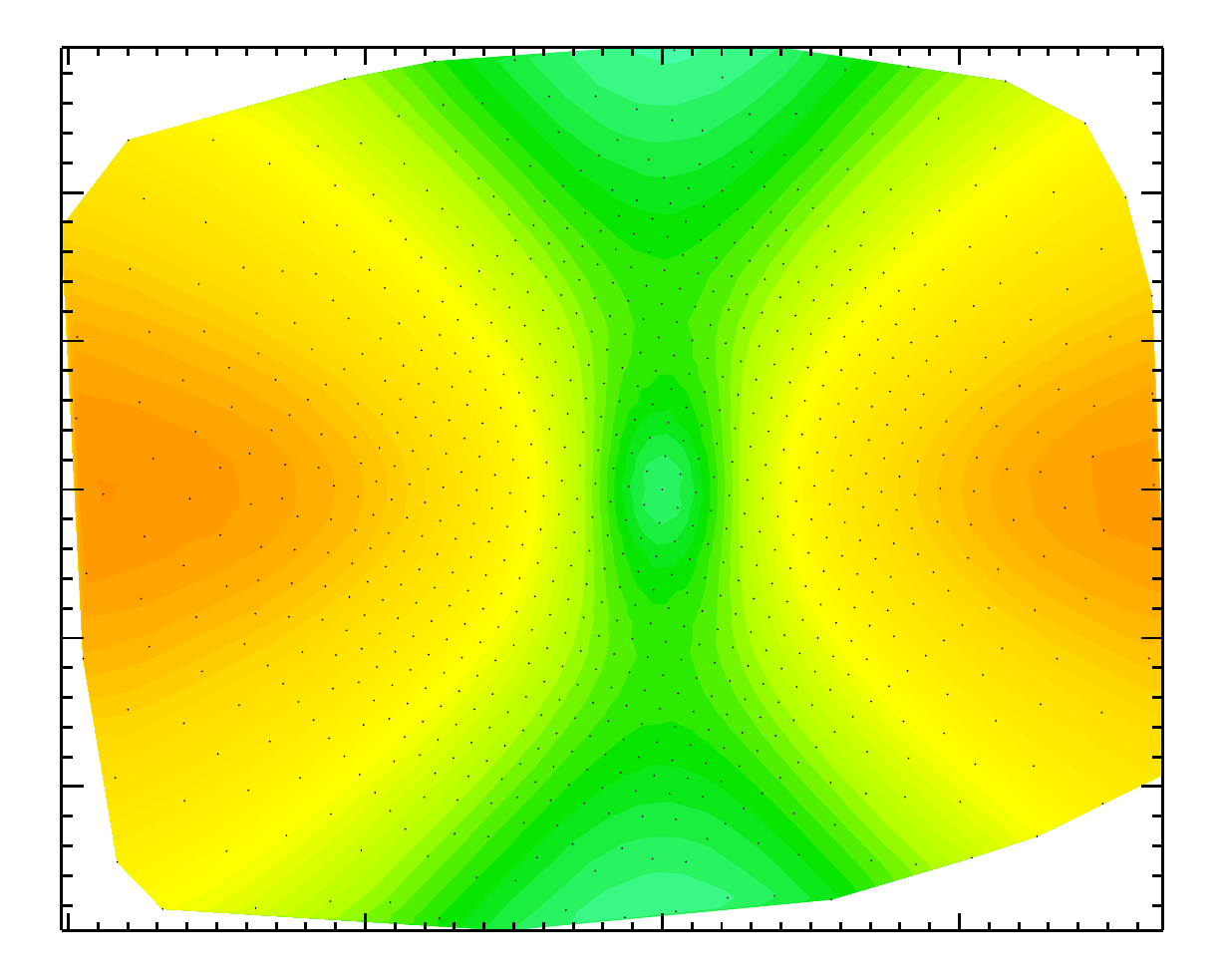}
\includegraphics[width=0.33\textwidth]{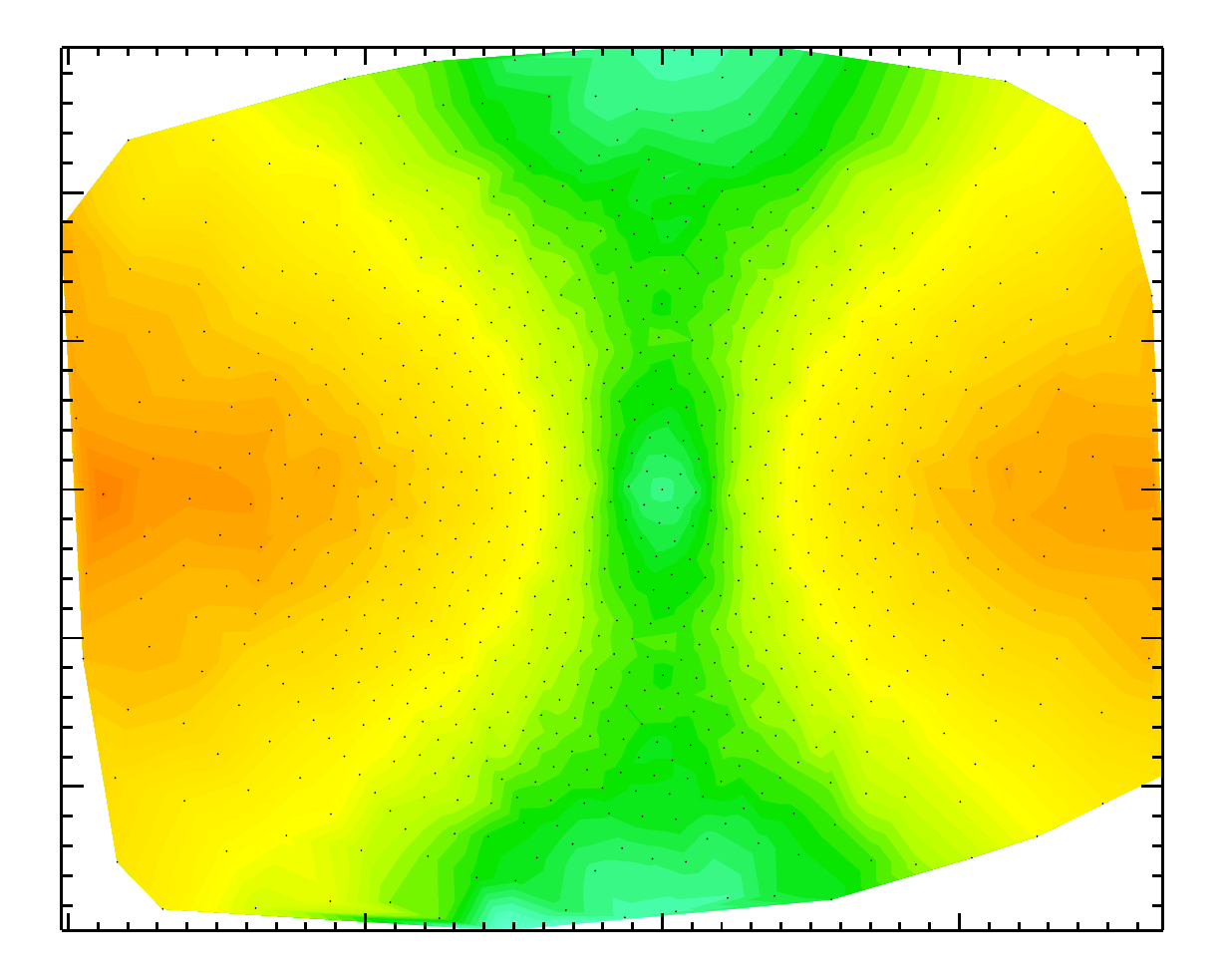}
\includegraphics[width=0.33\textwidth]{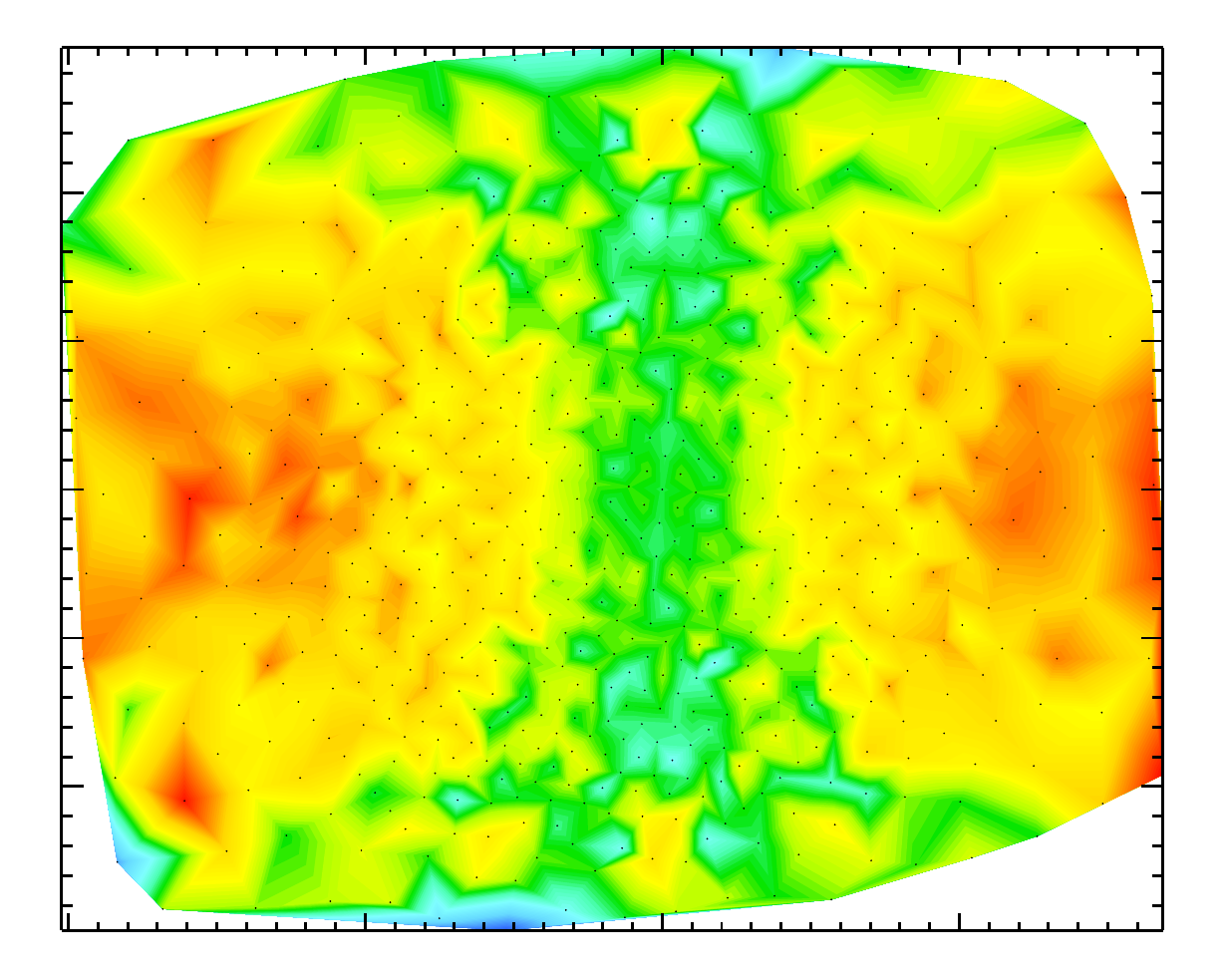}
\caption{As in Figure~\ref{mcmc-appendix_01} but for models with the input anisotropy $\beta$=-0.5.}
\label{mcmc-appendix_02}
\end{figure*}

\begin{figure*}
\centering
isotropic models, correct inclination\\
\includegraphics[width=0.41\textwidth,angle=0]{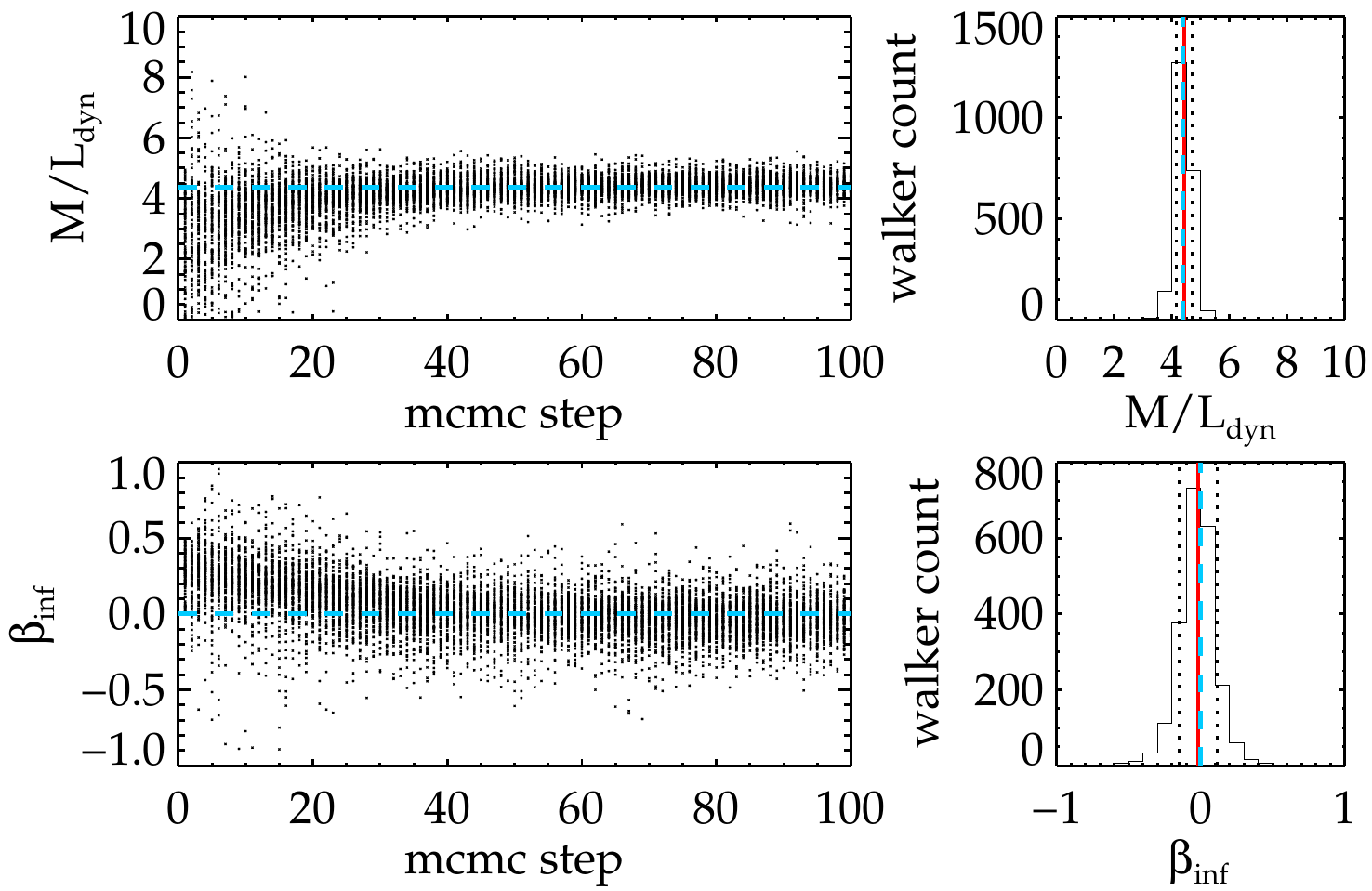}
\includegraphics[width=0.41\textwidth,angle=0]{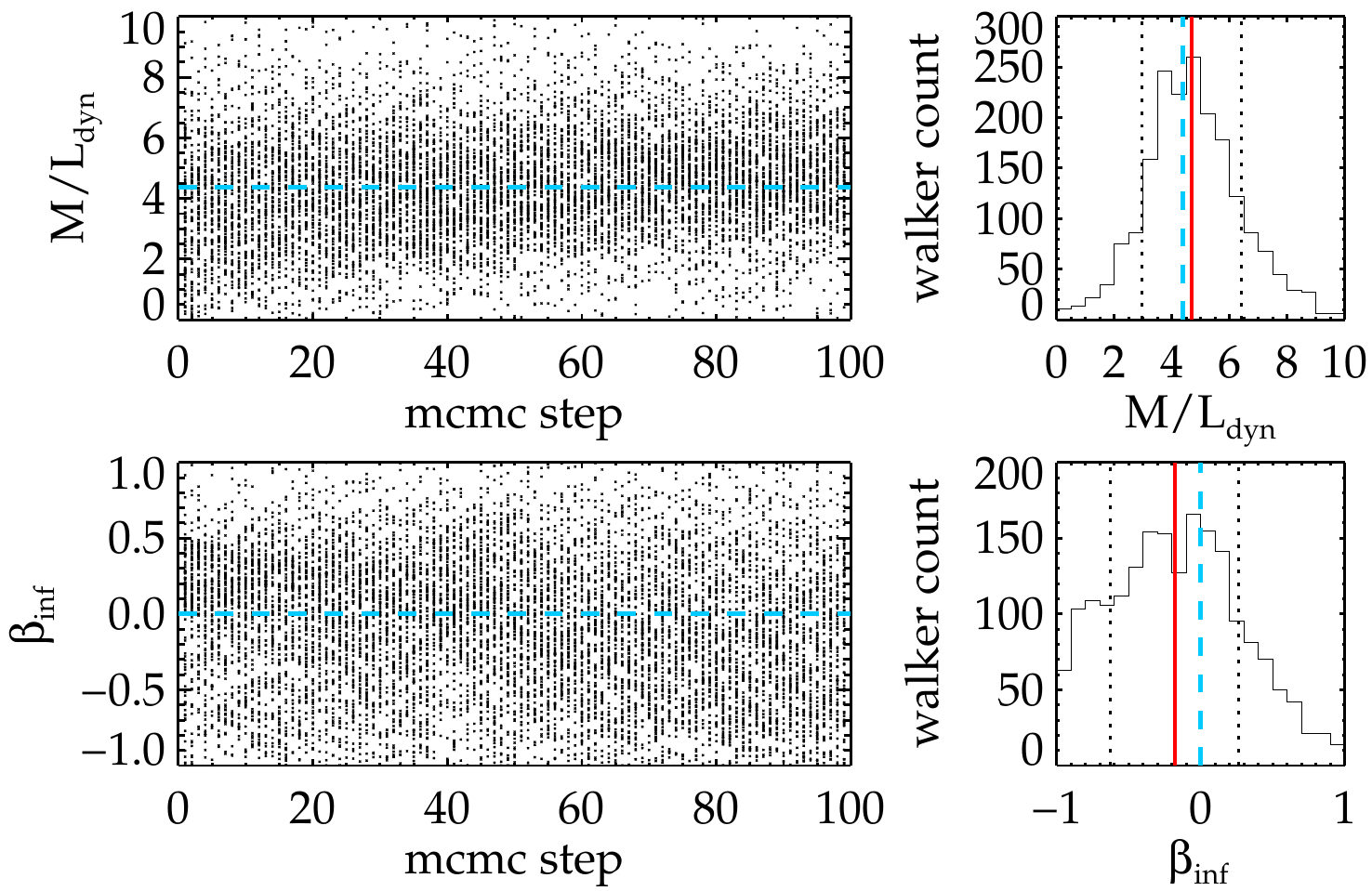}\\
isotropic models, edge-on\\
\includegraphics[width=0.41\textwidth,angle=0]{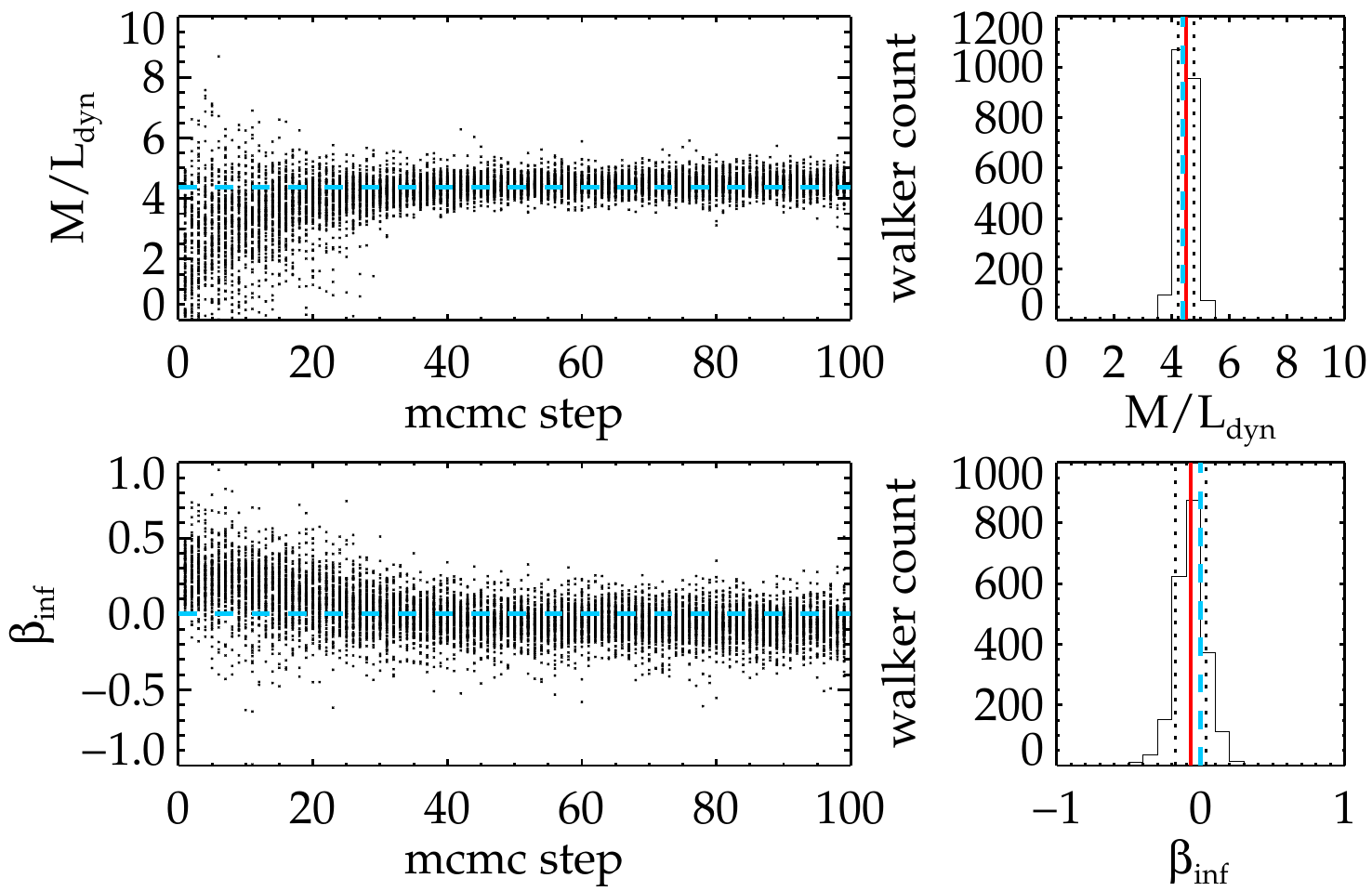}
\includegraphics[width=0.41\textwidth,angle=0]{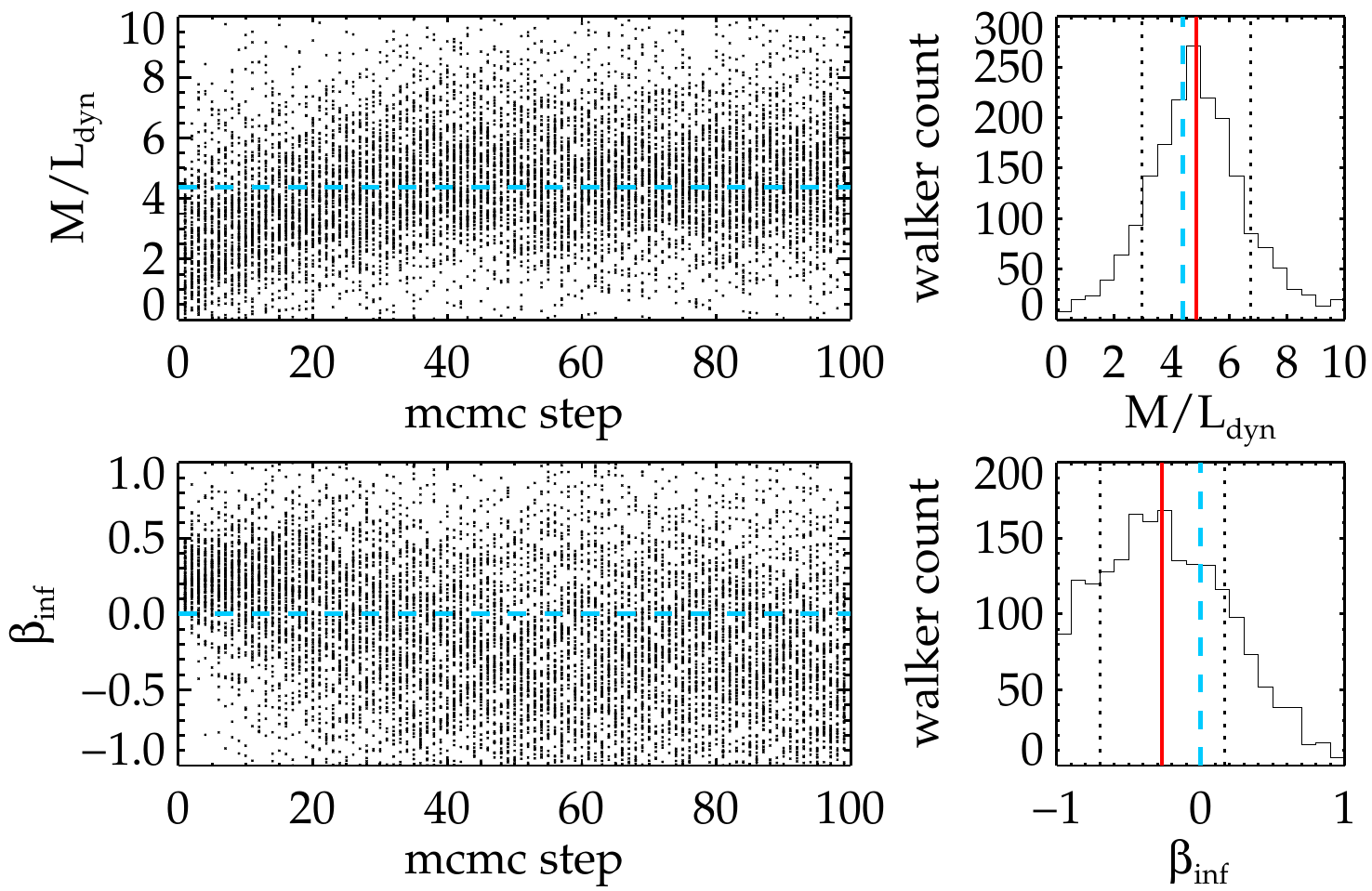}
\caption{MCMC test results for two of our model data: where in the recovery process we used the correct object inclination (\textit{top}) and where we deliberately assumed incorrect $i=90^o$ (\textit{bottom}). The results are shown for both ideal -- high $V_{rms}/\Delta\,V_{rms}$ -- models (\textit{left}) and those with $V_{rms}/\Delta\,V_{rms}$ resembling our real data (\textit{right}). In each of the scatter plots we show \textit{walker} positions in the paramter space (i.e. values of M/L and $\beta$ as a function of step/run of the simulations). The input values are shown with blue dashed lines and then in histogram plots the recovered values and their errors are shown with red solid lines and black dotted lines, respectively.}
\label{mcmc-appendix_1}
\end{figure*}

\begin{figure*}
\centering
anisotropic models, correct inclination\\
\includegraphics[width=0.41\textwidth,angle=0]{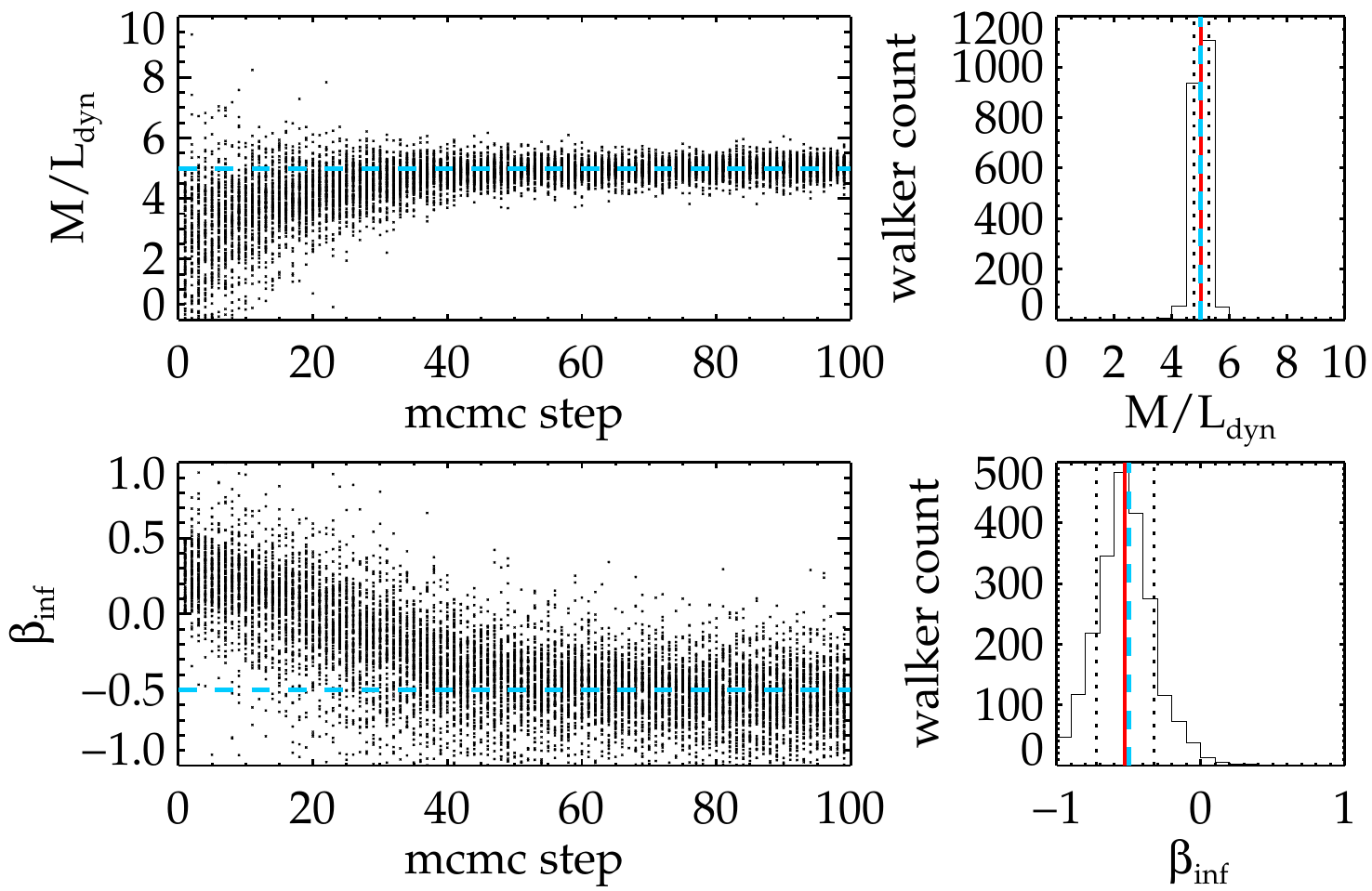}
\includegraphics[width=0.41\textwidth,angle=0]{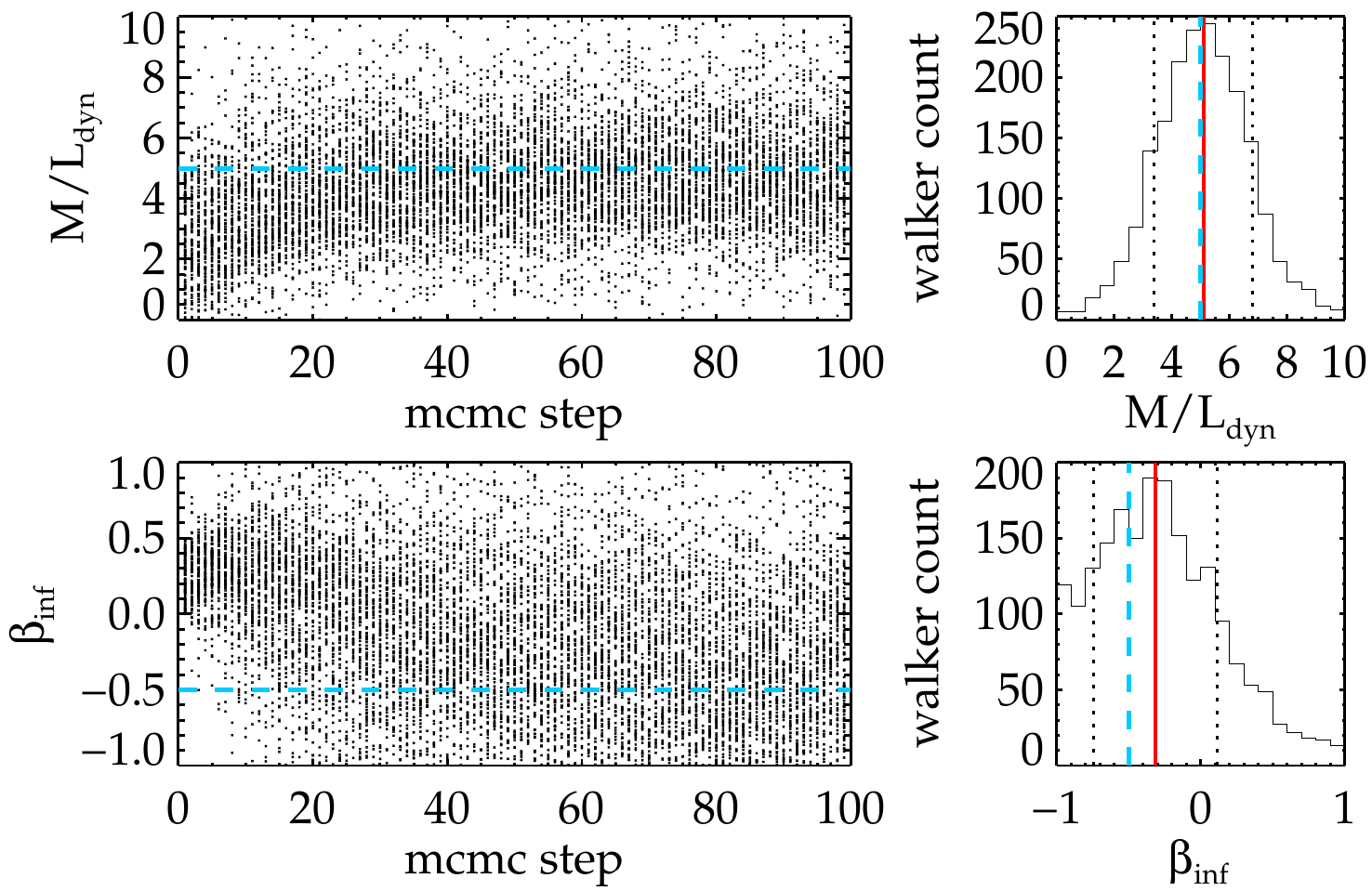}\\
anisotropic models, edge-on\\
\includegraphics[width=0.41\textwidth,angle=0]{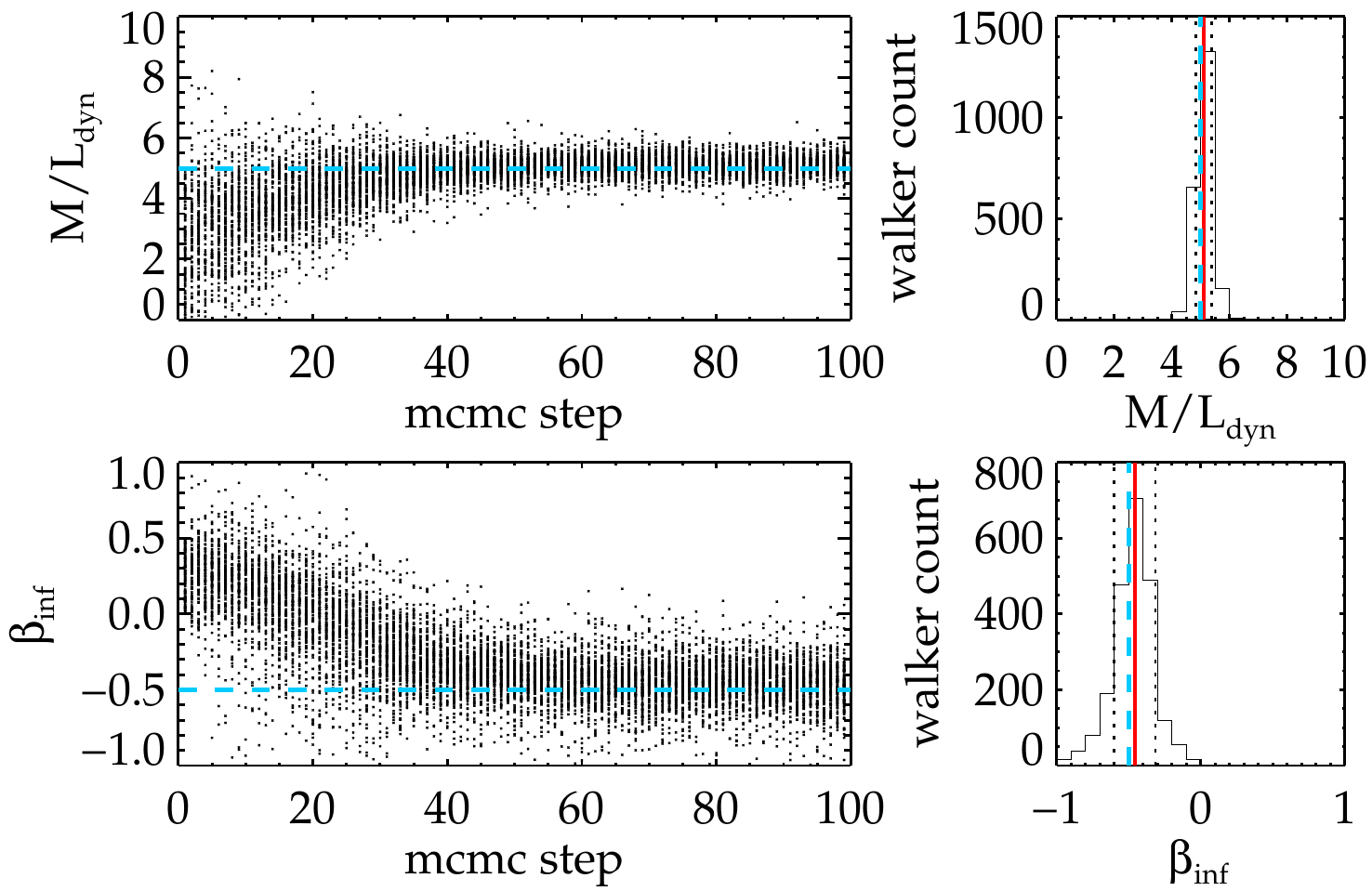}
\includegraphics[width=0.41\textwidth,angle=0]{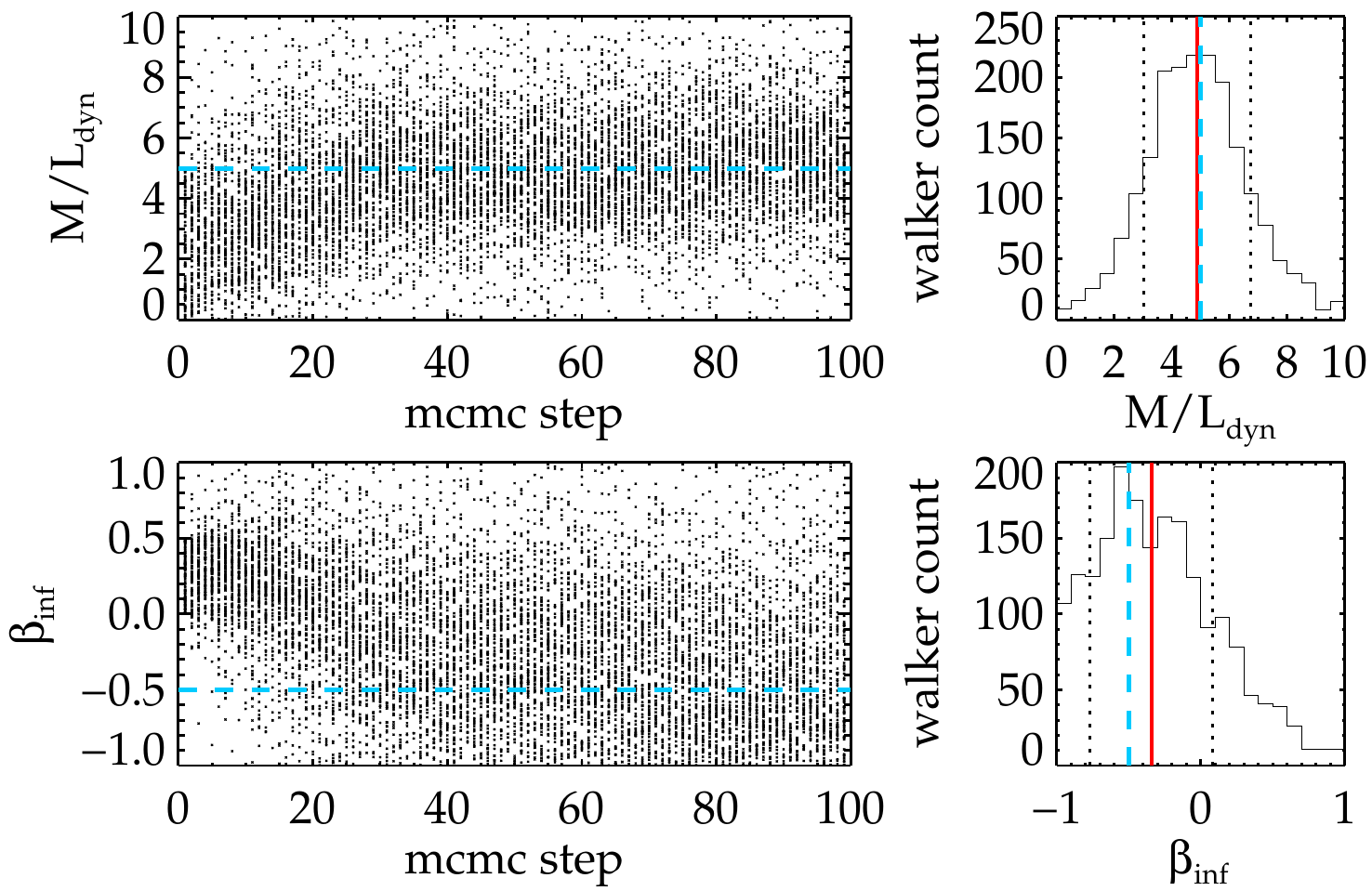}
\caption{As in Figure~\ref{mcmc-appendix_1} but for models with input anisotropy $\beta$=-0.5.}
\label{mcmc-appendix_2}
\end{figure*}

\begin{table*}
\caption{Tabulated results of the mcmc model tests. We created 2 models (with $\beta$=0.0 and $\beta$=-0.5) that were perturbed with relatively small (giant-like) and large (dwarf-like) error spectrum. Each of the resulting 4 model datasets was then used as input to our mcmc machinery and recovered twice, once assuming correct and once incorrect inclination.}
\centering
\begin{tabular}{|L|r|r|r|r|r|r|}
\hline
model type & \multicolumn{2}{c|}{model pars} & assumed $i$ & \multicolumn{2}{c|}{recovered values+errors} \\
\cline{2-3}
\cline{5-6}
           & M/L & $\beta$                  &             & M/L & $\beta$                         \\
\hline
high $V_{rms}/\Delta V_{rms}$ \newline($\sim$massive giants) &4.37& 0.0&correct&4.42$\pm$0.27&-0.02$\pm$0.13\\
						       &4.37& 0.0&90 deg &4.48$\pm$0.28&-0.07$\pm$0.11\\
						       &4.98&-0.5&correct&5.02$\pm$0.25&-0.53$\pm$0.20\\
						       &4.98&-0.5&90 deg &5.11$\pm$0.28&-0.46$\pm$0.14\\
\hline
low $V_{rms}/\Delta V_{rms}$ \newline ($\sim$ our dwarfs) &4.37& 0.0&correct&4.67$\pm$1.73&-0.18$\pm$0.45\\
						    &4.37& 0.0&90 deg &4.85$\pm$1.88&-0.27$\pm$0.48\\
						    &4.98&-0.5&correct&5.09$\pm$1.70&-0.31$\pm$0.43\\
						    &4.98&-0.5&90 deg &4.86$\pm$1.85&-0.34$\pm$0.42\\
\hline
 \end{tabular}
\label{mctable}
\end{table*}

\end{document}